\documentclass[prx,twocolumn,groupedaddress]{revtex4-2}
\usepackage{amsmath,amsfonts,amssymb,graphicx,xcolor,ulem}

\newcommand{\ve}[1]{\mathbf{#1}}

\begin{document}

% Use the \preprint command to place your local institutional report
% number in the upper righthand corner of the title page in preprint mode.
% Multiple \preprint commands are allowed.
% Use the 'preprintnumbers' class option to override journal defaults
% to display numbers if necessary
%\preprint{}

%Title of paper
%\title{Computing Spatially Dependent Memory Corrections to the Generalized Langevin Equation in the Presence of a Potential of Mean Force}
\title{ Derivation of the  non-equilibrium  generalized Langevin equation from a generic  time-dependent Hamiltonian }

% repeat the \author .. \affiliation  etc. as needed
% \email, \thanks, \homepage, \altaffiliation all apply to the current
% author. Explanatory text should go in the []'s, actual e-mail
% address or url should go in the {}'s for \email and \homepage.
% Please use the appropriate macro foreach each type of information

% \affiliation command applies to all authors since the last
% \affiliation command. The \affiliation command should follow the
% other information
% \affiliation can be followed by \email, \homepage, \thanks as well.
\author{Roland R. Netz}
%\email[]{Your e-mail address}
%\homepage[]{Your web page}
%\thanks{}
%\altaffiliation{}
\affiliation{Fachbereich Physik, Freie Universit\"at Berlin, 144195 Berlin, Germany}
\affiliation{Centre for Condensed Matter Theory, Department of Physics, Indian Institute of Science, Bangalore 560012, India}

%Collaboration name if desired (requires use of superscriptaddress
%option in \documentclass). \noaffiliation is required (may also be
%used with the \author command).
%\collaboration can be followed by \email, \homepage, \thanks as well.
%\collaboration{}
%\noaffiliation

\date{\today}

\begin{abstract}
% insert abstract here: PRX < 500 words, 248 words
It has been become standard practice to describe steady-state non-equilibrium phenomena 
by Langevin equations with colored noise and time-dependent  friction kernels
 that do not obey the fluctuation-dissipation theorem, but since these Langevin
 equations are typically not derived from first-principle Hamiltonian dynamics it is not clear
 whether they correspond to physically realizable scenarios.
 By exact Mori projection in phase space we  derive the non-equilibrium generalized Langevin equation (GLE) 
from a generic many-body Hamiltonian with a time-dependent  force h(t)  acting on  an arbitrary  
phase-space dependent  observable $A$. The GLE is obtained in explicit form to all orders in $h(t)$. 
For non-equilibrium  observables that correspond to  a Gaussian process, the resultant GLE has the same 
form as the equilibrium Mori GLE, in particular the memory kernel is proportional to the  total force autocorrelation function.
This means that the extraction and simulation methods 
developed for equilibrium GLEs can be used also for non-equilibrium Gaussian variables. 
This is a non-trivial and very useful  result, as many observables that characterize non-equilibrium systems 
 display Gaussian statistics. For non-Gaussian non-equilibrium variables 
correction terms appear in the GLE and in the relation between the complementary force autocorrelations
and the memory kernels,  which are explicitly given in term of cubic correlation functions of $A$. 
Interpreting the time-dependent  force h(t)  as a stochastic process, 
we  derive  non-equilibrium corrections to the
 fluctuation-dissipation theorem and 
 methods to extract all GLE parameters from experimental or simulation 
data, thus making our non-equilibrium GLE  a practical tool to study and model
general non-equilibrium systems.

  \end{abstract}

% insert suggested keywords - APS authors don't need toergebnisse erhalten do this
%\keywords{Statistical Physics, Non-Markovian Effects, Spatially Dependent Memory}

%\maketitle must follow title, authors, abstract, and keywords
\maketitle

%Popular Summary: Physical Review X requires authors to submit a nontechnical summary that conveys the context, the essential message(s), and the significance of the work to all readers. The summary should be concise (approximately 250 words), readable, objective, and have broad appeal. Please avoid including mathematical expressions.

% body of paper here - Use proper section commands
% References should be done using the \cite, \ref, and \label commands
\section{Introduction}

The statistical mechanics foundation of non-equilibrium phenomena has occupied physicists  for  many decades
\cite{Prigogine1947,Mazur1953,Lebowitz1959,zwanzig_ensemble_1960,deGroot,grabert_microdynamics_1980,Risken,zwanzig_nonequilibrium_2001}.
More recently, new experimental techniques, such as single-molecule and  optical methods,
applied to non-equilibrium biological systems have accented the need for theories that are able to deal with 
non-equilibrium experiments and data
\cite{Schmidt07,Gawedzki2009,Bechinger2010,Bocquet2012,Dinis2012,Bohec2013,Weitz2014}. 
At the same time, novel  theoretical approaches were developed and applied to  non-equilibrium driven lattice models
\cite{Krug,Zia,Derrida2001}, interacting non-equilibrium particle systems \cite{Barrat2007,Grosberg2015,Fodor2016},
 non-equilibrium barrier-crossing phenomena \cite{Carlon2018,Lavacchi2022}
and used to derive
non-equilibrium work and entropy relations \cite{Jarzynski2000,Hatano2001,Harada2005,Seifert2005,Godec2020},
generalized fluctuation-dissipation relations \cite{Prost2009,Wynants2009,Seifert2010,Lindner2017,Netz2018}
and  non-equilibrium entropy-production extremal principles  \cite{Netz2020}.
   
The generalized Langevin equation (GLE)  has played a key role
in the development of methods to deal with the dynamics of complex systems, as it is an exact equation of motion
for an  observable derived by projection
 from the many-body Hamiltonian, the GLE thus constitutes a method for exact coarse-graining of a 
Hamiltonian system
\cite{nakajima_quantum_1958,zwanzig_memory_1961,mori_transport_1965,Ciccotti1981,Straub_1987,lange_collective_2006,
kinjo_equation_2007,darve_computing_2009,hijon_morizwanzig_2010,izvekov_microscopic_2013,lee_multi_2019,
Ayaz2022,vroylandt_likelihood_2022}. 
The GLE was applied to protein folding 
\cite{ plotkin_non-markovian_1998,satija_generalized_2019,ayaz_non-markovian_2021,Dalton_2023},
barrier crossing dynamics
\cite{Bagchi_1983,Straub_1986,Pollak_1989,Carlon2018,Brunig_2022d},
motion of living cells \cite{Mitterwallner_2020},
spectroscopy \cite{Tuckerman1993,Gottwald2015,Brunig_2022a},
dynamical neworks \cite{Sollich2020} 
and data prediction \cite{chorin_optimal_2000}.
While the standard GLE formulations describe the motion of an observable in phase space 
and thus allow to quantify the approach of a non-equilibrium state  to equilibrium, they do not apply
to driven non-equilibrium system as described by a time-dependent Hamiltonian. Many works
dealt with generalizations of the  projection framework to time-dependent and
transient scenarios \cite{Robertson1966,Zwanzig1975,Picard1977,Uchiyama1999,Koide2002,Latz2002,
meyer_non-stationary_2017,Cui2018,Vrugt2019,meyer_non-markovian_2020}.
Non of these works dealt with the non-equilibrium  Hamiltonian system considered in this paper
and derived the non-equilibrium GLE in closed form. Since most of the aforementioned theories concerned
with non-equilibrium phenomena break the fluctuation-dissipation theorem more or less by hand, it 
is instructive to derive non-equilibrium  equations of motion for observables from 
time-dependent  Hamiltonians. This enables to check which non-equilibrium effective equations correspond
to an underlying Hamiltonian non-equilibrium dynamics and which do not. 
This is the vantage point of this paper.

  In the presence of a time-dependent force $h(t)$, a Hamiltonian  system is  generally out of equilibrium
  since the force performs work on the system. In fact,
even for constant force  $h(t)=h_0\neq 0$ such a  system is out of equilibrium if the observable $A$  is unconfined and thus
driven into a steady-state motion by the constant force, as  will be explained below.
For an  equilibrium system, i.e. for $h(t)=0$,  the  Mori GLE for the  observable $A(t)$  reads
\cite{mori_transport_1965} 

\begin{align} \label{eq_mori_GLE0}
 \ddot A(t)  = -  K    (A(t) - \langle  A \rangle)  
 - \int_{t_0}^{t} {\rm d}s\, \Gamma(t-s)  \dot A(s)  + F(t),
\end{align}
where the stiffness of the effective harmonic potential is denoted as $K$, 
the time-dependent  friction memory kernel as $\Gamma(t)$ and the complementary force
as $F(t)$.  
The time at which the projection is done is denoted as $t_0$ and the friction kernel
 $\Gamma(t)$  is related to the complementary force autocorrelation function via
 \cite{mori_transport_1965} 
\begin{align} \label{eq_mori_GLE0b}
\Gamma(t-s) =
 \frac{\langle F(s) F(t) \rangle}
 {\langle \dot A^2(t) \rangle}
\end{align}
(all averages are phase-space averages, as will be detailed below). 
Note that $F(t)$ is often denoted and treated as a random force, 
this is an approximation  since  $F(t)$ is in fact
  a phase-space dependent  deterministic function
and  fulfills well-defined  initial conditions at $t=t_0$,
Eq. \eqref{eq_mori_GLE0} is thus  deterministic and fully  time
reversible. The Mori GLE in Eq. \eqref{eq_mori_GLE0} is also exact
even for non-linear systems, unless the complementary force $F(t)$
is approximated as Gaussian. 

In this paper we derive the non-equilibrium generalized Langevin equation (GLE) for an arbitrary  phase-space dependent
observable $A$,  governed by a  general  many-body Hamiltonian that includes a time-dependent external force $h(t)$
 acting  on $A$, the derivation is exact to all orders of $h(t)$.  
 This  specific  time-dependent Hamiltonian is of high relevance, as it forms the starting point 
 for the   text-book derivation of the standard  fluctuation-dissipation theorem (FDT)  \cite{zwanzig_nonequilibrium_2001},
  %
%\begin{align}
%\label{intro}
%\langle  \Delta A(t) \rangle &=
 %\int_{-\infty}^\infty   {\rm d}s   h(s) \chi(t-s) +{\cal O}(h^2), 
   %  \end{align}%
which is one of the corner stones of statistical mechanics. 
Thus the standard FDT and our non-equilibrium GLE are  intimately
connected since they stem from the same Hamiltonian.

A key  question we address in this paper 
 is whether in the presence of an external time-dependent  force $h(t)$ acting on $A$,
the GLE in  Eq. \eqref{eq_mori_GLE0} and the relation between the friction kernel  and the 
complementary force autocorrelation in Eq. \eqref{eq_mori_GLE0b} still  hold.
Indeed, one  main result of this paper is that the simple forms of  Eqs. \eqref{eq_mori_GLE0} and \eqref{eq_mori_GLE0b}
indeed remain valid if the observable $A$ corresponds to  a Gaussian process
and if $F(t)$ is replaced by the combination of $F(t)$ and  $h(t)$. 
This 
is a non-trivial and relevant finding, since many biological non-equilibrium processes,
such as the motion of cells, are  Gaussian   to high accuracy \cite{Mitterwallner_2020}.
Conversely, for a non-Gaussian observable
$A$,  correction terms appear in  Eqs. \eqref{eq_mori_GLE0} and \eqref{eq_mori_GLE0b} 
that are explicitly given in terms of three-point (and higher-order)
correlation functions of $A$.
These explicit results allow for explicit prediction of the non-equilibrium correction terms based
on experimental or  simulation time-series data.
Our derivation of the GLE is non-perturbative and thus exact to all orders in the non-equilibrium force $h(t)$. 
For the special case of  a stochastic non-equilibrium force $h(t)$ that is  defined by its second moment, 
we derive a generalized non-equilibrium FDT which  
in the limit $h(t) \rightarrow 0$ simplifies to the standard FDT. 
 Finally, we give explicit formulas for extracting the parameters of our non-equilibrium 
GLE from simulation or experimental time series data, opening the route to the accurate 
and data-based modeling of non-equilibrium systems.

Sect.  \ref{GLE} contains the full derivation of the non-equilibrium GLE, this section can 
be skipped by a reader not interested in technical details. 
In Sect. \ref{NEQGLE} the non-equilibrium GLE is discussed and the role 
of non-Gaussian observable fluctuations is explained. 
In Sect. \ref{Sect_Stoch} the non-equilibrium force $h(t)$ is treated as a stochastic variable,
which restores time-homogeneity and simplifies the analysis of the GLE. Here 
the non-equilibrium FDT is derived. 
Sect. \ref{summary} presents a short discussion and an outlook.

\section{Derivation of the  non-equilibrium generalized  Langevin equation   \label{GLE}}

\subsection{ Definition of the time-dependent Hamiltonian   and solution of the Liouville equation}

We consider a  time-dependent  Hamiltonian for  a system of $N$ interacting particles or atoms  in three-dimensional space
with a  time-dependent force $h(t)$ that couples to a  generic phase-space dependent observable $A_S(\omega)$,
\begin{align}
\label{eq_Hamiltonian}
H (\omega,t) &= H_0 (\omega) - h(t) A_S(\omega),
\end{align}
 the subscript distinguishes this Schrödinger-like, i.e. time-independent, observable from the time-dependent 
 Heisenberg observable that will be introduced shortly. Although not really needed for our derivation, 
the time-independent Hamiltonian $H_0 (\omega)$  can be split into kinetic  and potential contributions according to 
\begin{align}
\label{eq_H0}
H_0 (\omega) &= \sum_{j=1}^{3N} \frac{P_j^2}{2 m_j} + V ( \ve R)
\end{align}
with coordinate-dependent masses $m_j$ 
and where the potential $V ( \ve R)$ contains all interactions between the particles and 
includes possible external potentials. 
A point in  6$N$-dimensional  phase space  is denoted by 
$\omega = (  \ve R ,  \ve P  )$, which is a $6N$-dimensional  vector containing  the Cartesian particle positions  $\ve R$ 
 and the  conjugate momenta  $ \ve P$ and fully specifies the microstate of the system.

Using  the time-dependent  Liouville operator
\begin{align}
\label{eq_Liouville}
{\cal L}(\omega,t)&= \sum_{j=1}^{3N}  
\left(   \frac{ \partial H(\omega,t) }{\partial P_j }  \frac{ \partial  }{\partial R_j }  -
 \frac{ \partial H (\omega,t)}{\partial R_j }  \frac{ \partial  }{\partial P_j } 
 \right),
\end{align}
the  $6N$-dimensional  Hamilton equation of motion can be compactly written as
%\begin{align}
%\label{eq_time_evolution}
$\dot{\omega}(t) = {\cal L}(\omega,t)  \omega(t)$,
%\end{align}
where $\omega(t)$ is the  phase-space  location of the system at time $t$ 
and $\dot{\omega}(t) = {\rm d} \omega(t) /   {\rm d} t$  is the corresponding phase-space velocity.
Instead of following   microstate trajectories in phase space, 
which is the Lagrangian description of the system dynamics,
it is much more convenient to switch to the Eulerian description and
consider the time dependent probability density distribution as a function of 
the time-independent phase-space position, $\rho(\omega,t)$,
which  obeys the  Liouville equation 
\begin{align}
\label{eq_Liouvilleeq}
\dot{\rho}(\omega,t) & = - {\cal L} (\omega,t)  \rho(\omega,t).
\end{align}
%
%From the Hamilton equation  and since the Liouville operator is time-independent, 
%it follows that the phase space position  is propagated in time by the  exponential operator 
%$e^{(t-t_0){\cal L}(\omega)}$, i.e., $e^{(t-t_0){\cal L}(\omega)}\omega(t_0) = \omega(t)$.
%\begin{subequations}
%\begin{align}
%\Phi(t): \Omega &\to \Omega\\
%\omega_0 &\mapsto \omega_t\\
%\Phi(t) &= e^{tL}.
%\end{align}
%\end{subequations}
In all what follows we suppress the dependence of the Liouville operator on phase space. 
We observe that a recursive solution of Eq. \eqref{eq_Liouvilleeq} can be 
written as 
 \begin{align}
\label{eq_LiouRec}
{\rho}(\omega,t) & =  \rho(\omega,t_0)  - \int_{t_0}^t  {\rm d} t_1    {\cal L} (t_1)  \rho(\omega,t_1).
\end{align}
By iteration  the following exact solution is obtained
 \begin{align}
\label{eq_LiouSol}
{\rho}(\omega,t) & = \exp_S\left( - \int_{t_0}^t  {\rm d}s  {\cal L} (s) \right)   \rho(\omega,t_0),
\end{align}
which depends on the initial density distribution at time $t_0$ and 
where  the time-ordered operator exponential in the Schrödinger picture has been introduced as 
 \cite{dyson_radiation_1949, feynman_operator_1951}
\begin{widetext}
\begin{align}
\label{eq_expS}
\exp_S\left( - \int_{t_0}^t  {\rm d}s  {\cal L} (s) \right) & \equiv
1+\sum_{n=1}^\infty (-1)^n 
 \int_{t_0}^t  {\rm d}t_1 {\cal L} (t_1)    \int_{t_0}^{t_1}  {\rm d}t_2 {\cal L} (t_2)   \int_{t_0}^{t_2}  {\rm d}t_3 {\cal L} (t_3)   \cdots 
   \int_{t_0}^{t_{n-1}}   {\rm d}t_n {\cal L} (t_n).
 \end{align}
\end{widetext}
For a time-independent Liouville operator ${\cal L} (t)= {\cal L}_0$, all time integrals can be done and one obtains
the solution in the terms of the  standard operator exponential 
 \begin{align} \label{opexp}
{\rho}(\omega,t) & = \exp \left( - (t-t_0)   {\cal L}_0  \right)   \rho(\omega,t_0),
\end{align}
where the exponential of an operator is defined by its ordinary  series expansion.

\subsection{From Schrödinger to Heisenberg observables}

A system observable can be generally written as a Schrödinger-type phase-space function $A_{S}(\omega)$,
it  can for example represent  the position of one particle,
the center-of-mass position of a group of particles or of a molecule, the reaction coordinate describing  a chemical reaction or the  folding of a protein.
To simplify the notation,  we consider a scalar observable but note that 
the   formalism can be straightforwardly 
extended  also to multi-dimensional observables. 
Using the  probability  density ${\rho}(\omega,t)$, 
the time-dependent expectation value (or mean) of the observable $A_{S}(\omega)$ can be written as
\begin{align}
\label{eq_mean}
a(t) &  \equiv \int {\rm d} \omega  \, A_{S}(\omega) \rho(\omega,t) \nonumber \\
&=  \int {\rm d} \omega  \, A_{S}(\omega) 
\exp_S\left( - \int_{t_0}^t  {\rm d}s  {\cal L} (s) \right)   \rho(\omega,t_0).
\end{align}
Since the Liouville operator is anti-self adjoint, it follows that   \cite{zwanzig_nonequilibrium_2001}
\begin{align}
\label{eq_mean2}
a(t) &=  \int {\rm d} \omega  \, \rho(\omega,t_0)     A(\omega, t),
\end{align}
where we have defined the Heisenberg observable as 
\begin{align}
\label{eq_Heisenberg}
A(\omega, t)  &\equiv   \exp_H\left(  \int_{t_0}^t  {\rm d}s  {\cal L} (s) \right)      A_{S}(\omega)
\end{align}
using the time-ordered operator exponential in the Heisenberg picture (or Heisenberg propagator)
\begin{widetext}
\begin{align}
\label{eq_expH}
\exp_H\left(  \int_{t_0}^t  {\rm d}s  {\cal L} (s) \right) & \equiv
1+\sum_{n=1}^\infty 
 \int_{t_0}^t  {\rm d}t_1    \int_{t_0}^{t_1}  {\rm d}t_2  \cdots  \int_{t_0}^{t_{n-1}}  {\rm d}t_n   {\cal L} (t_n)  \cdots   {\cal L} (t_2)  {\cal L} (t_1).
  \end{align}
\end{widetext}
Obviously, as follows from Eqs.~\eqref{eq_Heisenberg} and \eqref{eq_expH},
the Heisenberg observable  satisfies the equation of motion 
\begin{align}
\label{eq_Heisenberg2}
\dot{A}(\omega, t) & =  \frac{ {\rm d}A(\omega, t) }{ {\rm d}t}=   \exp_H\left(  \int_{t_0}^t  {\rm d}s  {\cal L} (s) \right)     {\cal L}(t) A_{S}(\omega)
\end{align}
with the initial condition    $A(\omega, t_0)=A_{S}(\omega)$.
As  derived in Appendix   \ref{sec_App_Propagator}, the Heisenberg observable
 also satisfies the initial differential boundary condition 
\begin{align}
\label{eq_Heisenberg3}
 \frac{ {\rm d}A(\omega, t) }{ {\rm d}t_0} & = -  {\cal L} (t_0)    \exp_H\left(  \int_{t_0}^t  {\rm d}s   {\cal L}(s)  \right)    A_{S}(\omega),
\end{align}
which will be later needed to derive operator expansions. 

To understand the meaning of a Heisenberg observable, we for the moment consider the 
initial density distribution $\rho(\omega,t_0)=\delta(\omega-\omega_0)$,
which describes a system that at time $t_0$ is in the microstate $\omega_0$. 
Inserting this into Eq.~\eqref{eq_mean2},  we obtain $a(t) = A(\omega_0, t)$.
In other words, $A(\omega_0, t)$ describes the time-dependent mean of an observable for a system  that at time $t=t_0$ was 
in the microstate $\omega_0$, i.e., it describes the temporal evolution of the conditional mean of the  observable  $A_{S}(\omega)$. 
It transpires that if we  derive an equation of motion for $A(\omega, t)$,
we have an equation for how this conditional mean changes in time.  
This is the central idea of projection and of GLEs \cite{nakajima_quantum_1958,zwanzig_memory_1961,mori_transport_1965}.

Taking another time derivative of Eq.~\eqref{eq_Heisenberg2}, we obtain for the acceleration of the observable 
\begin{align}
\label{eq_Heisenberg4}
\ddot{A}(\omega, t) & =   \exp_H\left(  \int_{t_0}^t  {\rm d}s  {\cal L} (t) \right)  
\left(    {\cal L}^2(t)  + \dot{\cal L}(t) \right)   A_{S}(\omega).
\end{align}
Up to now the discussion applied to a general time-dependent Hamiltonian; for the specific Hamiltonian 
Eq.~\eqref{eq_Hamiltonian}, where a time dependent force $h(t)$ multiplies the observable $A_S(\omega)$,
the Liouville operator splits into two parts
\begin{align}
\label{eq_Liouville2}
{\cal L}(t)&= {\cal L}_0-h(t) \Delta {\cal L},
\end{align}
with the unperturbed Liouville operator given by 
\begin{align}
\label{eq_Liouville3}
{\cal L}_0&= \sum_{j=1}^{3N}  
\left(   \frac{ \partial H_0(\omega) }{\partial P_j }  \frac{ \partial  }{\partial R_j }  -
 \frac{ \partial H_0 (\omega)}{\partial R_j }  \frac{ \partial  }{\partial P_j } 
 \right),
\end{align}
and the perturbation Liouville operator given by 
\begin{align}
\label{eq_Liouville4}
\Delta {\cal L}&= \sum_{j=1}^{3N}  
\left(   \frac{ \partial A_S(\omega) }{\partial P_j }  \frac{ \partial  }{\partial R_j }  -
 \frac{ \partial A_S (\omega)}{\partial R_j }  \frac{ \partial  }{\partial P_j }   \right).
\end{align}
These operators have the important properties 
${\cal L}_0 H_0(\omega)=0$,
$\Delta {\cal L}  A_S(\omega)=0$, and
$\Delta {\cal L} H_0(\omega)=- {\cal L}_0 A_S(\omega)$,
from which we derive, using   
Eqs.~\eqref{eq_Heisenberg2}  and \eqref{eq_Heisenberg4},
 the simplified expressions for the observable velocity and acceleration
\begin{align}
\label{eq_Heisenberg5}
\dot{A}(\omega, t) & =    \exp_H\left(  \int_{t_0}^t  {\rm d}s  {\cal L} (s) \right)     {\cal L}_0 A_{S}(\omega),
\end{align}
\begin{align}
\label{eq_Heisenberg6}
\ddot{A}(\omega, t) & =   \exp_H\left(  \int_{t_0}^t  {\rm d}s  {\cal L} (s) \right)  
   {\cal L}(t)  {\cal L}_0   A_{S}(\omega).
\end{align}
The fact that the velocity Eq.~\eqref{eq_Heisenberg5}
exhibits no time dependence to the right of the operator exponential is crucial, as it will 
later on allow us to use time-independent projection for the derivation of the  non-equilibrium
GLE.

\subsection{ Projection}

Here we follow standard procedures  \cite{nakajima_quantum_1958,zwanzig_memory_1961,mori_transport_1965,zwanzig_nonequilibrium_2001}. 
We introduce a time-independent  projection operator ${\cal P}$ that acts on a phase space function
 and its complementary operator ${\cal Q}$ via the relation $1= {\cal Q} + {\cal P}$.
 Inserting this unit operator into the acceleration  Eq.~\eqref{eq_Heisenberg6},  we obtain 
\begin{align}
\label{eq_GLE1}
  \ddot A(\omega, t) &= \exp_H\left(  \int_{t_0}^t  {\rm d}s  {\cal L} (s) \right)  ( {\cal P}+ {\cal Q})   
  {\cal L}(t)  {\cal L}_0   A_{S}(\omega)  \nonumber \\
&= \exp_H\left(  \int_{t_0}^t  {\rm d}s  {\cal L} (s) \right)  {\cal P}   {\cal L}(t)  {\cal L}_0   A_{S}(\omega) \nonumber \\
&+ \exp_H\left(  \int_{t_0}^t  {\rm d}s  {\cal L} (s) \right)  {\cal Q}   {\cal L}(t)  {\cal L}_0   A_{S}(\omega),
\end{align}
where we used that the Heisenberg propagator is a linear operator. 
The projection is performed at  time $t_0$ at which the time propagation  starts 
(the relevance of this will become clear  later).
By inserting the time-dependent Dyson operator 
expansion  \cite{dyson_radiation_1949, feynman_operator_1951,nakajima_quantum_1958,zwanzig_memory_1961,mori_transport_1965}
 for  the Heisenberg propagator (see Appendix \ref{sec_App_opexp} for a derivation) 
\begin{align}
\label{eq_GLE2}
&\exp_H\left(  \int_{t_0}^t  {\rm d}s  {\cal L} (s) \right)  
= \exp_H\left(  {\cal Q} \int_{t_0}^t  {\rm d}s  {\cal L} (s) \right) +  \\
&\int_{t_0}^t  {\rm d}s  \exp_H\left(  \int_{t_0}^s  {\rm d}s'  {\cal L} (s') \right) 
 {\cal P}  {\cal L} (s)
\exp_H\left(  {\cal Q} \int_{s}^t  {\rm d}s'  {\cal L} (s') \right) \nonumber
\end{align}
into the second term on the right hand side  in Eq.~\eqref{eq_GLE1},
we obtain the GLE in general form 
\begin{align}
\label{eq_GLE3}
\ddot A(\omega, t) & = 
\exp_H\left(  \int_{t_0}^t  {\rm d}s  {\cal L} (s) \right)  {\cal P}   {\cal L}(t)  {\cal L}_0   A_{S}(\omega)
+F(\omega,t_0,t)   \nonumber \\
&+ \int_{t_0}^t  {\rm d}s   \exp_H\left(  \int_{t_0}^s  {\rm d}s'  {\cal L} (s') \right) 
 {\cal P}  {\cal L} (s) F(\omega,s,t),
\end{align}
where the complementary  force is defined as
\begin{align}
\label{eq_F_operator}
F(\omega,t_0,t)&  \equiv  
\exp_H\left(  {\cal Q} \int_{t_0}^t  {\rm d}s'  {\cal L} (s') \right)
 {\cal Q}   {\cal L}(t)  {\cal L}_0   A_{S}(\omega).
\end{align}
The first term on the right-hand side in Eq.~\eqref{eq_GLE3}
will turn out to represent the  conservative force from a  potential,
the third term represents  friction and non-Markovian effects  and the force $F(\omega,t_0,t)$ 
represents all effects that are not included in the other two terms.   
 $F(\omega,t_0,t)$  is a function of  phase space and evolves in the complementary space,
 i.e. it satisfies $ {\cal P} F(\omega,t_0,t)=0$ (as will be explained further below). 
 While we suppress the $t_0$ dependence of the observable $A(\omega, t)$ and its derivatives,
 which can cause no confusion since this argument is invariant throughout most of  the  calculation, 
 the complementary  force $F(\omega,t_0,t)$  needs both  time arguments since
 both arguments are varied in Eq.~\eqref{eq_GLE3}.

 Clearly, the explicit form of Eq.~\eqref{eq_GLE3} depends on the specific  projection operator 
 $\cal P$.  Here we choose the Mori projection, because it is most straightforward to implement
 and our main result concerning the effect of non-Gaussian observables on the structure of the non-equilibrium GLE 
 is accurately and transparently produced by Mori projection. We note in passing that
 the Mori GLE is exact even for non-Gaussian observables, unless one approximates the complementary
 force distribution. 
The Mori projection applied on a general Heisenberg  observable $B(\omega,t)$ using
the  Schrödinger observable $A_S(\omega) $ as a projection function is given by \cite{mori_transport_1965}
\begin{align}
\label{eq_mori_projection}
 &{\cal P}  B(\omega,t) = \langle   B(\omega,t)  \rangle
  +  \frac{\langle   B(\omega,t)  {\cal L}_0 A_S(\omega)  \rangle}  {\langle ( {\cal L}_0 A_S(\omega) )^2 \rangle} 
  {\cal L}_0 A_S (\omega)  \nonumber \\
 &  +\frac{\langle   B(\omega,t)  (A_S (\omega) - \langle  A_S \rangle )  \rangle}  
 {\langle (A_S (\omega) - \langle  A_S \rangle )^2 \rangle} 
 (A_S (\omega) - \langle  A_S \rangle ).
  \end{align}
Here we have defined the expectation value of an arbitrary phase-space function $X(\omega) $ 
  with respect to a time-independent projection  distribution 
$ \rho_{\rm p}(\omega)$ as
\begin{align}  \label{eq_average}
\langle   X(\omega)  \rangle = 
\int {\rm d}\omega   X (\omega) \rho_{\rm p}(\omega),
  \end{align}
  which we here take to be the equilibrium canonical distribution
  of the  time-independent Hamiltonian 
 \begin{align} \label{HamP}
  &   \rho_{\rm p}(\omega)= e^{- \beta H_0(\omega) + \beta h_p A_S(\omega) }/ Z,
      \end{align}
  where  $Z$ is the partition function. The factor $\beta$ has units of inverse energy and can 
  be thought of as the inverse thermal energy characterizing the projection distribution. 
  Note that for generality  we added a linear force $h_p$ in the projection Hamiltonian.
 % For clarity of presentation we occasionally use the short-hand notation $\hat X = \langle X(\omega) \rangle$ to denote
%phase-space independent averages over the projection distribution. 
Time-dependent projection has been  used to derive generic non-equilibrium GLEs
\cite{Robertson1966,Zwanzig1975,Picard1977,Uchiyama1999,Koide2002,Latz2002,
meyer_non-stationary_2017,Cui2018,Vrugt2019,meyer_non-markovian_2020},
but is not needed here because of the specific form of our time-dependent Hamiltonian. 
The  time-independent Mori  projection in Eq.~\eqref{eq_mori_projection} 
projects onto a constant,  the Schrödinger observable $A(\omega, t_0)= A_S (\omega)$ 
and its time derivative $ \dot A(\omega, t_0)= {\cal L}_0 A_S (\omega)$. 
Thus the projection in Eq.~\eqref{eq_mori_projection} maps any observable $B(\omega,t)$
 onto the subspace of all functions linear in the observables 
 1, $A_S (\omega)$ and $ {\cal L}_0 A_S (\omega)$, meaning that 
 ${\cal P}1=1$, ${\cal P}A_S (\omega)= A_S (\omega)$ and 
 ${\cal P}{\cal L}_0 A_S (\omega)= {\cal L}_0 A_S (\omega)$. From this follows immediately 
 that ${\cal Q}1= {\cal Q} A_S (\omega)={\cal Q}{\cal L}_0 A_S (\omega)= 0$, which are important properties.
 In particular, it follows that 
 several expectation values involving  the  complementary   force vanish, namely 
  $\langle F(\omega,t_0,t)\rangle = \langle   F(\omega,t_0,t) A_S(\omega) \rangle =
  \langle   F(\omega,t_0,t){\cal L}_0 A_S(\omega,) \rangle =0$, which will
  be used to extract GLE parameters from non-equlibrium time-series data,

The Mori projection is  linear,
 i.e., for two arbitrary observables $B(\omega,t)$ and $C(\omega,t')$ it satisfies
 ${\cal P} (c_1 B(\omega,t) + c_2 C(\omega,t'))=c_1 {\cal P} B(\omega,t)  + c_2 {\cal P} C(\omega,t')$,
 it is idempotent, i.e.,  ${\cal P}^2= {\cal P}$, and it is self-adjoint, i.e. it satisfies 
  the relation
\begin{align}
\langle C(\omega,t) {\cal P} B(\omega,t')  \rangle = \langle B(\omega,t') {\cal P} C(\omega,t)   \rangle.
\label{eq_projection_orthogonal}
\end{align} 
From these properties it follows that the complementary  projection
 operator ${\cal Q}=1-{\cal P}$ is also linear,  idempotent and self-adjoint.
Thus, ${\cal P}$ and ${\cal Q}$ are orthogonal to each other, i.e. 
${\cal P}{\cal Q}= 0 = {\cal Q}{\cal P}$,
details  are shown in Appendix \ref{sec_App_idempotency}.

\section{Properties of the non-equilibrium Langevin equation} \label{NEQGLE}
\subsection{General properties}

Using the projection Eq.  \eqref{eq_mori_projection} in
 the generic  GLE  Eq.~\eqref{eq_GLE3}, we obtain the explicit GLE 
\begin{align} \label{eq_mori_GLE}
 & \ddot A(\omega, t)  = -  K(t)    (A(\omega,t) - \langle  A_S \rangle)  
 - \int_{t_0}^{t} {\rm d}s\, \Gamma(s,t)  \dot A(\omega, s)   \nonumber \\
& + \int_{t_0}^{t} {\rm d}s\, \Gamma_A (s,t) ( A(\omega, s) -  \langle  A_S \rangle)    \nonumber \\
& + F(\omega,t_0,t) +(h(t)-h_p) /M,
\end{align}
the details of the derivation are shown in Appendix \ref{sec_App_GLE}.
Eq.~\eqref{eq_mori_GLE} is an exact and explicit equation of motion for the Heisenberg observable
 $A(\omega, t)$ and is time-reversible, which is a consequence of the time-reversibility of the underlying Hamilton 
and Liouville equations.
% For practical applications, one typically models the  force $F(\omega,t)$ as a stochastic process with 
 %zero mean and a second moment given by eq.~\eqref{eq_mori_memory},
 % higher-order moments of $F(\omega,t)$ 
 % are typically neglected and the distribution of $F(\omega,t)$ is  assumed to be Gaussian. 
 %For non-linear  systems, however, this assumption can not hold, 
 Inspection of the GLE shows that $F(\omega,t_0,t)$ is   the only term (except $h(t)$)  in the GLE that accounts for  possible
   non-linearities (i.e. non-Gaussian contributions) 
 in $ A(\omega, t)$.
 Thus, imposing $F(\omega,t_0,t)$ to be a Gaussian variable   corresponds to a
 severe    approximation for non-linear systems. On the other hand,
 keeping the full  non-Gaussian contributions of $F(\omega,t_0,t)$ makes
 Eq.~\eqref{eq_mori_GLE} an exact description of the observable dynamics. 
 %which reflects a fundamental short-coming of the Mori projection scheme in conjunction with 
 %replacing $F(\omega,t)$ by a random Gaussian process.
 %Alternative methods to derive GLEs with non-linear potential and friction terms have been recently proposed
  %\cite{Ayaz2022,vroylandt_position_2022,rotenberg_use_2020,shi_new_2003}.
  %But we will stay here with the Mori scheme because it simplifies analytical calculations and because
  %some filtering applications will produce data that is Gaussian to a very good approximation. 

 The first term in Eq.~\eqref{eq_mori_GLE}  is a force 
 due to an effective  harmonic  potential with a 
   time-dependent potential stiffness  $K(t)$  given by 
\begin{align} \label{GLEK1}
K(t)  &= K_0 +  K_1(t)
\end{align}
with
\begin{align} \label{GLEK2}
 & K_0 = 
\frac{\langle ( {\cal L}_0 A_S(\omega) )^2 \rangle}
{\langle (A_S (\omega) - \langle  A_S \rangle )^2 \rangle},
\end{align}
\begin{align} \label{GLEK3}
 &  K_1(t) = 
\frac{  - \beta (h(t)-h_p)     \langle  (A_S (\omega) - \langle  A_S \rangle )  ({\cal L}_0 A_S(\omega) )^2 \rangle}
{\langle (A_S (\omega) - \langle  A_S \rangle )^2 \rangle}.
\end{align}
The second term  in Eq.~\eqref{eq_mori_GLE} accounts for linear friction and depends on the memory kernel  given by 
\begin{align}  \label{eq_mori_memory1}
\Gamma(s,t) = \Gamma_0(s,t) +   \Gamma_1(s,t)
\end{align}
with
\begin{align} \label{eq_mori_memory2}
\Gamma_0(s,t) =
 \frac{\langle F(\omega,s,s) F(\omega,s, t) \rangle}
 {\langle ( {\cal L}_0 A_S(\omega) )^2 \rangle},
\end{align}
\begin{align} \label{eq_mori_memory3}
 \Gamma_1(s,t) =
 \frac{  - \beta (h(s)-h_p)  \langle       F(\omega,s, t)  ( {\cal L}_0 A_S(\omega) )^2  \rangle}
 {\langle ( {\cal L}_0 A_S(\omega) )^2 \rangle}.
\end{align}
There is also a positional  memory term which is not present in the equilibrium
GLE in  Eq. \eqref{eq_mori_GLE0} and which 
 involves the kernel function
\begin{align} \label{eq_mori_memory4}
& \Gamma_A(s,t) = \\
& \frac{  \beta (h(s)-h_p)  \langle     F(\omega,s, t)   (A_S (\omega) - \langle  A_S \rangle )    {\cal L}_0 A_S(\omega)  \rangle}
 {\langle (A_S (\omega) - \langle  A_S \rangle )^2 \rangle}. \nonumber
\end{align}
The last two terms  in Eq.~\eqref{eq_mori_GLE}
are the complementary force $F(\omega,t_0,t) $ defined in Eq.  \eqref{eq_F_operator}
and the time-dependent force $h(t)$, where the mass is given by
\begin{align} \label{eq_mass}
& M=  \frac{1} {\beta \langle ( {\cal L}_0 A_S(\omega) )^2 \rangle}.
\end{align}
For vanishing force  $h(t)=0$, which renders the  equilibrium scenario,  and choosing $h_p=0$, we see that 
$ K_1(t) =  \Gamma_1(s,t) = \Gamma_A(s,t)=0$ and we thus  recover the standard
form of the equilibrium Mori GLE in Eq. \eqref{eq_mori_GLE0},
where  in particular the friction kernel $\Gamma_0(s,t) $ 
is via Eq. \eqref{eq_mori_GLE0b} related to the complementary force autocorrelation function
(note that in Eqs.  \eqref{eq_mori_GLE0} and  \eqref{eq_mori_GLE0b} we have suppressed the phase-space
dependence of the observable $A$ and of the complementary force $F$). 
In contrast, if 
$h(t)-h_p \neq 0$, we see that additional terms are present in the GLE and that
for $  \Gamma_1(s,t)  \neq 0$ the 
friction kernel   $ \Gamma(s,t)$  does not equal the complementary force autocorrelation function.
We first want to discuss whether
 a non-zero  $  \Gamma_1(s,t) $  necessarily  indicates  the presence of non-equilibrium effects.

An insightful scenario to address this question is one where the force $h(t)=h_0$ is constant, in which case the 
complementary force and all memory kernels become time homogeneous
and can be written as $F(\omega,s,t) = F(\omega,t-s)$, $\Gamma(s,t)=\Gamma(t-s)$, 
$\Gamma_A(s,t)=\Gamma_A(t-s)$. 
Let us first discuss  an unconfined system, i.e., a system characterized by a diverging second moment, 
$ \langle (A_S (\omega) - \langle  A_S \rangle )^2 \rangle=\infty$. In this case
$K=0= \Gamma_A(t-s)$ and  we must take $h_p=0$ in order
to have a bounded projection distribution $\rho_p(\omega)$  in  Eq. \eqref {HamP}. 
From Eq.  \eqref{eq_mori_memory3} we see that 
 $  \Gamma_1(t-s)$ can in general be non zero (as we will discuss in more detail
 in the next section), in which case the total  friction memory kernel  $  \Gamma(t-s)$
 does not equal the complementary force autocorrelation, reflecting that an unconfined system under
the influence of a constant force dissipates energy and thus is a non-equilibrium system. 
 In contrast, a confined system that is  characterized by a finite  second moment 
$ \langle (A_S (\omega) - \langle  A_S \rangle )^2 \rangle$, is in the presence of a constant force $h(t)=h_0$ an equilibrium system. 
But we see that   for $h_0 \neq h_p$  the terms 
$ K_1(t)$, $ \Gamma_1(s,t)$ and $ \Gamma_A(s,t)$  do not necessarily vanish.
In other words, unless we choose as the projection distribution in Eq. \eqref {HamP} the equilibrium
distribution, the  friction kernel in the Mori GLE $ \Gamma(s,t)$  not necessarily equals the complementary force autocorrelation. 
Thus, a non-zero $  \Gamma_1(s,t) $ not necessarily  indicates a driven non-equilibrium system,
but can also be produced by choosing a specific projection distribution $\rho_p(\omega)$
and characterize the approach of the system towards equilibrium. 
As we will discuss next,
 $  \Gamma_1(s,t) $ is predicted to  vanish for Gaussian non-equilibrium systems, 
 so the interpretation of   non-equilibrium GLEs has to be done with care. 
 Having made this important point, we from now on put $h_p=0$. 

  Note that  for equilibrium systems, the GLE parameters and the complementary   force   can
  be extracted from simulation or experimental time series data by various well-established techniques
 \cite{Straub_1987,carof_two_2014,jung_iterative_2017,daldrop_external_2017,daldrop_butane_2018,
 klippenstein_cross-correlation_2021,vroylandt_likelihood_2022,Ayaz2022}. 
 Similar  extraction techniques for  non-equlibrium time-series data
  will be discussed in Sec. \ref{Sect_Stoch}.

\subsection{Gaussian versus non-Gaussian observables } \label{Gauss}

In order to highlight the role played  by non-Gaussian fluctuations of the observable $A(\omega,t)$, 
we  slightly rewrite the  friction memory kernel in  
Eq. \eqref{eq_mori_memory1}  as
\begin{align}  \label{eq_mori_memory5}
 \Gamma(s,t) =  \Gamma_0(s,t) + \beta h(s) h(t)  /M +  \Gamma_2(s,t) 
\end{align}
with
\begin{align} \label{eq_mori_memory6}
 \Gamma_2(s,t) = 
  \frac{  - \beta h(s)  \langle     (  F(\omega,s, t)+h(t) /M)   ( {\cal L}_0 A_S(\omega) )^2  \rangle}
 {\langle ( {\cal L}_0 A_S(\omega) )^2 \rangle}.  
\end{align}
For the potential memory term in Eq.  \eqref{eq_mori_memory4} we choose  the modified form 
\begin{align} \label{eq_mori_memory7}
& \Gamma_A(s,t) = \\
& \frac{  \beta h(s)   \langle      (  F(\omega,s, t)+h(t)/M)  (A_S (\omega) - \langle  A_S \rangle )    {\cal L}_0 A_S(\omega)  \rangle}
 {\langle (A_S (\omega) - \langle  A_S \rangle )^2 \rangle}. \nonumber
\end{align}
From the GLE \eqref{eq_mori_GLE} we see that $F(\omega,s, t)+h(t)/M$ is linear in 
$A(\omega,\cdot)-\langle  A_S \rangle$ with a projection time 
given by $t_0=s$, we conclude from Eqs.  \eqref{eq_mori_memory6} and \eqref{eq_mori_memory7}
that $ \Gamma_2(s,t) $ and $ \Gamma_A(s,t)$ are proportional to expectation values that are 
(at least) cubic in  $A(\omega,\cdot)-\langle  A_S \rangle $. 
In other words, for an observable $A(\omega,t)-\langle  A_S \rangle$  that is Gaussian, the kernel functions
$ \Gamma_2(s,t) $ and $ \Gamma_A(s,t)$  (and also the potential stiffness correction
$K_1(t)$) vanish. In this case,
  we are thus led to the simplified   GLE, valid for Gaussian observables, 
\begin{align} \label{eq_mori_GLE5}
 & \ddot A(\omega, t)  = -  K_0    (A(\omega,t) - \langle  A_S \rangle)  
 - \int_{t_0}^{t} {\rm d}s\, \Gamma_G(s,t)  \dot A(\omega, s)   \nonumber \\
& + F(\omega,t_0,t) + h(t)/M
\end{align}
where the  Gaussian friction kernel is given by the autocorrelation of  the sum of the complementary and non-equilibrium forces 
according to 
\begin{align} \label{eq_mori_memory8}
&\Gamma_G (s,t) =
 \frac{\langle (F(\omega,s,s) +h(s)/M)(F(\omega,s, t) \rangle  + h(t) /M)}
 {\langle ( {\cal L}_0 A_S(\omega) )^2 \rangle}  \nonumber \\
&=  \frac{\langle F(\omega,s,s) F(\omega,s, t) \rangle  + h(t) h(s)/M^2}
 {\langle ( {\cal L}_0 A_S(\omega) )^2 \rangle} \nonumber \\
 &=  \frac{\langle F(\omega,s,s) F(\omega,s, t) \rangle  }
 {\langle ( {\cal L}_0 A_S(\omega) )^2 \rangle}
 + \beta h(t) h(s)/M.
\end{align}
In the derivation of Eq.  \eqref{eq_mori_memory8} we used that the non-equilibrium force $h(t)$ is phase-space independent. 
Thus,  we conclude that a Gaussian  non-equilibrium variable  is described by a GLE with a friction memory kernel
that via Eq. \eqref{eq_mori_memory8}  is related to the  autocorrelation of the total force acting on the observable. 
One notes that Eqs. \eqref{eq_mori_GLE5} and \eqref{eq_mori_memory8} are equivalent to the standard
Mori  GLE formulation, Eqs.  \eqref{eq_mori_GLE0} and \eqref{eq_mori_GLE0b}, provided the  force in Eq.  \eqref{eq_mori_GLE0} 
is interpreted as the sum of the complementary force and the  non-equilibrium force. 
It should be noted that an observable can be Gaussian while  the entire many-body system is non-Gaussian, meaning
that other observables and system coordinates may very well exhibit non-Gaussian fluctuations. Thus
the class of systems exhibiting Gaussian observables is a rather large one and includes for example
moving cells \cite{Mitterwallner_2020}.

Another useful equation is derived by averaging the entire GLE Eq. \eqref{eq_mori_GLE} over phase space, resulting in 
\begin{align} \label{eq_mori_GLE6}
&   \ddot  a(t)    =  \int {\rm d} \omega  \, \rho_p(\omega)     \ddot A(\omega, t) \\
&=   -  K(t)    (a(t)- \langle  A_S \rangle)  
 - \int_{t_0}^{t} {\rm d}s\, \Gamma(s,t)  \dot a(s)  \nonumber \\
& + \int_{t_0}^{t} {\rm d}s\, \Gamma_A (s,t) (  a(s) -  \langle  A_S \rangle)   + h(t)  /M,  \nonumber 
\end{align}
where the  phase-space averaged Heisenberg observable is denoted as 
$a(t)=\langle A(\omega,t)\rangle$ and 
$\langle  F(\omega,t_0,t) \rangle =0$ was used. 
This equation describes how the mean observable $ a(t) $ evolves in time under the influence
of the force $h(t)$, it therefore establishes the relation between the mean observable $a(t)$
and the external force $h(t)$.
It is exact and valid beyond the linear-response approximation
and  can therefore be viewed 
 as a generalization of the
linear-response FDT, as will be explored in detail in Sect. \ref{NEQFDT}. 

To derive yet another GLE, we define the  deviation of the Heisenberg observable $A(\omega, t)$ around its mean as 
\begin{align} 
\Delta  A(\omega, t) =    A(\omega, t) -  a(t),  
\end{align}
by subtracting Eqs. \eqref{eq_mori_GLE} and  \eqref{eq_mori_GLE6}  the GLE for 
$\Delta  A(\omega, t)$  follows as 
\begin{align} \label{eq_mori_GLE7}
 & \Delta \ddot  A(\omega, t)  = -  K(t)   \Delta  A(\omega,t) 
 - \int_{t_0}^{t} {\rm d}s\, \Gamma(s,t) \Delta \dot A(\omega, s)   \nonumber \\
& + \int_{t_0}^{t} {\rm d}s\, \Gamma_A (s,t) \Delta  A(\omega, s)   + F(\omega,t_0,t).
\end{align}
From Eq. \eqref{eq_mori_GLE7} we see that the complementary force $F(\omega,t_0,t)$
is linear in  $\Delta  A(\omega,\cdot)$; 
 this means that if $\Delta  A(\omega,t)$ is a Gaussian variable,
the kernel functions 
$ \Gamma_1(s,t) $ and $ \Gamma_A(s,t)$ in Eqs. \eqref{eq_mori_memory3} and \eqref{eq_mori_memory4}
vanish and we are thus led to the  GLE for the  Gaussian deviatory non-equilibrium variable $\Delta  A(\omega, t)$
\begin{align} \label{eq_mori_GLE8}
 & \Delta \ddot A(\omega, t)  = -  K_0     \Delta A(\omega,t)   
 - \int_{t_0}^{t} {\rm d}s\, \Gamma_0(s,t) \Delta  \dot A(\omega, s)   \nonumber \\
& + F(\omega,t_0,t),
\end{align}
where the friction kernel  $\Gamma_0(s,t) $  is given by the complementary force 
autocorrelation via Eq.  \eqref{eq_mori_memory2}.
Thus, the deviations of a  Gaussian  non-equilibrium variable 
from its mean are described by a GLE of the form of Eq.  \eqref{eq_mori_GLE0} that satisfies the standard FDT relation
Eq. \eqref{eq_mori_GLE0b}.

We have demonstrated  in this section that a Gaussian observable is  described by a GLE that 
has the same form as  the equilibrium GLE  Eq. \eqref{eq_mori_GLE0},
in other words, the stochastic behavior of an observable of a  non-equilibrium system
differs from an equilibrium system only if the observable is  non-Gaussian. 
This is a very important finding, since many experimental observables are Gaussian to a very good degree,
for all such variables the standard equilibrium Mori GLE in the form
of Eq.  \eqref{eq_mori_GLE0}, or, more precisely,  Eq.  \eqref{eq_mori_GLE5},
 is a valid  description of the dynamics.

 \section{Stochastic non-equilibrium force} \label{Sect_Stoch}
 
 \subsection{Infinitesimal response} \label{Response}
 
 The GLE discussed so far is difficult to deal with in practice  since it is inhomogeneous in time;
this  is utterly expected and  reflects the presence of  the time-dependent force $h(t)$ in the Hamiltonian
but  complicates the further analysis. 
   In many
 experimental scenarios the time evolution of the force $h(t)$   is not known or unimportant, 
 it  therefore becomes useful to interpret $h(t)$ as a stochastic variable that is only characterized
 by its first moments
\begin{align} \label{eq_force}
\overline{1}=1, \,\,\,\, \overline{h (t)}=0,  \,\,\,\, \overline{h(t)h(s)}=\sigma(t-s).
\end{align} 
The assumption of a vanishing first moment does not restrict the generality of the model since (at least 
for bounded systems) we can  subtract
a constant from $h(t)$ and move it into the equilibrium part of the Hamiltonian in Eq. \eqref{eq_Hamiltonian}. 
Here $\sigma(t-s)$ denotes  the force autocorrelation function,  by only defining the first two moments of $h(t)$ we 
are not necessarily  implying that the force is a Gaussian stochastic variable, as will become  clearer later on. 
By averaging the non-equilibrium GLE Eq. \eqref{eq_mori_GLE} over the non-equilibrium
force $h(t)$ we obtain the  GLE
\begin{align} 
\label{eq_mori_GLE9a}
&   \ddot  A(\omega,t)    = -  K_0    (A(\omega,t)- \langle  A_S \rangle)   - \int_{t_0}^{t} {\rm d}s\, \overline{\Gamma}(t-s)  \dot A(\omega,s)  \nonumber \\
& + \int_{t_0}^{t} {\rm d}s\, \overline{\Gamma}_A (t-s) (  A(\omega,s) -  \langle  A_S \rangle) + 
\overline{F}(\omega,t-t_0)  + F_\epsilon(t),
\end{align}
where we have added an infinitesimal generating force $F_\epsilon(t)$
by the substitution $h(t) \rightarrow h(t)  + F_\epsilon(t)$ in Eq. \eqref{eq_mori_GLE} prior to averaging,
 which later will be used to 
derive the infinitesimal response of a non-equilibrium system. When performing the force averaging,
   we use that  the
observable $A(\omega,t)$ in  Eq. \eqref{eq_mori_GLE}  has no explicit dependence on $h(t)$, 
which reflects that  its time evolution is completely determined by its initial value, its initial  velocity and the GLE parameters,
which  do explicitly depend on $h(t)$ (note that, in contrast, the solution $A(\omega,t)$ of the GLE 
Eq. \eqref{eq_mori_GLE} does depend on $h(t)$ and the average of this solution over $h(t)$
obviously  is not the same
as the solution of Eq. \eqref{eq_mori_GLE9a}). 
The force-averaged memory kernels $ \overline{\Gamma}(t-s) $ and $ \overline{\Gamma}_A (t-s)$ 
and the force-averaged complementary force $\overline{F}(\omega,t-t_0)$ are homogeneous in time, 
as is shown by perturbative operator expansion
 to leading order in powers of $\sigma$  in Appendix \ref{sec_App_GLE_time_homogeneity}.
 The friction kernel follows from Eq. \eqref{eq_mori_memory5}  explicitly as
\begin{align}  \label{eq_mori_memory9}
\overline{ \Gamma}(t-s) =  \overline{\Gamma}_0(t-s) + \beta \sigma(t-s)  /M +  \overline{\Gamma}_2(t-s),
\end{align}
where $ \overline{\Gamma}_0(t-s)$ and
$ \overline{\Gamma}_2(t-s) $ denote the force averages over Eqs. 
 \eqref{eq_mori_memory2} and \eqref{eq_mori_memory6}.

By additionally averaging over the phase space variable $\omega$
 we obtain the  GLE for  the  phase-space and force-averaged Heisenberg observable  $a(t)$,
\begin{align} 
\label{eq_mori_GLE9}
&   \ddot  a(t)    = -  K_0    (a(t)- \langle  A_S \rangle)   - \int_{t_0}^{t} {\rm d}s\, \overline{\Gamma}(t-s)  \dot a(s)  \nonumber \\
& + \int_{t_0}^{t} {\rm d}s\, \overline{\Gamma}_A (t-s) (  a(s) -  \langle  A_S \rangle) + F_\epsilon(t).
\end{align}
%
%which has a structure  similar to Eq. \eqref{eq_mori_GLE6}
%where the force $h(t)$ has not been averaged over. 
By Fourier transforming Eq. \eqref{eq_mori_GLE9} according to 
$\tilde a(\nu)   = \int_{-\infty}^{\infty} dt e^{- \imath t \nu}a(t)$, 
we obtain the infinitesimal response relation to first order in $F_\epsilon(t)$ as
\begin{align}  \label{eq_response}
\tilde a(\nu) - 2\pi \delta(\nu) \langle  A_S \rangle = \tilde \chi(\nu) \tilde F_\epsilon(\nu),
\end{align}
where the  Fourier-transformed  response function is determined  by 
\begin{align}  \label{eq_response2}
& 1/ \tilde \chi(\nu)  = K_0 - \nu^2 + \imath \nu  \tilde{\overline{ \Gamma}}^+(\nu) - \tilde{ \overline{\Gamma}}^+_A(\nu)    \nonumber \\
& = K_0 - \nu^2 + \imath \nu ( \tilde{\overline{ \Gamma}}_0^+(\nu) + \beta  \tilde{\sigma}^+(\nu) /M 
+  \tilde{\overline{ \Gamma}}_2^+(\nu))  - \tilde{ \overline{\Gamma}}^+_A(\nu).
\end{align}
 In the last step we have inserted Eq.  \eqref{eq_mori_memory9}.
Note that the  response function $\tilde \chi(\nu)$ accounts for  the full non-linear dependence
of the observable  on the non-equilibrium force $h(t)$, so it describes
the  response to an infinitesimal force $F_\epsilon(t)$ in the presence of a finite 
(not necessarily small) back-ground  force  $h(t)$.
In deriving Eq. \eqref{eq_response2} we have shifted the projection time into the far past, $t_0 \rightarrow - \infty$,  
and have introduced  causal or single-sided memory kernels and correlations, i.e. 
$\overline{\Gamma}^+(t)=0$,  $\overline{\Gamma}_A^+(t)=0$,  $\overline{\Gamma}_0^+(t)=0$,  $\overline{\Gamma}_2^+(t)=0$,  
$\sigma^+(t)=0$  for $t<0$. In fact,
$ \tilde \chi(\nu)$ can be obtained from the Volterra equation 
for the correlation function that follows from the GLE,
as explained  in Sec.  \ref{sec_Volterra}. 
For a Gaussian observable, 
in which case $\overline{\Gamma}_A(t)=0= \overline{\Gamma}_2(t)$ as discussed before,
 but in the presence of a non-equilibrium stochastic force, $h(t) \neq 0$, we obtain for the response function 
\begin{align}  \label{eq_response4}
&1/   \tilde \chi(\nu)  =
  K_0 - \nu^2 + \imath \nu  \tilde{\overline{ \Gamma}}_0^+(\nu)  +  \imath \nu \beta  \tilde{\sigma}_+(\nu) /M  .
\end{align}
In the absence of a non-equilibrium force, i.e. for  $h(t)=0$, 
Eq. \eqref{eq_response2} reduces to the standard equilibrium response function 
\begin{align}  \label{eq_response3}
& 1/  \tilde \chi(\nu)  =
K_0 - \nu^2 + \imath \nu  \tilde{\overline{ \Gamma}}_0^+(\nu) .
\end{align}
By comparison of Eqs. \eqref{eq_response2}, \eqref{eq_response4} and  \eqref{eq_response3} 
we see that the presence of a stochastic non-equilibrium force modifies the response function significantly
and adds terms that are determined by  the force autocorrelation function $\sigma(t)$.

%%%%%%%%%%%%%%%%%%%%%%%%%%%%%%%%%%%%%%%%%%%%%
\subsection{Fluctuation-dissipation theorem for finite non-equilibrium force $h(t)$}
\label{NEQFDT}

In order to derive the non-equilibrium version of the FDT,  we need 
to calculate the two point correlation function 
\begin{align} \label{eq_corr}
C(t-t') = \overline{ \langle (A(\omega,t)-\langle A_S \rangle)(A(\omega,t')  -\langle A_S \rangle) \rangle}
\end{align}
for general times $t,t' \geq t_0$,
which  is obtained by simultaneous averaging over phase space and  the non-equilibrium force $h(t)$. 
In Appendix \ref{sec_App_2pCorrelations}  we derive Eq. \eqref{eq_corr} and show that two-point correlation functions are 
generally given
 by a phase-space average over products of Heisenberg variables,
 for this we use   the  product propagation relation derived in Appendix  \ref{sec_App_ProductPropagatorRelation}.
%Secondly, one has to perform the force average over the product of two Heisenberg observable at two different times. 
%This is done in Appendix \ref{sec_App_ForceAverage2pCorr}
 % it is shown  that the force averaging decouples in a systematic expansion to leading order in the non-Gaussian
%fluctuation strength of the observable, so that we can instead write
%\begin{align} \label{eq_corr2}
%C(t-t') \approx   \langle ( \overline{  A(\omega,t)-\langle A_S \rangle} ( \overline{ A(\omega,t')  -\langle A_S \rangle} ) \rangle
%\end{align}
In App. \ref{sec_App_ForceAverage2pCorr}
we show that the Fourier-transformed correlation function is  to first order in an  expansion in powers
of the  second-order  force moments $ \tilde{\overline{ \Gamma}}_0(\nu)$ and  $ \tilde{\sigma}(\nu)$ given by 
\begin{align}  \label{eq_corr3}
&\tilde C(\nu) = \tilde \chi(\nu) \tilde \chi(-\nu) \left[ 
\langle ( {\cal L}_0 A_S(\omega) )^2 \rangle  \tilde{\overline{ \Gamma}}_0(\nu)  + 
 \tilde{\sigma}(\nu) /M^2 \right],
\end{align}
where we note that the force moments 
 $ \tilde{\overline{ \Gamma}}_0(\nu)  =  \tilde{\overline{ \Gamma}}_0^+(\nu) +\tilde{\overline{ \Gamma}}_0^+(-\nu)$
and $   \tilde{\sigma}(\nu) =  \tilde{\sigma}^+(\nu) +\tilde{\sigma}^+(-\nu) $ are  Fourier transforms of  time-symmetrized  functions. 
Using the expression for the response function $ \tilde \chi(\nu) $ in Eq. \eqref{eq_response2}, we can rewrite the correlation function as 
\begin{align}  \label{eq_corr4}
&\frac{ \tilde C(\nu)}{ \langle ( {\cal L}_0 A_S(\omega) )^2 \rangle}  = 
\frac{ \tilde \chi(-\nu)}{\imath \nu} - \frac{ \tilde \chi(\nu)}{\imath \nu} \\
&- \tilde \chi(\nu) \tilde \chi(-\nu) \left[ 
  \tilde{\overline{ \Gamma}}_2^+(\nu) +  \tilde{\overline{ \Gamma}}_2^+(-\nu)
   - \frac{   \tilde{\overline{ \Gamma}}_A^+(\nu) -  \tilde{\overline{ \Gamma}}_A^+(-\nu)  }{\imath \nu}  
  \right].  \nonumber 
\end{align}
Since all time-domain  kernel functions are real, we have
$ \Re( \tilde \Gamma^+(\nu))  = \Re( \tilde \Gamma^+ (-\nu))$ and $ \Im( \tilde \Gamma^+ (\nu))  = - \Im( \tilde \Gamma^+ (-\nu))$, 
where $ \Re$ and $ \Im$ denote the real and imaginary parts of a complex number $X$ according to 
$X= \Re X + \imath \Im X$. With this,
we can rewrite the correlation function as
\begin{align}  \label{eq_corr5}
\frac{ \nu  \tilde C(\nu) }{ 2 \langle ( {\cal L}_0 A_S(\omega) )^2 \rangle} & = 
  -   \Im \left (  \tilde \chi(\nu)\right)   \\
&- \tilde \chi(\nu) \tilde \chi(-\nu) \left[ 
 \nu  \Re ( \tilde{\overline{ \Gamma}}_2^+(\nu) ) 
   - \Im (   \tilde{\overline{ \Gamma}}_A^+(\nu) ) 
    \right], \nonumber 
\end{align}
which is the   FDT in the presence of a 
non-equilibrium stochastic force.  As discussed in Sec. \ref{Gauss},
if the observable $A(\omega,t)$ is Gaussian, the memory kernel contributions  
$  \tilde{\overline{ \Gamma}}_2(\nu)$ and $\tilde{\overline{ \Gamma}}_A(\nu)$ vanish
and we recover an equation  that   resembles the standard FDT
\cite{zwanzig_nonequilibrium_2001}
\begin{align}  \label{eq_corr6}
\frac{ \nu  \tilde C(\nu) }{ 2 \langle ( {\cal L}_0 A_S(\omega) )^2 \rangle} & = 
  -   \Im \left (  \tilde \chi(\nu)\right).
 \end{align}
However, the correlation function $ \tilde C(\nu)$ and the response function $ \tilde \chi(\nu)$
do depend on the non-equilibrium force $h(t)$, so Eq.   \eqref{eq_corr6}
is a non-trivial generalization of the standard FDT (which does not depend on the non-equilibrium force $h(t)$) 
to non-equilibrium Gaussian systems. 
 In Appendix \ref{sec_App_FDT}
 we derive the standard FDT by leading-order perturbation  analysis of the 
 Heisenberg observable in Eq.  \eqref{eq_Heisenberg}. 
 The comparison with this derivation is instructive, since it deviates slightly from the text-book derivation of the FDT 
 and highlights the non-perturbative character of our non-equilibrium GLE. 
Comparison of Eqs.  \eqref{eq_corr5} and \eqref{eq_corr6}
shows that our  generalized FDT  for  non-Gaussian
non-equilibrium systems  in Eq.  \eqref{eq_corr5}  contains additional  terms that are essentially proportional 
to the non-equilibrium force moment $\sigma$ and which  result from a systematic first-order expansion
in powers of  $\sigma$, as put forward in Appendix \ref{sec_App_ForceAverage2pCorr}.

A particularly transparent formulation of  the  non-equilibrium  fluctuation dissipation theorem Eq.  \eqref{eq_corr5} is given by
\begin{align}  \label{eq_corr7}
& - \frac{ \nu  \tilde C(\nu)/   \langle ( {\cal L}_0 A_S(\omega) )^2 \rangle  }{ 2  \Im \left (  \tilde \chi(\nu)\right) }  =
   \nonumber   \\
& \frac{  \tilde{\overline{ \Gamma}}_0(\nu)  +  \beta  \tilde{\sigma}(\nu) /M}
{2 \left ( \Re ( \tilde{\overline{ \Gamma}}^+(\nu))  - \Im (  \tilde{\overline{ \Gamma}}_A^+(\nu)  ) /\nu \right) }= 
1 - \Xi(\nu),
\end{align}
where in the first equation we used Eqs.  \eqref{eq_corr3} and \eqref{eq_response2}.
Eq.  \eqref{eq_corr7} contains  two alternative forms of our non-equilibrium FDT, the first in terms of the correlation function
and the imaginary part of the response function, which is preferred when dealing with experimental data
and when the non-equilibrium force $h(t)$ is not known, 
the second  in terms of the force correlations $ \tilde{\overline{ \Gamma}}_0(\nu)$, $\tilde{\sigma}(\nu)$
and the real part of the memory function, which is useful when 
 the non-equilibrium force $h(t)$ is known and which 
 will be further explained in Sec.  \ref{extract}.
In Eq.  \eqref{eq_corr7}  we defined the frequency-dependent non-equilibrium correction 
\begin{align}  \label{eq_corr8}
 \Xi(\nu)= 
 \frac{ \tilde \chi(\nu) \tilde \chi(-\nu) }{  \Im \left (  \tilde \chi(\nu)\right) }
\left[ 
 \Re ( \nu  \tilde{\overline{ \Gamma}}_2^+(\nu) ) 
   - \Im (   \tilde{\overline{ \Gamma}}_A^+(\nu) ) 
    \right],
\end{align}
which has been previously introduced to quantify the departure from equilibrium in 
biological non-equilibrium data  \cite{Netz2018}.
Obviously, for Gaussian observables, i.e. for  $ \tilde{\overline{ \Gamma}}_2^+(\nu) =0= \tilde{\overline{ \Gamma}}_A^+(\nu) $
we have $ \Xi(\nu)=0$. 
%This happens if the non-equilibrium force $h(t)$ is zero or if the observable $A(\omega,t$ is Gaussian,
%so we see that the equilibrium FDT can also hold for non-equilibrium systems. 
The expression for $ \Xi(\nu)$ is proportional to the factor 
$ \tilde \chi(\nu) \tilde \chi(-\nu) /  \Im ( \tilde \chi(\nu))$, which depends
on the response function $ \chi(\nu)$ and can be described in terms of 
 rather generic models. 
The non-Gaussian memory contributions 
$ \tilde{\overline{ \Gamma}}_2^+(\nu)$ and $\tilde{\overline{ \Gamma}}_A^+(\nu) $
are conversely rather  system-specific and  defy a generic approach.

%%%%%%%%%%%%%%%%%%%%%%%%%%%%%%%%%%%%%%%%%%%%%%%%%%%%%
\subsection{ Response function from the Volterra equation}   \label{sec_Volterra}

The standard way of extracting the memory kernel from time-series data is by turning the stochastic  
GLE for the phase-space dependent observable  $A(\omega,t)$  into a non-stochastic integro-differential 
equation for the two-point correlation function,  which can be  solved by Fourier transformation or 
 recursively after discretization in the time domain \cite{Straub_1987}.
Here we show that the same recipe also works for our
non-equilibrium GLE. To proceed, we multiply the GLE in Eq. \eqref{eq_mori_GLE} by 
$\dot A(\omega, t_0)= {\cal L}_0 A_S(\omega)$ and average over phase  space $\omega$ and the 
non-equilibrium  force $h(t)$, 
by which we obtain the equation
\begin{align}  \label{eq_Volt1}
&- \dddot C_0(t-t_0) = K_0 \dot C_0(t-t_0) +\int_0^{t-t_0}  {\rm d}s 
\overline{\Gamma}(s) \ddot C_0(t-t_0-s)  \nonumber \\
& - \int_0^{t-t_0}  {\rm d}s \overline{\Gamma}_A(s) \dot C_0(t-t_0-s) 
\end{align}
for the two point correlation function 
\begin{align} \label{eq_Volt2}
C_0(t-t_0) = \overline{ \langle (A(\omega,t_0)-\langle A_S \rangle)(A(\omega,t)  -\langle A_S \rangle) \rangle}.
\end{align}
Note that the two-point correlation function $C_0(t)$ defined here differs from the one defined in 
Eq. \eqref{eq_corr} in that  one of the times coincides with  the projection time $t_0$.
This makes a fundamental difference, as will become clear shortly. 
When deriving Eq. \eqref{eq_Volt1},  we used  that the phase-space average over the product of
$\dot A(\omega, t_0)$ and the forces $F(\omega,t_0,t)$ or $h(t)$ in the GLE Eq. \eqref{eq_mori_GLE} 
vanishes and that the resulting equation becomes homogeneous in time due to the average over the stochastic 
force $h(t)$,  as explained in Appendix \ref{sec_App_GLE_time_homogeneity}.
   
By performing a single-sided Fourier transform of  Eq. \eqref{eq_Volt1} while 
assuming $t\geq t_0$ and defining   
$\tilde C_0^+(\nu)   = \int_{0}^{\infty} dt e^{- \imath t \nu}C_0(t)$, we obtain
the solution
\begin{align}  \label{eq_Volt3}
\tilde C_0^+(\nu)  = \frac{\ddot C_0(0) \tilde \chi(\nu)}{\imath \nu} + \frac{C_0(0)}{\imath \nu} 
\end{align}
in terms of the response function $ \tilde \chi(\nu)$ defined in Eq. \eqref{eq_response2}.
This expression for $\tilde C_0^+(\nu) $ is equivalent to the expression Eq.  \eqref{eq_corr6}
for  $\tilde C(\nu)$ in the Gaussian limit but  for a non-Gaussian observable differs from the full expression
for $\tilde C(\nu)$, in other words, $C_0(t)$ and $C(t)$ are completely different correlation function
for non-Gaussian non-equilibrium systems.
%details of the derivation are shown in Appendix  \ref{sec_App_Volterra}. 
It transpires that from the time-domain correlation function $C_0(t)$, which can be 
straightforwardly  obtained in experiments or simulations 
by turning on the non-equilibrium force at  time $t_0$ (using that the distribution at time $t_0$  prior
to application of the non-equilibrium force equals
the canonical projection distribution $\rho_p(\omega)$ in Eq. \eqref{HamP}), 
 the single-sided Fourier transform $\tilde C_0^+(\nu)$
and the values $\ddot C_0(0)$ and $C_0(0)$ follow, from  which $\tilde \chi(\nu)$ is determined via
Eq. \eqref{eq_Volt3} by direct  inversion. 
The zero-frequency part of the  response function can be furthermore extracted by noting that 
$\int_{-\infty}^{\infty} {\rm d} \nu \tilde C_0(\nu) = 2 \pi \ddot C_0(0) /(K_0 - \tilde {\overline{\Gamma}}_A^+(0))$
where $\tilde C_0(\nu)= \tilde C_0^+(\nu) + \tilde C_0^+(-\nu)$,
as follows by residual calculus. Alternatively, instead of Fourier transformation, 
Eq. \eqref{eq_Volt1} can be recursively   solved by discretization  \cite{Straub_1987,carof_two_2014,Ayaz2022}.

\subsection{Extracting non-equilibrium GLE parameters from time series data} \label{extract}

In the previous  section we showed that  the correlation function $C_0(t)$, defined in Eq. \eqref{eq_Volt2},
% and obtained from the GLE  Eq. \eqref{eq_mori_GLE} by multiplying by $ \dot A(\omega,t)$  at projection time $t_0$, 
is the solution of the differential equation Eq. \eqref{eq_Volt1} and can be used to calculate
 the response function  $ \tilde \chi(\nu)$
via inversion of Eq. \eqref{eq_Volt3}. This is a very practical method to obtain $ \tilde \chi(\nu)$
from an experimental or simulated correlation function. In fact, from $ \tilde \chi(\nu)$
all parameters of the non-equilibrium-force-averaged GLE in Eq. \eqref{eq_mori_GLE9a} can be obtained. To see this,
we introduce the running  integral over the memory function 
$ \overline{G}_A (s)= \int_{s}^{t-t_0} {\rm d}s'\, \overline{\Gamma}_A (s')$,
with which Eq. \eqref{eq_mori_GLE9a} can after partial integration be rewritten as 
\begin{align} 
\label{eq_extract1}
&   \ddot  A(\omega,t)    = - \left  ( K_0 -   \overline{G}_A (0) \right )   (A(\omega,t)- \langle  A_S \rangle)    \\
& - \int_{t_0}^{t} {\rm d}s\, \left ( \overline{\Gamma}(t-s) + \overline{G}_A (t-s) \right )   \dot A(\omega,s)  
 + \overline{F}(\omega,t-t_0). \nonumber
\end{align}
The GLE in Eq. \eqref{eq_extract1} now depends on a single combined kernel function 
$ \overline{\Gamma}(t) + \overline{G}_A (t)$.  
Shifting the projection time into the far past, $t_0 \rightarrow - \infty$,    the single-sided Fourier transform of 
this kernel function  
can be related to the  Fourier-transformed response function $\tilde \chi(\nu)$ given  in Eq.  \eqref{eq_response2}
according to 
\begin{align}  \label{eq_extract2}
\tilde{ \overline{\Gamma}}^+(\nu) + \tilde{\overline{G}}^+_A (\nu) =
\frac{1}{\imath \nu} \left (
\frac{1} {  \tilde \chi(\nu)} - \frac{1} {  \tilde \chi(0)} + \nu^2 \right ).
\end{align}
The stiffness  of the effective harmonic potential that appears in Eq. \eqref{eq_extract1}
is  determined by the  zero-frequency limit of the response function  according to 
\begin{align}  \label{eq_extract3}
K_0 -   \overline{G}_A (0) = 
 \frac{1} {  \tilde \chi(0)},
 \end{align}
as follows from Eq.  \eqref{eq_response2}.
It transpires that all parameters of the GLE in Eq. \eqref{eq_extract1}
can be derived from the response function $\tilde \chi(\nu)$
according to Eqs. \eqref{eq_extract2} and \eqref{eq_extract3}.
Furthermore, by inverting Eq. \eqref{eq_extract1}, the complementary
force trajectory $F(\omega,t_0,t)$  can be calculated from a trajectory of 
the observable $A(\omega,t)$ (which becomes the force-averaged
 complementary force trajectory $\overline{F}(\omega,t-t_0)$ only after averaging
 over different realizations of $h(t)$).
From that and using the  definition  Eq.  \eqref{eq_mori_memory2},
we can  by averaging over $h(t)$ calculate the complementary force correlation $\overline{ \Gamma}_0(t)$,
 which allows to check whether time series data is of equilibrium or non-equilibrium nature.
 For this we rewrite the FDT in Eq.  \eqref{eq_corr7} using the definition of $\overline{G}_A (t)$ as
\begin{align}  \label{eq_extract4}
 \frac{  \tilde{\overline{ \Gamma}}_0(\nu)  +  \beta  \tilde{\sigma}(\nu) /M}
{2  \Re \left ( \tilde{\overline{ \Gamma}}^+(\nu)+   \tilde{\overline{ G}}_A^+(\nu)  \right) }= 
1 - \Xi(\nu),
\end{align}
which for an equilibrium system simplifies to 
\begin{align}  \label{eq_extract5}
 \frac{  \tilde{\overline{ \Gamma}}_0(\nu)  }
{2  \Re \left ( \tilde{\overline{ \Gamma}}^+(\nu)    \right) }= 
1.
\end{align}
Knowing $ \tilde{\overline{ \Gamma}}_0(\nu)$ from the determined complementary force trajectory and 
$ \tilde{ \overline{\Gamma}}^+(\nu) + \tilde{\overline{G}}^+_A (\nu)$ from Eq.  \eqref{eq_extract2},
we can check whether Eq. \eqref{eq_extract5} is violated and in that case infer that the 
system is of non-equilibrium nature (the inverse conclusion cannot be easily drawn since
Eq. \eqref{eq_extract5}
could  be satisfied even if the system is out of equilibrium due to  fortunate cancellation
of terms in Eq. \eqref{eq_extract4}).
In conclusion, knowledge of the response function  $\tilde \chi(\nu)$,
which can be obtained from the correlation function $C_0(t)$ (determined
by Eq.  \eqref{eq_Volt1}) via  Eq.  \eqref{eq_Volt3}, allows to detect whether
a system is in equilibrium or not. 
In addition, if the non-equilibrium  force trajectory $h(t)$ is known,  the 
 correction terms 
$  \tilde{\overline{ \Gamma}}_2(\nu)$ and $\tilde{\overline{ \Gamma}}_A(\nu)$,
that appear in the non-equilibrium 
FDT  in Eq. \eqref{eq_extract4} using
the definition of $\Xi(\nu)$ in   Eq. \eqref{eq_corr8},
can, according to Eqs.  \eqref{eq_mori_memory6} and \eqref{eq_mori_memory7},
be calculated explicitly. This  allows for an independent check of the 
non-equilibrium FDT  in Eq. \eqref{eq_extract4}.

More suitable for certain experimental systems where the non-equilibrium force $h(t)$ can not
be turned on and off at will, 
knowing  the correlation function $C(t)$  defined in Eq.  \eqref{eq_corr} and the
 response function   $\tilde \chi(\nu)$ defined via  the infinitesimal response relation in Eq. \eqref{eq_response},
 violation of the equilibrium
FDT in Eq. \eqref{eq_corr6}  can be checked and  the non-Gaussian memory  terms 
 in the non-equilibrium 
FDT in Eq. \eqref{eq_corr5}  can be extracted
in an alternative fashion.

\subsection{Joint observable distribution from path integrals} 

It remains to elucidate under which conditions the Heisenberg variable $A(\omega, t)$ is described by a Gaussian process
and thus the GLE Eq.  \eqref{eq_mori_GLE5} is valid.
For this we consider  the two-point joint probability distribution of $A(\omega, t)$, which is defined as
\begin{align} \label{dist1}
\rho(A_2,t_2;A_1,t_1) = \overline{ \langle \delta (A_2-A(\omega,t_2)   \delta (A_1-A(\omega,t_1) \rangle}
\end{align}
and involves  averages over phase space $\omega$ and the non-equilibrium force $h(t)$ according to 
Eqs. \eqref{eq_average} and   \eqref{eq_force}.
For the delta functions we use the Fourier representation and obtain 
\begin{align} \label{dist2}
& \rho(A_2,t_2;A_1,t_1) = 
 \int_{-\infty}^{\infty} \frac{{\rm d} q_1}{2\pi} \int_{-\infty}^{\infty} \frac{{\rm d} q_2}{2\pi}  \nonumber  \\
& \times e^{\imath q_1 A_1+\imath q_2 A_2} \overline{ \langle \exp\left(  - \imath q_1 A(\omega,t_1) - \imath q_2 A(\omega,t_2)\right) \rangle}.
\end{align}
Using the Fourier-transformed  linear response relation in Eq. \eqref{eq_response}
and the response function  defined in Eq. \eqref{eq_response2},
we write the  time-domain solution of the GLE Eq. \eqref{eq_mori_GLE} as
\begin{align}  \label{eq_resp1}
 A(\omega,t) -  \langle  A_S \rangle =  \int_{-\infty}^{\infty} {\rm d} s \chi(s) (F(\omega, t-s)+ h(t-s)/M).
\end{align}
Note that in deriving Eq. \eqref{eq_resp1} we have  only partially   averaged Eq. \eqref{eq_mori_GLE} over 
$h(t)$ so that the equation becomes time-homogeneous but still has the $h(t)$  source term. 
We also neglect the $h(t)$ dependence of the complementary force 
so that it becomes homogeneous in time and can be written as F$(\omega, t-s)$,
the calculation is therefore first order in the force correlation functions. 
The averages over the forces $F(\omega, t)$ and $ h(t)$ in Eq. \eqref{dist2}
we express by Gaussian path integrals as
\begin{align}  \label{eq_resp2}
 & \langle  X(F, h)  \rangle =  \int_{-\infty}^{\infty} \frac{ {\cal D} F(\omega,\cdot)}{{\cal N}_F}
 X(F, h) \nonumber \\
 & \exp\left( -  \int_{-\infty}^{\infty} {\rm d} s {\rm d} s' 
 \frac{ F(\omega,s) F(\omega,s') 
  \Gamma_0^{-1}(s-s') }{ 2  {\langle ( {\cal L}_0 A_S(\omega) )^2 \rangle}}
 \right),
 \end{align}
\begin{align}  \label{eq_resp3}
 & \overline{  X(F, h)} =  \int_{-\infty}^{\infty} \frac{ {\cal D} h(\cdot)}{{\cal N}_h}
 X(F, h) \nonumber \\
 & \exp\left( -  \int_{-\infty}^{\infty} {\rm d} s {\rm d} s' 
 \frac{ h(s) h(s') 
  \sigma^{-1}(s-s') }{ 2}
 \right),
 \end{align}
where ${{\cal N}_F}$ and ${{\cal N}_h}$ are normalization constants
and $  \Gamma_0^{-1}(s)$ and $  \sigma^{-1}(s)$ are the  inverse functions
of the Gaussian kernels defined in Eqs.  \eqref{eq_mori_memory2}
and  \eqref{eq_force}
according to $\int {\rm d}s \Gamma_0^{-1}(t-s) \Gamma_0(s-t')= \delta(t'-t)$.
  Non-Gaussian fluctuations of $F(\omega, t)$ and $ h(t)$ 
need  not be explicitly considered here as by construction  they do not change their two-point correlations.
Performing the Gaussian path integrals, the two-point distribution follows in matrix
notation as
\begin{align} \label{dist3}
& \rho(A_2,t_2;A_1,t_1) = \nonumber \\
& \frac{  \exp \left(-  (A_j -  \langle  A_S \rangle) I_{jk}^{-1}  (A_k -  \langle  A_S \rangle)/2  \right)}
 { \sqrt{ \det 2 \pi I}},
\end{align}
where the indices  $j,k=1,2$ are summed over and  the entries of the  two-by-two matrix
\begin{align} \label{dist4}
 I_{jk} = C(t_j - t_k)
\end{align}
 are given by the two-point correlation function defined in Eq. \eqref{eq_corr3},
 details of the derivation are given in Appendix  \ref{sec_App_PathIntegral}.
From Eq.  \eqref{dist3} we see that if the forces $F(\omega, t)$ and $ h(t)$ are described by general
Gaussian processes, as assumed in Eqs. \eqref{eq_resp2} and \eqref{eq_resp3}, 
then also the observable $A(\omega, t)  - \langle  A_S \rangle $
is a Gaussian process determined by the correlation function $C(t)$ defined in  Eq. \eqref{eq_corr3}.
In this case, the GLE  Eq. \eqref{eq_mori_GLE5} is valid, which has the structure of an equilibrium GLE. 
Conversely, if one of the two forces $F(\omega, t)$ or $ h(t)$ is non-Gaussian, 
then also the  observable $A(\omega, t) -  \langle  A_S \rangle $
is non-Gaussian and the GLE in Eq. \eqref{eq_mori_GLE}  applies,
which does not satisfy the FDT in Eq.  \eqref{eq_mori_memory8}.
To reiterate this point, 
if the FDT in Eq.  \eqref{eq_mori_memory8}
is violated, this can be due to non-Gaussian contributions in the complementary force
$F(\omega, t)$ or in the non-equilibrium force $ h(t)$. It will in general
be difficult to tell in an experiment which of the forces generates the non-Gaussian
behavior of the observable $A(\omega, t)$, in particular since 
$F(\omega, t)$ depends on $ h(t)$,
 unless the non-equilibrium force $ h(t)$
is generated externally and thus explicitly  known.

\section{Summary and Discussion} \label{summary}

In the paper  we have derived the non-equilibrium GLE from a many-body Hamiltonian which contains a non-equilibrium 
time-dependent force $h(t)$ that acts on a general phase-space-dependent observable $A_S(\omega)$. 
This is the same Hamiltonian one uses for deriving the standard FDT, in other words, 
the GLE we derive is conjugate to the standard FDT with one important distinction:
While the standard FDT describes the first-order response of the time-dependent mean of the observable to the 
external force $h(t)$ and thus the relation between the response function $\chi(t)$ and
the two-point correlation function $C(t)$ is independent of the force $h(t)$,  we derive the non-equilibrium GLE
Eq. \eqref{eq_mori_GLE}  non-perturbatively, i.e.  exactly  to all orders in $h(t)$. 
Having pointed out the exact nature of our derived non-equilibrium GLE,
one should add that  the kernel functions and the complementary force
that appear in the GLE Eq. \eqref{eq_mori_GLE} depend implicitly on the force trajectory via the propagators.
While linear and non-linear response theory describes how the mean of an observable depends
on the external force $h(t)$, the GLE is an equation of motion for the fluctuating observable, 
it therefore  opens up a complementary field of applications and is particular relevant for the description
of time series data.

From the exact GLE Eq. \eqref{eq_mori_GLE} we infer  that  Gaussian non-equilibrium observables are 
described by a GLE that takes  the form of the 
equilibrium GLE  in Eq.  \eqref{eq_mori_GLE0}. This is an important and rather non-trivial finding, 
as key observables of many
non-equilibrium systems are in fact Gaussian. For example, the motion of cancer cells and
algae \cite{Mitterwallner_2020} has been shown to be described by a Gaussian process provided one looks at single-cell data.
The correction terms in the non-equilibrium GLE  Eq. \eqref{eq_mori_GLE} that
account for  non-Gaussian effects turn out to be three-point (and higher-order) correlation functions that involve the 
complementary force. 

 The non-equilibrium GLE  in Eq.  \eqref{eq_mori_GLE} breaks time-homogeneity and thus is difficult to deal with 
 in practice. We therefore derive a GLE assuming that the external force $h(t)$ is stochastic and defined by its
 second moment. The preaveraging over $h(t)$ reinstalls time homogeneity of the GLE  and allows us to derive
 the non-equilibrium FDT to first order in a cumulant expansion in terms of  the external force and the complementary force.
 Similar to the GLE, the non-equilibrium FDT for Gaussian observable has the same form as the equilibrium FDT, only for non-Gaussian
 observables correction terms appear in the  non-equilibrium FDT that again are related to
 three-point (and higher-order)  correlation functions
 that involve the complementary force. 
 
 We also  introduce different methods for extracting the parameters of our non-equilibrium GLE from time series data.
 Here we distinguish between methods that require knowledge of the external force trajectory $h(t)$, 
 as is the case for simulation data and for experiments where a system is perturbed by an externally applied time-dependent force,
 for example by laser-traps or atomic-force microscopes, and methods where the force trajetcory $h(t)$ is not known, 
 as is the case in most experiments on biological systems.  
 
%In summary, 
%knowing the two-point correlation function $ \tilde C(\nu)$ and the response
%function $ \tilde \chi(\nu)$, 
%for a non-Gaussian non-equilibrium system the correction function 
%$ \Re ( \nu  \tilde{\overline{ \Gamma}}_2^+(\nu) ) 
%   - \Im (   \tilde{\overline{ \Gamma}}_A^+(\nu) ) $ can be extracted from the trajectory.
%   The response function in turn can be extracted from a system by applying an external 
%   force or, which might more straightforward to implement in practice,
%by solving the Volterra equation that can be derived from the GLE , as we will explain in the next section.

It is a widespread misconception 
%that Eq. \eqref{eq_mori_GLE0b} is a consequence of the  fluctuation-dissipation theorem (FDT)  and
that deviations from the equality in  Eq. \eqref{eq_mori_GLE0b}
would signal a break-down of the standard FDT.
The FDT  relates the first-order response function 
of the  observable $A$  to the force  $h(t)$ with  the two-point correlation function of $A$,
thus, by definition,  the standard FDT is independent of $h(t)$,
similar to Eq. \eqref{eq_mori_GLE0b} for an equilibrium system.
For finite  force  $h(t)\neq 0$
 the relation between the friction kernel 
$\Gamma(t-s)$ and the complementary force autocorrelation in Eq.  \eqref{eq_mori_GLE0b}
is  replaced by a relation that explicitly depends on $h(t)$.
Likewise, the  relation between the non-linear response of $A$ to an infinitesimal  field increase
 $h(t) + M F_\epsilon(t)$ 
and the fluctuations of $A$ becomes dependent on $h(t)$ and in fact  involves three-point 
correlation functions of $A$, as we  show  in this paper. 

A second widespread misconception is that the Mori GLE Eq.  \eqref{eq_mori_GLE0} is approximate
and only holds on the Gaussian level. The opposite is true, Eq.  \eqref{eq_mori_GLE0} is exact 
and any non-linear properties that the observable $A$ might have are accurately  represented by 
 non-Gaussian contributions of $F$. This means that the non-equilibrium GLE we derive,
 Eq.  \eqref{eq_mori_GLE},  also 
 applies to non-Gaussian non-equilibrium observables, provided the non-Gaussian contributions
 from the complementary force are correctly included.

One conclusion from our work is that for Gaussian systems, it does not make sense to 
consider a GLE that violates the FDT, this nicely explains our previous finding that 
for Gaussian systems, there is no way of detecting non-equilibrium properties from time-series data
\cite{Mitterwallner_2020}.

We mention in passing that the external force h(t) in general performs work on the system, meaning
that the total energy of the system will in general increase with time. Since the GLE 
Eq.  \eqref{eq_mori_GLE} is exact, it correctly takes into account these transient energetic effects.

The time-dependent Hamiltonian Eq. \eqref{eq_Hamiltonian} includes   only the linear coupling between
 the time-dependent force $h(t)$ and the  observable $A_S(\omega)$, more  complicated coupling terms
 are conceivable.  The motivation for studying this simple time-dependent Hamiltonian is that it 
  leads to the standard FDT, our non-equilibrium GLE thus is conjugate to the standard 
 FDT. In the future, it will be interesting to derive GLEs from
  more complex time-dependent Hamiltonians.

\begin{acknowledgments}
We acknowledge support by Deutsche Forschungsgemeinschaft Grant CRC 1449 "Dynamic Hydrogels at Biointerfaces", 
ID 431232613 Project A03,
by the ERC Advanced Grant 835117 NoMaMemo
and by the Infosys Foundation. 
\end{acknowledgments}

\appendix
%%%%%%%%%%%%%%%%%%%%%%%%%%%%%%%%%%%%%%%%%%%%%%%%%%%%%%%%%%%
%%%%%%%%%%%%%%%%%%%%%%%%%%%%%%%%%%%%%%%%%%%%%%%%%%%%%%%%%%%
\section{Derivative of Heisenberg propagator with respect to initial time  \label{sec_App_Propagator}}

Using the Heavyside function $\theta(t)$,  defined as $\theta(t)=1$ for $t>0$ and  
 $\theta(t)=0$ for $t<0$, the time-ordered operator exponential in the Heisenberg picture 
Eq. \ref{eq_expH} can be rewritten as 
\begin{widetext}
\begin{align}
\label{eq_expH2}
\exp_H\left(  \int_{t_0}^t  {\rm d}s  {\cal L} (s) \right) & =
1+\sum_{n=1}^\infty 
 \int_{t_0}^t  {\rm d}t_1    \int_{t_0}^{t}  {\rm d}t_2 \theta(t_1-t_2) \int_{t_0}^{t}  {\rm d}t_3 \theta(t_2-t_3) 
  \cdots  \int_{t_0}^{t}  {\rm d}t_n   \theta(t_{n-1}-t_n)   {\cal L} (t_n)  \cdots   {\cal L} (t_3) {\cal L} (t_2)  {\cal L} (t_1).
  \end{align}
We now reorder the integration variables to obtain
\begin{align}
\label{eq_expH3}
& \exp_H\left(  \int_{t_0}^t  {\rm d}s  {\cal L} (s) \right) = \nonumber \\
& 1+\sum_{n=1}^\infty 
\int_{t_0}^{t}  {\rm d}t_n   
\int_{t_0}^{t}  {\rm d}t_{n-1}   \theta(t_{n-1}-t_n)      
 \int_{t_0}^{t}  {\rm d}t_{n-2} \theta(t_{n-2}-t_{n-1})     \cdots
 \int_{t_0}^t  {\rm d}t_1   \theta(t_1-t_2)   
    {\cal L} (t_n)  {\cal L} (t_{n-1})  {\cal L} (t_{n-2}) \cdots   {\cal L} (t_1),
  \end{align}
note that the Liouville operators cannot be reordered since 
$ {\cal L} (t_1)$ and ${\cal L} (t_2)$ in general do not commute for $t_1 \neq t_2$.
We now replace the Heavyside functions by the appropriate integration boundaries and obtain
\begin{align}
\label{eq_expH4}
& \exp_H\left(  \int_{t_0}^t  {\rm d}s  {\cal L} (s) \right) = 
 1+\sum_{n=1}^\infty 
\int_{t_0}^{t}  {\rm d}t_n   
\int_{t_n}^{t}  {\rm d}t_{n-1}   
 \int_{t_{n-1}}^{t}  {\rm d}t_{n-2}   \cdots
 \int_{t_2}^t  {\rm d}t_1   
    {\cal L} (t_n)  {\cal L} (t_{n-1})  {\cal L} (t_{n-2}) \cdots   {\cal L} (t_1).
  \end{align}
From this expression  Eq. \eqref{eq_Heisenberg3} follows directly.

\section{Derivation of the time-dependent Dyson operator expansion \label{sec_App_opexp}}

The  operator expansion for two general time-dependent operators ${\cal V}(t)$ and ${\cal W}(t)$ reads
\begin{align}
\label{eq_opexp1}
\exp_H\left(  \int_{t_0}^t   {\cal V} (s)+{\cal W} (s)  {\rm d}s  \right)  
= \exp_H\left(  \int_{t_0}^t  {\rm d}s  {\cal V} (s)  \right) + 
\int_{t_0}^t  {\rm d}s  \exp_H\left(  \int_{t_0}^s  {\rm d}s'   {\cal V} (s') \right) 
 {\cal W} (s)
\exp_H\left(  \int_{s}^t   {\cal V} (s')+{\cal W} (s')  {\rm d}s'   \right).
\end{align}
To prove this relation, we use Eqs. \eqref{eq_Heisenberg2}  and \eqref{eq_Heisenberg3}  to obtain from Eq. \eqref{eq_opexp1}
\begin{align}
\label{eq_opexp2}
\exp_H\left(  \int_{t_0}^t   {\cal V} (s)+{\cal W} (s)  {\rm d}s  \right)  
= \exp_H\left(  \int_{t_0}^t  {\rm d}s  {\cal V} (s)  \right) - 
\int_{t_0}^t  {\rm d}s \frac{{\rm d}}{ {\rm d}s}
 \exp_H\left(  \int_{t_0}^s  {\rm d}s'   {\cal V} (s') \right) 
\exp_H\left(  \int_{s}^t   {\cal V} (s')+{\cal W} (s')  {\rm d}s'   \right). 
\end{align}
Now the integral can be performed and the equality  is obtained.
An alternative operator expansion relation reads
\begin{align}
\label{eq_opexp3}
\exp_H\left(  \int_{t_0}^t   {\cal V} (s)+{\cal W} (s)  {\rm d}s  \right)  
= \exp_H\left(  \int_{t_0}^t  {\rm d}s  {\cal V} (s)  \right) + 
\int_{t_0}^t  {\rm d}s  \exp_H\left(  \int_{t_0}^s {\cal V} (s)+{\cal W} (s)  {\rm d}s'     \right) 
 {\cal W} (s)
\exp_H\left(  \int_{s}^t     {\rm d}s'   {\cal V} (s')        \right),
\end{align}
\end{widetext}
which can be proven analogously. 
Choosing $ {\cal V} (s)=  {\cal Q}  {\cal L} (s)$ and ${\cal W} (s) = {\cal P} {\cal L} (s) $
in Eq. \eqref{eq_opexp3} we obtain Eq. \eqref{eq_GLE2}.

\section{Derivation of essential Mori  projection properties \label{sec_App_idempotency}}

For the following derivations it is useful to split the Mori projection operator in Eq. \eqref{eq_mori_projection}
into  three parts according to 
\begin{align}
\label{app_mori1}
 {\cal P}  B(\omega,t) = {\cal P}_1  B(\omega,t) +{\cal P}_2 B(\omega,t) +{\cal P}_3  B(\omega,t)
   \end{align}
with
 \begin{align}
{\cal P}_1  B(\omega,t) =
 \langle   B(\omega,t)  \rangle,
   \end{align}
 \begin{align}
{\cal P}_2  B(\omega,t) =
  \frac{\langle   B(\omega,t)  {\cal L}_0 A_S(\omega)  \rangle}  {\langle ( {\cal L}_0 A_S(\omega) )^2 \rangle} 
  {\cal L}_0 A_S (\omega),
  \end{align}
 \begin{align}
{\cal P}_3  B(\omega,t) =
 \frac{\langle   B(\omega,t)  (A_S (\omega) - \langle  A_S \rangle ) \rangle }  
 {\langle (A_S (\omega) - \langle  A_S \rangle )^2 \rangle} 
 (A_S (\omega) - \langle  A_S \rangle ).
  \end{align}

The linearity of the Mori projection,
 i.e., the fact that for two arbitrary observables $B(\omega,t)$ and $C(\omega,t')$ the property 
 ${\cal P} (c_1 B(\omega,t) + c_2 C(\omega,t'))=c_1 {\cal P} B(\omega,t)  + c_2 {\cal P} C(\omega,t')$ holds,
 is self-evident, ${\cal Q}$ is also easily seen to be linear. 

\begin{widetext}
The idempotency of  ${\cal P}$, i.e., the fact that  ${\cal P}^2= {\cal P}$, is not self-evident and will be proven.
We split the proof in three parts. First,
 \begin{align} \label{app_mori2}
{\cal P} {\cal P}_1  B(\omega,t) & =
 \langle   \langle  B(\omega,t)  \rangle  \rangle +
    \langle  B(\omega,t)  \rangle   \frac{  \langle  {\cal L}_0 A_S(\omega)  \rangle}  {\langle ( {\cal L}_0 A_S(\omega) )^2 \rangle} 
  {\cal L}_0 A_S (\omega)
 +  \langle  B(\omega,t)  \rangle  \frac{   \langle   (A_S (\omega) - \langle  A_S \rangle ) \rangle }  
 {\langle (A_S (\omega) - \langle  A_S \rangle )^2 \rangle} 
 (A_S (\omega) - \langle  A_S \rangle ) \nonumber \\
 & =   {\cal P}_1    B(\omega,t) .
   \end{align}
For  the first and third terms  we used that $ \langle   \langle  B(\omega,t)  \rangle  \rangle =  \langle  B(\omega,t)  \rangle$, which holds since the 
probability distribution in Eq.  \eqref{HamP} is normalized. 
For the second term we assume that $A_S$ is a function of position only, i.e.  $A_S (\omega) =A_S (\bf R)$, 
such that  $ {\cal L}_0 A_S(\omega) $ is linear in the momenta and the average $  \langle  {\cal L}_0 A_S(\omega)  \rangle$
vanishes. 

Second,
 \begin{align} \label{app_mori3}
{\cal P} {\cal P}_2  B(\omega,t) & =
  \frac{\langle   B(\omega,t)  {\cal L}_0A_S(\omega)  \rangle}  {\langle ( {\cal L}_0 A_S(\omega) )^2 \rangle} 
\left[
    \langle  {\cal L}_0A_S (\omega)  \rangle +
   \frac{  \langle ( {\cal L}_0A_S(\omega) )^2 \rangle}  {\langle ( {\cal L}_0A_S(\omega) )^2 \rangle} 
  {\cal L}_0 A_S (\omega)
 +    \frac{   \langle   (A_S (\omega)-\langle  A_S \rangle )    {\cal L}_0 A_S (\omega)\rangle }  
 {\langle (A_S (\omega) - \langle  A_S \rangle )^2 \rangle} 
 (A_S (\omega) - \langle  A_S \rangle ) \right]  \nonumber \\
 & =   {\cal P}_2   B(\omega,t)  ,
   \end{align}
where again we used  that $A_S (\omega) =A_S (\bf R)$
such that  $ {\cal L}_0 A_S(\omega) $ is linear in the momenta and the averages $  \langle  {\cal L}_0 A_S(\omega)  \rangle$
and $  \langle   (A_S (\omega)-\langle  A_S \rangle )    {\cal L}_0 A_S (\omega)\rangle$ vanish. 

Third,
 \begin{align} \label{app_mori4}
{\cal P} {\cal P}_3  B(\omega,t) & =
 \frac{\langle   B(\omega,t)  (A_S (\omega) - \langle  A_S \rangle ) \rangle }  
 {\langle (A_S (\omega) - \langle  A_S \rangle )^2 \rangle} \nonumber \\
& \times  \left[
    \langle  A_S (\omega) - \langle  A_S \rangle   \rangle +
   \frac{  \langle (A_S (\omega) - \langle  A_S \rangle )    {\cal L}_0A_S(\omega) \rangle}  
   {\langle  ( {\cal L}_0A_S(\omega) )^2 \rangle} 
  {\cal L}_0 A_S (\omega)
 +    \frac{   \langle   (A_S (\omega)-\langle  A_S \rangle )^2  \rangle }  
 {\langle (A_S (\omega) - \langle  A_S \rangle )^2 \rangle} 
 (A_S (\omega) - \langle  A_S \rangle ) \right]  \nonumber \\
 & =   {\cal P}_3   B(\omega,t)  ,
   \end{align}
where again we used  that $A_S (\omega) =A_S (\bf R)$.
Adding Eqs.  \eqref{app_mori2},  \eqref{app_mori3},  \eqref{app_mori4} we see
that ${\cal P}^2= {\cal P}( {\cal P}_1 +{\cal P}_2 + {\cal P}_3)= {\cal P}_1 +{\cal P}_2 + {\cal P}_3= {\cal P}$
and thus 
${\cal P}$ is idempotent. From the idempotency of ${\cal P}$ it follows that ${\cal Q}$ is also idempotent, to prove this one writes
\begin{align}
{\cal Q}^2 B(\omega,t)= (1- {\cal P})^2 B(\omega,t) = (1-2 {\cal P} + {\cal P}^2) B(\omega,t) = 
(1-{\cal P}) B(\omega,t) ={\cal Q} B(\omega,t).
\end{align}

The self-adjointedness of ${\cal P}$, Eq. \eqref{eq_projection_orthogonal}, is straightforwardly proven by writing
\begin{align}
&\langle C(\omega,t) {\cal P} B(\omega,t')  \rangle = \\
& \langle C(\omega,t) \rangle \langle B(\omega,t')  \rangle
+    \langle C(\omega,t)   {\cal L}_0A_S(\omega)  \rangle
   \frac{ \langle  B(\omega,t')  {\cal L}_0A_S(\omega) \rangle}  
   {\langle  ( {\cal L}_0A_S(\omega) )^2 \rangle} 
 +    \langle C(\omega,t) ( A_S (\omega) - \langle  A_S \rangle )  \rangle
\frac{   \langle B(\omega,t')   (A_S (\omega)-\langle  A_S \rangle )  \rangle }  
 {\langle (A_S (\omega) - \langle  A_S \rangle )^2 \rangle} \\
&= \langle B(\omega,t') {\cal P} C(\omega,t)   \rangle.
\end{align} 
By using ${\cal Q}=1-{\cal P}$ we see straightforwardly that ${\cal Q}$ is also self-adjoint.

Using similar arguments as above, one can show that  ${\cal P} c$= c, 
  ${\cal P}  (A_S (\omega) - \langle  A_S \rangle )=  (A_S (\omega) - \langle  A_S \rangle )$,
   ${\cal P}    {\cal L}_0A_S(\omega) =    {\cal L}_0A_S(\omega) $,
   from which follows that also   ${\cal P}  A_S (\omega) =  A_S (\omega)  $.
   From these relations  we can directly conclude that 
   ${\cal Q} c$= 0, 
  ${\cal Q}  (A_S (\omega) - \langle  A_S \rangle )=  0$,
   ${\cal Q}    {\cal L}_0A_S(\omega) =   0$,
   and also   ${\cal Q}  A_S (\omega) =  0 $.
   
   From the idempotency of ${\cal P}$ or ${\cal Q}$ we follow that 
   ${\cal P} {\cal Q}= {\cal P} (1-{\cal P})=  {\cal P} -{\cal P}^2= 0$
   and, similarly,  ${\cal Q} {\cal P}=0$, thus, the operators  ${\cal P}$ and $ {\cal Q}$ are orthogonal to 
   each other.
\end{widetext}

\section{Derivation of non-equilibrium GLE }
 \label{sec_App_GLE}
We consider the first term in Eq. \eqref{eq_GLE3}, which  can be split into two terms
and reads, apart from the propagator in front, 
\begin{align}
\label{app_GLE1}
  {\cal P}   {\cal L}(t)  {\cal L}_0   A_{S}(\omega)
=  {\cal P}   {\cal L}_0^2   A_{S}(\omega) -h(t)   {\cal P}   \Delta {\cal L}  {\cal L}_0   A_{S}(\omega).
\end{align} 
We apply the projection operator Eq. \eqref{eq_mori_projection}
on the first term in Eq. \eqref{app_GLE1},
which generates three contributions. The first contribution is given by 
\begin{align}
\label{app_GLE2}
\langle  {\cal L}_0^2   A_{S}(\omega) \rangle = 
- \beta h_p \langle (   {\cal L}_0   A_{S}(\omega))^2 \rangle,
\end{align} 
where we used that $ {\cal L}_0$ is anti-self-adjoint and that 
$ {\cal L}_0 \rho_p(\omega) = \beta h_p \rho_p(\omega) {\cal L}_0 A_{S}(\omega)$.
Using the same properties of $ {\cal L}_0$, the second contribution can be written as
(apart from the normalization factor)
\begin{align}
\label{app_GLE3}
\langle  (   {\cal L}_0   A_{S}(\omega))  {\cal L}_0^2   A_{S}(\omega) \rangle = 
- \beta h_p \langle (   {\cal L}_0   A_{S}(\omega))^3 \rangle/2. 
\end{align} 
 The third contribution (apart from the normalization factor)
follows  as
\begin{align}
\label{app_GLE4}
& \langle  (  ( A_{S}(\omega)- \langle A_S \rangle )  {\cal L}_0^2   A_{S}(\omega) \rangle 
\nonumber \\
&=  - \langle (   {\cal L}_0   A_{S}(\omega))^2 \rangle
- \beta h_p \langle  ( A_{S}(\omega)- \langle A_S \rangle )  (   {\cal L}_0   A_{S}(\omega))^2 \rangle. 
\end{align} 

We now apply the projection operator Eq.  \eqref{eq_mori_projection}
on the second term in Eq. \eqref{app_GLE1},
which again generates three contributions. The first contribution is given by 
\begin{align}
\label{app_GLE5}
-h(t) \langle  \Delta  {\cal L}  {\cal L}_0   A_{S}(\omega) \rangle = 
 \beta h(t) \langle (   {\cal L}_0   A_{S}(\omega))^2 \rangle,
\end{align} 
where we used that $   \Delta{\cal L}$ is anti-self-adjoint and that 
$  \Delta  {\cal L} \rho_p(\omega) = \beta \rho_p(\omega) {\cal L}_0 A_{S}(\omega)$.
Using the same properties of $ \Delta {\cal L}$, 
the second contribution can be written as (apart from the normalization factor)
\begin{align}
\label{app_GLE6}
- h(t) \langle  (   {\cal L}_0   A_{S}(\omega))  \Delta  {\cal L} {\cal L}_0   A_{S}(\omega) \rangle = 
 \beta h(t)  \langle (   {\cal L}_0   A_{S}(\omega))^3 \rangle/2. 
\end{align} 
 The third contribution  (apart from the normalization factor) follows  as
\begin{align}
\label{app_GLE7}
&- h(t)  \langle  (  ( A_{S}(\omega)- \langle A_S \rangle )  \Delta  {\cal L}  {\cal L}_0   A_{S}(\omega) \rangle 
\nonumber \\
&=  \beta h(t)  \langle  ( A_{S}(\omega)- \langle A_S \rangle )  (   {\cal L}_0   A_{S}(\omega))^2 \rangle. 
\end{align} 

Again assuming that the observable is a function of position only,   $A_S (\omega) =A_S (\bf R)$,
as we did in Appendix. \ref{sec_App_idempotency} when we derived the idempotency of the projection operator
${\cal P}$, we see that Eqs. \eqref{app_GLE3} and  \eqref{app_GLE6} vanish because 
$ \langle (   {\cal L}_0   A_{S}(\omega))^3 \rangle$ is odd in the momenta. 

Combining the results in Eqs. \eqref{app_GLE2},  \eqref{app_GLE4},  \eqref{app_GLE5},  \eqref{app_GLE7}, 
the first term in Eq. \eqref{eq_GLE3} reads
\begin{align}
\label{app_GLE8}
& \exp_H\left(  \int_{t_0}^t  {\rm d}s  {\cal L} (s) \right)  {\cal P}   {\cal L}(t)  {\cal L}_0   A_{S}(\omega)
\nonumber \\
& = -  K(t)    (A(\omega,t) - \langle  A_S \rangle)  +(h(t)-h_p) /M,
\end{align} 
where $K(t)$ and $M$ are defined in Eqs.  \eqref{GLEK1}  and  \eqref{eq_mass}.

We now consider the last term in Eq. \eqref{eq_GLE3}, which 
 reads, without the time integral and the propagator in front, 
$  {\cal P}   {\cal L}(s)  F(\omega,s,t)$.
The projection operator Eq. \eqref{eq_mori_projection}
 generates three contributions. The first contribution is given by 
\begin{align}
\label{app_GLE9}
&\langle  {\cal L}(s)    F(\omega,s,t)  \rangle = 
\beta (h(s)-h_p) \langle    F(\omega,s,t)   {\cal L}_0 A_S(\omega)    \rangle \nonumber \\
&= \beta (h(s)-h_p) \langle    F(\omega,s,t)    {\cal Q}  {\cal L}_0 A_S(\omega)    \rangle \nonumber \\
&= 0,
\end{align} 
where in the first line we used that $ {\cal L}(s)$ is anti-self-adjoint and that 
$ {\cal L}(s) \rho_p(\omega) = \beta ( h_p-h(s))  \rho_p(\omega) {\cal L}_0 A_{S}(\omega)$,
in the second line that ${\cal Q}$ is idempotent and self-adjoint, 
and in the third line that $  {\cal Q}  {\cal L}_0 A_S(\omega) =0$
(as derived in  Appendix  \ref{sec_App_idempotency}).
The second contribution can be written as
(apart from the normalization factor)
\begin{align}
\label{app_GLE10}
&\langle  ( {\cal L}_0 A_S(\omega) )   {\cal L}(s)    F(\omega,s,t)  \rangle \nonumber \\
&= - \langle F(\omega,s,t)  \rangle    \rangle  {\cal L}(s)   {\cal L}_0 A_S(\omega)  \nonumber \\
&+ \beta (h(s)-h_p) \langle  (  {\cal L}_0 A_S(\omega)  )^2   F(\omega,s,t)   \rangle \nonumber \\
&= - \langle F(\omega,s,s)   F(\omega,s,t)  \rangle    \rangle \nonumber \\
&+ \beta (h(s)-h_p) \langle  (  {\cal L}_0 A_S(\omega)  )^2   F(\omega,s,t)   \rangle,
\end{align} 
where in the first equation  we used the same properties of  $ {\cal L}(s)$ as before
and in the second equation  that ${\cal Q}$ is idempotent and self-adjoint
and the definition of the complementary force in Eq. \eqref{eq_F_operator}.

The third contribution can be written as
(apart from the normalization factor)
\begin{align}
\label{app_GLE11}
&\langle  ( A_{S}(\omega)- \langle A_S \rangle )    {\cal L}(s)    F(\omega,s,t)  \rangle \nonumber \\
&= - \langle  F(\omega,s,t)      {\cal L}(s)   ( A_{S}(\omega)- \langle A_S \rangle )         \rangle \nonumber \\
&+ \beta (h(s)-h_p) \langle  F(\omega,s,t)   ( A_{S}(\omega)- \langle A_S \rangle )         {\cal L}_0 A_S(\omega)    \rangle \nonumber \\
&= \beta (h(s)-h_p) \langle  F(\omega,s,t)   ( A_{S}(\omega)- \langle A_S \rangle )         {\cal L}_0 A_S(\omega)    \rangle,
\end{align} 
where in the first equation  we used the same properties of  $ {\cal L}(s)$ as before
and in the second equation  that ${\cal Q}$ is idempotent and self-adjoint
and that $\Delta  {\cal L} A_S(\omega)=0$. 

Combining the results in Eqs. \eqref{app_GLE8},  \eqref{app_GLE9},  \eqref{app_GLE10},  \eqref{app_GLE11}, 
the  expression for the general GLE in Eq. \eqref{eq_GLE3} leads to the explicit GLE  in Eq.  \eqref{eq_mori_GLE}.

\begin{widetext}
\section{Derivation of time-homogeneous memory kernels for stochastic force \label{sec_App_GLE_time_homogeneity}}

In this section we show that the memory kernels of the GLE in Eq.  \eqref{eq_mori_GLE}
become time-homogeneous for a stochastic non-equilibrium  force as defined by Eq.  \ref{eq_force}.
The proof is done to quadratic order in powers of the force $h(t)$. 
We start with the  operator expansion for two general time-dependent operators ${\cal V}(t)$ and ${\cal W}(t)$,
Eq. \ref{eq_opexp1}, and note that we can construct a systematic perturbative expansion in the operator ${\cal W}(t)$
by writing
\begin{align}
\label{homo1}
\exp_H\left(  \int_{t_0}^t   {\cal V} (s)+{\cal W} (s)  {\rm d}s  \right)  
&= \exp_H\left(  \int_{t_0}^t  {\rm d}s  {\cal V} (s)  \right) + 
\int_{t_0}^t  {\rm d}s  \exp_H\left(  \int_{t_0}^s  {\rm d}s'   {\cal V} (s') \right) 
 {\cal W} (s)
\exp_H\left(  \int_{s}^t   {\cal V} (s')+{\cal W} (s')  {\rm d}s'   \right) \nonumber \\
&= \exp_H\left(  \int_{t_0}^t  {\rm d}s  {\cal V} (s)  \right) + 
\int_{t_0}^t  {\rm d}s  \exp_H\left(  \int_{t_0}^s  {\rm d}s'   {\cal V} (s') \right) 
 {\cal W} (s)
\exp_H\left(  \int_{s}^t   {\cal V} (s')  {\rm d}s'   \right) +{\cal O} ({\cal W}^2).
\end{align}
Expressions valid to higher  order in ${\cal W}$ can be constructed by recursively  inserting
the second line of   Eq. \eqref{homo1}   into the last operator exponential in the first line. 

Following this method, the complementary force can be expanded in powers of the non-equilibrium force
as 
\begin{align}
\label{homo2}
F(\omega,t_0,t)&  \equiv  
\exp_H\left(  {\cal Q} \int_{t_0}^t  {\rm d}s'  {\cal L} (s') \right)
 {\cal Q}   {\cal L}(t)  {\cal L}_0   A_{S}(\omega) \nonumber \\
& = \exp_H\left(  {\cal Q} \int_{t_0}^t  {\cal L}_0 - h(s')  \Delta   {\cal L}  {\rm d}s'  \right)
 {\cal Q}   ( {\cal L}_0 - h(t)  \Delta   {\cal L} )   {\cal L}_0   A_{S}(\omega)\nonumber \\
 & =\left[ e^{(t-t_0){\cal QL}_0} -  
 \int_{t_0}^t  {\rm d}s'  e^{(s'-t_0){\cal QL}_0} h(s') {\cal Q}  \Delta {\cal L} e^{(t-s'){\cal QL}_0} \right]  
 {\cal Q}   ( {\cal L}_0 - h(t)  \Delta   {\cal L} )   {\cal L}_0   A_{S}(\omega) + {\cal O} (h^2),
\end{align}
where higher-order terms can be recursively derived. We calculate here exemplarily the 
force average over the non-equilibrium friction memory kernel defined in Eq.  \eqref{eq_mori_memory3}
\begin{align} \label{homo3}
\overline { \Gamma}_1(s,t) =
 \frac{  - \beta  \overline {  h(s)  \langle       F(\omega,s, t)  ( {\cal L}_0 A_S(\omega) )^2  \rangle}}
 {\langle ( {\cal L}_0 A_S(\omega) )^2 \rangle} = 
  \frac{  - \beta  (X_1(s,t) + X_2(s,t))}
 {\langle ( {\cal L}_0 A_S(\omega) )^2 \rangle},
\end{align}
which we split into two terms using the expansion in Eq. \eqref{homo2}.
The first term reads, after performing the average over $h(t)$, 
\begin{align} \label{homo4}
X_1(s,t) = - \sigma(t-s) 
\left \langle  ( {\cal L}_0 A_S(\omega) )^2 e^{(t-s){\cal QL}_0}
 {\cal Q}  \Delta   {\cal L}   {\cal L}_0   A_{S}(\omega)  \right \rangle,
\end{align}
which clearly is a function of $t-s$ only.
The second term reads
\begin{align} \label{homo5}
X_2(s,t) =&  -
\left \langle  ( {\cal L}_0 A_S(\omega) )^2 
 \int_{s}^t  {\rm d}s'    e^{(s'-s){\cal QL}_0} \sigma(s'-s)  
 {\cal Q}  \Delta   {\cal L}  e^{(t-s'){\cal QL}_0} 
  {\cal Q}  {\cal L}_0^2   A_{S}(\omega)  \right \rangle \nonumber \\
  & 
  =  - \left \langle  ( {\cal L}_0 A_S(\omega) )^2 
 \int_{0}^{t-s}  {\rm d}\tilde s'    e^{\tilde s'{\cal QL}_0} \sigma(\tilde s')  
 {\cal Q}  \Delta   {\cal L}  e^{(t-s-\tilde s'){\cal QL}_0} 
  {\cal Q}  {\cal L}_0^2   A_{S}(\omega)  \right \rangle,
\end{align}
which also clearly is a function of $t-s$ only
and  where we have used the change of integration variable according to $ \tilde s'= s'-s$.
Thus, $\overline { \Gamma}_1(t-s) $ defined in Eq.  \eqref{eq_mori_memory3}
is a function of $t-s$ only, the same holds for $\overline { \Gamma}_2(t-s) $ defined in Eq.  \eqref{eq_mori_memory6}.
Using similar techniques, it can be also shown that $\overline { \Gamma}_0(t-s) $ defined in Eq.  \eqref{eq_mori_memory2},
 $\overline { \Gamma}_A(t-s) $ defined in Eq.  \eqref{eq_mori_memory4} and 
 $\overline { F}(\omega,t-s) $ are functions of $t-s$ only, 
which proves the functional dependencies assumed in Eq.  \eqref{eq_mori_GLE9a}.

\section{Derivation of two-point correlation functions from Heisenberg observables}  \label{sec_App_2pCorrelations}

We  derive the two-point correlation function of the Heisenberg observable, to reduce the notational complexity
we here use a time-independent Liouville operator $ {\cal L}_0(\omega)$.
Splitting the time  propagation of the  density distribution  in Eq. \eqref{opexp} into two steps we obtain
 \begin{align} \label{HeiCorr1}
{\rho}(\omega,t) & =  e^{ - (t-t')   {\cal L}_0 (\omega)  }  e^{ - (t'-t_0)   {\cal L}_0(\omega) } \rho(\omega,t_0).
\end{align}
Introducing delta functions we obtain
 \begin{align} \label{HeiCorr2}
{\rho}(\omega,t)  &=
\int {\rm d}\omega'\int {\rm d}\omega_0 
 e^{ - (t-t')   {\cal L}_0(\omega) } \delta (\omega-\omega') 
   e^{ - (t'-t_0)   {\cal L}_0(\omega') }\delta (\omega'-\omega_0) 
    \rho(\omega_0,t_0) \nonumber \\ 
 &   \equiv  \int {\rm d}\omega'  \int {\rm d}\omega_0 {\rho}(\omega,t;\omega',t';\omega_0,t_0 ),
\end{align}
where in the second line we defined the three-point joint distribution function ${\rho}(\omega,t;\omega',t';\omega_0,t_0 )$.
With this distribution,  the two-point correlation function defined in Eq.  \eqref{eq_corr} can be written as
 \begin{align} \label{HeiCorr3}
C(t,t')   &=
 \int {\rm d}\omega   \int {\rm d}\omega'  \int {\rm d}\omega_0
\left(  A_S(\omega)  - \langle A_S \rangle  \right) 
\left(  A_S(\omega') - \langle A_S \rangle  \right)    {\rho}(\omega,t;\omega',t';\omega_0,t_0 ).
\end{align}
Inserting the definition of ${\rho}(\omega,t;\omega',t';\omega_0,t_0 )$ from Eq.  \eqref{HeiCorr2}
and using that  $ {\cal L}_0(\omega)$ is  anti-self-adjoint,
we obtain
 \begin{align} \label{HeiCorr4}
C(t,t')   &=
 \int {\rm d}\omega   {\rho}(\omega,t_0 ) 
  e^{  (t'-t_0)   {\cal L}_0(\omega) }  \left(  A_S(\omega)  - \langle A_S \rangle  \right) 
   e^{  (t-t')   {\cal L}_0(\omega) }   \left(  A_S(\omega)  - \langle A_S \rangle  \right). 
\end{align}
We now use the product propagation relation proved in Appendix  \ref{sec_App_ProductPropagatorRelation}
and obtain
 \begin{align} \label{HeiCorr5}
C(t,t')   &=
 \int {\rm d}\omega   {\rho}(\omega,t_0 ) 
 \left(  e^{  (t'-t_0)   {\cal L}_0(\omega) } \left(  A_S(\omega)  - \langle A_S \rangle  \right)  \right ) 
   e^{  (t-t_0)   {\cal L}_0(\omega) }   \left(  A_S(\omega)  - \langle A_S \rangle  \right),
\end{align}
which can, using the definition of the Heisenberg observable, be  rewritten as 
 \begin{align} \label{HeiCorr6}
C(t,t')   &=
 \int {\rm d}\omega   {\rho}(\omega,t_0 ) 
 \left(   A(\omega,t_0,  t) - \langle A_S \rangle  \right)
 \left(   A(\omega,t_0,  t')- \langle A_S \rangle  \right).
\end{align}
In this expression $t_0$ denotes the projection time and $t \geq t_0$ and $t' \geq t_0$ 
denote arbitrary moments in  time, note that we suppress the dependence of the Heisenberg
observable on the projection time in the rest of the paper.

\end{widetext}

\section{Derivation of product propagation   relation} \label{sec_App_ProductPropagatorRelation}

To reduce the notational complexity, we here use a time-independent Liouville operator ${\cal L}_0(\omega)$
and suppress the phase-space dependence of ${\cal L}_0(\omega)$ and of the two general
phase-space function $A(\omega)$ and $B(\omega)$ in the following.
The Liouville propagator acting on the product of two phase-space functions follows from the
series definition of the exponential as 
 \begin{align} \label{OpProd1}
   e^{t  {\cal L}_0}  AB=  \sum_{n=0}^{\infty} \frac{ t^n {\cal L}_0^n}{n!} AB.
\end{align}
Since ${\cal L}_0$  is a linear differential operator the product rule applies, which reads 
 \begin{align} \label{OpProd2}
 {\cal L}_0  AB=  A {\cal L}_0 B +B {\cal L}_0 A.
\end{align}
Applying the product rule recursively, we find
 \begin{align} \label{OpProd3}
 {\cal L}_0^n  A B=  \sum_{m=0}^n \frac{n!}{m! (n-m)!} \left( {\cal L}_0^m A \right)  {\cal L}_0^{n-m} B.
\end{align}
Combining Eqs.  \eqref{OpProd1} and  \eqref{OpProd3} we find
 \begin{align} \label{OpProd4}
 e^{t  {\cal L}_0}  AB & =  \sum_{n=0}^{\infty}  \sum_{m=0}^n \frac{ t^n}{m! (n-m)!} 
  \left( {\cal L}_0^m A \right)  {\cal L}_0^{n-m} B \nonumber \\ 
  &=  \sum_{m=0}^{\infty}  \sum_{n=m}^\infty  \frac{ t^n}{m! (n-m)!} 
  \left( {\cal L}_0^m A \right)  {\cal L}_0^{n-m} B \nonumber \\ 
&=  \sum_{m=0}^{\infty}  \sum_{n=0}^\infty  \frac{ t^{n+m} }{m! n!} 
  \left( {\cal L}_0^m A \right)  {\cal L}_0^n B.
  \end{align}
Using Eq.  \eqref{OpProd1}  we finally arrive at the product popagation  relation
 \begin{align} \label{OpProd5}
   e^{t  {\cal L}_0}  AB=   \left(  e^{t  {\cal L}_0}  A \right) e^{t  {\cal L}_0} B.
\end{align}

\begin{widetext}

\section{Force averaging of two-point  correlation function} \label{sec_App_ForceAverage2pCorr}

In this section we derive the two-point correlation function   Eq.  \eqref{eq_corr} that results from an average over phase space 
and over the stochastic non-equilibrium force $h(t)$. 
We start from the non-equilibrium GLE Eq. \eqref{eq_mori_GLE}. To reduce the notational burden we consider
an unconfined system  that is characterized by a diverging second moment, 
$ \langle (A_S (\omega) - \langle  A_S \rangle )^2 \rangle=\infty$, in which case
$K(t)=0= \Gamma_A(s,t)$. Eq.  \eqref{eq_mori_GLE} in this scenario simplifies to 
\begin{align} \label{ForceAve1}
 F(\omega,t_0,t) + h(t) /M = 
  \ddot A(\omega, t)   + 
  \int_{t_0}^{t} {\rm d}s\, \Gamma(s,t)  \dot A(\omega, s).
  \end{align}
Averaging  the product of Eq.  \eqref{ForceAve1} at two different times $t$ and $t'$ 
over phase space  and $h(t)$ we obtain the expression
\begin{align} \label{ForceAve2}
& \overline{ \left \langle  \left (F(\omega,t_0,t) + \frac{h(t) }{M} \right ) 
 \left ( F(\omega,t_0,t') + \frac{h(t') }{M} \right )    \right \rangle }  = 
  \overline{ \left \langle  F(\omega,t_0,t)  F(\omega,t_0,t')  \right \rangle }
 +  \overline{    \frac{h(t) h(t')  }{M^2}  }\nonumber \\ 
% & = 
%  \left \langle   \overline{ F(\omega,t_0,t) }  \overline{ F(\omega,t_0,t') } \right \rangle 
% +  \overline{    \frac{h(t) h(t')  }{M^2}  } +\Delta X_F \nonumber \\ 
&=  \overline{  \left \langle 
   \left ( \ddot A(\omega, t) + \int_{t_0}^{t} {\rm d}s\, \Gamma(s,t)  \dot A(\omega, s) \right )
  \left ( \ddot A(\omega, t') + \int_{t_0}^{t'} {\rm d}s'\, \Gamma(s',t')  \dot A(\omega, s')  \right)
 \right \rangle } \nonumber \\ 
&  = 
 \left \langle 
  \left ( \overline{     \ddot A(\omega, t) + \int_{t_0}^{t} {\rm d}s\, \Gamma(s,t)  \dot A(\omega, s)}  \right )
  \left (  \overline{  \ddot A(\omega, t') + \int_{t_0}^{t'} {\rm d}s'\, \Gamma(s',t')  \dot A(\omega, s') }  \right)
 \right \rangle + \Delta X,
  \end{align}
where in the first equation we used that $h(t)$ is independent of phase space and that $ \langle  F(\omega,t_0,t) \rangle =0$. 
We defined a  correction term as
\begin{align} \label{ForceAve3}
 \Delta X = & \overline{  \left \langle 
   \left ( \ddot A(\omega, t) + \int_{t_0}^{t} {\rm d}s\, \Gamma(s,t)  \dot A(\omega, s) \right )
  \left ( \ddot A(\omega, t') + \int_{t_0}^{t'} {\rm d}s'\, \Gamma(s',t')  \dot A(\omega, s')  \right)
 \right \rangle } \nonumber \\ 
&-
 \left \langle 
  \left ( \overline{     \ddot A(\omega, t) + \int_{t_0}^{t} {\rm d}s\, \Gamma(s,t)  \dot A(\omega, s)}  \right )
  \left (  \overline{  \ddot A(\omega, t') + \int_{t_0}^{t'} {\rm d}s'\, \Gamma(s',t')  \dot A(\omega, s') }  \right)
 \right \rangle.
  \end{align}
%
%and
%
%\begin{align} \label{ForceAve3b}
 %\Delta X_F =  \overline{ \left \langle  F(\omega,t_0,t)  F(\omega,t_0,t')  \right \rangle } - 
  % \left \langle   \overline{ F(\omega,t_0,t) }  \overline{ F(\omega,t_0,t') } \right \rangle 
  % \end{align}
%
Using that the observable $A(\omega,t)$  as the function that is determined by the differential equation 
\eqref{ForceAve1} does not explicitly depend on the
non-equilibrium force $h(t)$, the correction term can be rewritten as 
\begin{align} \label{ForceAve3b}
 \Delta X = & 
  \int_{t_0}^{t} {\rm d}s\, \int_{t_0}^{t'} {\rm d}s'\, 
  \left \langle  \dot A(\omega, s)  \dot A(\omega, s')   \right \rangle
   \left (  \overline{  \Gamma(s,t)  \Gamma(s',t') }-
   \overline{  \Gamma(s,t) }  \overline{  \Gamma(s',t') }
  \right)
  \end{align}
in terms of the  second non-equilibrium force cumulant of the  memory kernel.
Performing the average over $h(t)$ and neglecting the correction term $\Delta X$ in Eq. \eqref{ForceAve3} we obtain 
\begin{align} \label{ForceAve4}
 \overline{  \left \langle   F(\omega,t- t_0)    F(\omega,t'- t_0) \right \rangle } 
  + \frac{\sigma(t-t') }{M^2}  =
 \left \langle 
  \left (    \ddot A(\omega, t) + \int_{t_0}^{t}{\rm d}s\,  \overline   \Gamma(t-s)  \dot A(\omega, s)  \right )
  \left (  \ddot A(\omega, t') + \int_{t_0}^{t'}{\rm d}s'\,  \overline  \Gamma(t'-s')  \dot A(\omega, s')   \right)
 \right \rangle.
   \end{align}
 Performing a double Fourier integral over $t$ and $t'$ and using Eqs. \eqref{eq_mori_memory2} 
and  \eqref{eq_response2}, we obtain
\begin{align} \label{ForceAve5}
2 \pi \delta(\nu - \nu') \left[ 
\langle ( {\cal L}_0 A_S(\omega) )^2 \rangle  \tilde{\overline{ \Gamma}}_0(\nu)  + 
 \tilde{\sigma}(\nu) /M^2 \right] 
 =  \tilde \chi^{-1} (\nu) \tilde \chi^{-1} (\nu') 
  \left \langle  \tilde A(\omega,\nu) \tilde A(\omega,\nu') 
 \right \rangle.
   \end{align}
Realizing that the double Fourier integral over $t$ and $t'$  of the two-point correlation function
$C(t-t')$ is given by 
\begin{align} \label{ForceAve6}
2 \pi \delta(\nu - \nu') \tilde C(\nu) = 
  \left \langle  \tilde A(\omega,\nu) \tilde A(\omega,\nu') 
 \right \rangle ,
   \end{align}
 Eq.    \eqref{eq_corr3} follows. Inspection of Eq.    \eqref{eq_corr3}
 shows that it is linear in the complementary force and non-equilibrium force second moments
 $\tilde{\overline{ \Gamma}}_0(\nu)$ and $ \tilde{\sigma}(\nu)$.
 The correction term $\Delta X$ contains higher-order powers of these moments and is neglected.

%\section{GLE parameter extraction by solving the Volterra equation}  \label{sec_App_Volterra}

\section{Derivation of the standard FDT in the  Heisenberg picture} \label{sec_App_FDT}

Here we derive the standard  (i.e. non-equilibrium) FDT by leading-order perturbation  analysis of the 
 Heisenberg observable in Eq.  \eqref{eq_Heisenberg}, which deviates
 from the text-book derivation of the standard FDT. To proceed,
 using the expression for the Heisenberg observable Eq.  \eqref{eq_Heisenberg}
and the expression for the time-dependent Liouville operator  Eq. \eqref{eq_Liouville2},
we obtain

\begin{align}
\label{DFT1}
A(\omega, t) =   \exp_H\left(  \int_{t_0}^t  {\rm d}s  {\cal L} (s) \right)      A_{S}(\omega)
=   \exp_H\left(  \int_{t_0}^t  {\rm d}s  \left(  {\cal L}_0  -h(s) \Delta  {\cal L} \right) \right)      A_{S}(\omega),
\end{align}
where the Heisenberg propagator is defined in Eq. \eqref{eq_expH}.
Expanding the operator exponential to first order in $h(t)$, as shown in   Eq. \eqref{homo1}, we obtain
\begin{align}
\label{DFT2}
A(\omega, t) = \left[ e^{(t-t_0)  {\cal L}_0} - 
\int_{t_0}^t  {\rm d}s  h(s)  e^{(s-t)  {\cal L}_0}  \Delta  {\cal L}  e^{(t-s)  {\cal L}_0} \right] 
      A_{S}(\omega).
\end{align}
The mean observable follows using the definition Eq.  \eqref{eq_average} as
\begin{align}
\label{DFT3}
a(t) = \left  \langle A(\omega, t) \right  \rangle
= \left  \langle e^{(t-t_0)  {\cal L}_0}A_{S}(\omega)\right  \rangle - 
\left \langle \int_{t_0}^t  {\rm d}s  h(s)  e^{(s-t)  {\cal L}_0}  \Delta  {\cal L}  e^{(t-s)  {\cal L}_0}  A_{S}(\omega)
\right \rangle,
     \end{align}
     which can be rewritten, using the anti-self-adjointedness of $ {\cal L}_0$ and $ \Delta  {\cal L}$, as 
\begin{align}
\label{DFT4}
a(t) &= \left  \langle A_{S}(\omega)\right  \rangle -
 \int_{t_0}^t  {\rm d}s \beta  h(s)  \left \langle   A_{S}(\omega)   {\cal L}_0  e^{(t-s)  {\cal L}_0}  A_{S}(\omega)
\right \rangle \nonumber \\
&= \left  \langle A_{S}(\omega)\right  \rangle -
 \int_{t_0}^t  {\rm d}s \beta  h(s) \frac{   {\rm d}}{ {\rm d} t} \left \langle   A_{S}(\omega)     e^{(t-s)  {\cal L}_0}  A_{S}(\omega)
\right \rangle  \nonumber \\
 &= \left  \langle A_{S}(\omega)\right  \rangle +
 \int_{t_0}^t  {\rm d}s   h(s) \chi(t-s)/M.
     \end{align}
Here we defined the response function as 
\begin{align}
\label{DFT5}
 \chi(t) = - \beta M  \frac{{\rm d}}{ {\rm d} t}   \left \langle   A_{S}(\omega)     e^{t  {\cal L}_0}  A_{S}(\omega)
\right \rangle =  - \beta M  \frac{{\rm d}}{ {\rm d} t} C(t),
     \end{align}
which is the standard FDT as one finds it in text books, where the derivation is typically done in the Schrödinger picture
as there is no advantage of the Heisenberg picture for this derivation.  We in this paper use
the Heisenberg picture since it is  needed when
deriving the GLE. Note that Eq. \eqref{DFT5} has no dependence on the non-equilibrium force $h(t)$ whatsoever 
and thus is equivalent to Eq.  \eqref{eq_corr6}, which is  the Fourier-transformed version of the FDT, in the limit $h(t) =0$. 

To make the equivalence
between Eqs. \eqref{DFT5} and  \eqref{eq_corr6} obvious, 
we  Fourier transform Eqs. \eqref{DFT4} and \eqref{DFT5}. For this, we define
the response function $\chi(t)$ as single-sided and choose $t_0 \rightarrow - \infty$,
 after which  Eq. \eqref{DFT4}  reads 
\begin{align}
\label{DFT6}
a(t) &=
 \left  \langle A_{S}(\omega)\right  \rangle +
 \int_{-\infty}^\infty   {\rm d}s   h(s) \chi(t-s)/M.
     \end{align}
Fourier transformation yields 
\begin{align}  \label{DFT7}
\tilde a(\nu)  =  2\pi \delta(\nu) \langle  A_S \rangle  + \tilde \chi(\nu) \tilde h(\nu)/ M,
\end{align}
which is equivalent to Eq. \eqref{eq_response}.
Accounting for the single-sidedness of  $\chi(t)$,  Eq. \eqref{DFT5} can be written as 
\begin{align}
\label{DFT8}
 \chi(t) =   - \beta M  \theta(t) \frac{{\rm d}}{ {\rm d} t} C(t),
     \end{align}
where $\theta(t)$ denotes the Heavyside function.
After Fourier transformation  Eq.\eqref{DFT8} reads
\begin{align}
\label{DFT9}
\tilde  \chi(\nu ) = 
\beta M C(0)    - \imath \nu  \beta M \tilde C^+(\nu).
     \end{align}
The odd part of this equation reads
\begin{align}
\label{DFT10}
\tilde  \chi(\nu ) - \tilde  \chi(- \nu ) =     - \imath \nu \beta M  \tilde C(\nu),
     \end{align}
where we used that the Fourier transform of the symmetric correlation
function is $ \tilde C(\nu) =  \tilde C^+(\nu) +  \tilde C^+(-\nu)$.
Since the time-domain response function $\chi(t)$ is a real function, we finally obtain
\begin{align}
\label{DFT11}
2 \Im \left(  \tilde  \chi(\nu ) \right)  =     -  \nu \beta M  \tilde C(\nu) = - \frac{ \nu  \tilde C(\nu)}
{  \langle ( {\cal L}_0 A_S(\omega) )^2 \rangle},
     \end{align}
     where we used the definition of the mass M in Eq. \eqref{eq_mass} and 
which is equivalent to Eq.  \eqref{eq_corr6},  provided we take the equilibrium limit $h(t)=0$
in Eq.  \eqref{eq_corr6}.

\section{Derivation of joint distribution function via path integrals} \label{sec_App_PathIntegral}

Inserting the Fourier-transform of Eq.  \eqref{eq_resp1} into Eq. \eqref{dist2}   we obtain
\begin{align} \label{app_path1}
 \rho(A_2,t_2;A_1,t_1) = 
 \int_{-\infty}^{\infty} \frac{{\rm d} q_1}{2\pi} \int_{-\infty}^{\infty} \frac{{\rm d} q_2}{2\pi} 
  e^{\imath q_1 (A_1 - \langle A_S \rangle) + \imath q_2 (A_2 - \langle A_S \rangle)} 
\overline{ \left \langle \exp\left(  - \imath  \int_{-\infty}^{\infty} \frac{{\rm d} \nu }{2\pi}
\left( \tilde F(\omega, \nu) + \tilde h(\nu)/M \right) \tilde g(\nu) \right) \right \rangle }
\end{align}
where we have defined 
\begin{align} \label{app_path2}
\tilde g(\nu) \equiv  \tilde \chi(\nu) \left( q_1 e^{\imath \nu t_1} + q_2 e^{\imath \nu t_2} \right).
\end{align}
We diagonalize   the path integrals in Eqs.  \eqref{eq_resp2} and  \eqref{eq_resp3} 
by Fourier transformation and obtain 
\begin{align}  \label{app_path3}
 & \langle  X(\tilde F, \tilde h)  \rangle =  \int_{-\infty}^{\infty} 
 \frac{ {\cal D} \tilde F(\omega,\cdot)}{\tilde {\cal N}_F}
 X(\tilde F, \tilde h) 
  \exp\left( -  \int_{-\infty}^{\infty} \frac{ {\rm d} \nu}{2 \pi}  
 \frac{ \tilde F(\omega,\nu ) \tilde F(\omega,-\nu) 
  }{ 2  {  \langle ( {\cal L}_0 A_S(\omega) )^2 \rangle}    \tilde \Gamma_0(\nu)  }
 \right),
 \end{align}
\begin{align}  \label{app_path4}
  \overline{  X(\tilde F, \tilde h)} =  \int_{-\infty}^{\infty} 
 \frac{ {\cal D} \tilde h(\cdot)}{\tilde{\cal N}_h}
 X(\tilde F, \tilde h) 
  \exp\left( -  \int_{-\infty}^{\infty}\frac{ {\rm d} \nu}{2 \pi} 
 \frac{ \tilde h(\nu) \tilde h(- \nu) 
 }{ 2 \tilde  \sigma(\nu)}
 \right),
 \end{align}
where $\tilde {\cal N}_F$ and $\tilde {\cal N}_h$ are normalization constants.
Now the averages over $ \tilde F(\omega, \nu)$ and $ \tilde h(\nu)$ in 
Eq. \eqref{app_path1} can be done by explicitly performing the path integrals, after which we obtain
\begin{align} \label{app_path5}
 \rho(A_2,t_2;A_1,t_1) &= 
 \int_{-\infty}^{\infty} \frac{{\rm d} q_1}{2\pi} \int_{-\infty}^{\infty} \frac{{\rm d} q_2}{2\pi} 
  e^{\imath q_1 (A_1 - \langle A_S \rangle) + \imath q_2 (A_2 - \langle A_S \rangle)} 
  \nonumber \\ & \times
 \exp\left(  -   \int_{-\infty}^{\infty} \frac{{\rm d} \nu }{4\pi}
  \tilde g(\nu)  \tilde g(-\nu) \left( 
 \langle ( {\cal L}_0 A_S(\omega) )^2 \rangle   \tilde \Gamma_0(\nu)
 + \tilde  \sigma(\nu)/M^2
  \right) \right).
\end{align}
Inserting the definition of $\tilde g(\nu)$ from Eq.  \eqref{app_path2}
we obtain 
\begin{align} \label{app_path6}
 \rho(A_2,t_2;A_1,t_1) &= 
 \int_{-\infty}^{\infty} \frac{{\rm d} q_1}{2\pi} \int_{-\infty}^{\infty} \frac{{\rm d} q_2}{2\pi} 
  e^{\imath q_1 (A_1 - \langle A_S \rangle) + \imath q_2 (A_2 - \langle A_S \rangle)} 
 \exp\left(  -   \int_{-\infty}^{\infty} \frac{{\rm d} \nu }{4\pi}
 \tilde C(\nu) \sum_{j,k=1}^2 q_j q_k e^{\imath \nu(t_j - t_k)}
   \right),
\end{align}
where we used the explicit expression for the Fourier-transformed correlation function 
$ \tilde C(\nu)$ from Eq.  \eqref{eq_corr3}. Now the Fourier integral  in the exponent can be done
and we obtain 
\begin{align} \label{app_path7}
 \rho(A_2,t_2;A_1,t_1) &= 
 \int_{-\infty}^{\infty} \frac{{\rm d} q_1}{2\pi} \int_{-\infty}^{\infty} \frac{{\rm d} q_2}{2\pi} 
  e^{\imath q_1 (A_1 - \langle A_S \rangle) + \imath q_2 (A_2 - \langle A_S \rangle)} 
 \exp\left(  -  \frac{1}{2}  \sum_{j,k=1}^2 q_j q_k 
 C(t_k - t_j)    \right),
\end{align}
which can be slightly rewritten as 
\begin{align} \label{app_path8}
 \rho(A_2,t_2;A_1,t_1) &= 
 \int_{-\infty}^{\infty} \frac{{\rm d} q_1}{2\pi} \int_{-\infty}^{\infty} \frac{{\rm d} q_2}{2\pi} 
 \exp\left(  -  \frac{1}{2}  \sum_{j,k=1}^2 q_j q_k   C(t_k - t_j) 
 + \imath  \sum_{j=1}^2   q_j (A_j - \langle A_S \rangle)
    \right).
\end{align}
Now we use the definition from the main text Eq.  \eqref{dist4},  $I_{jk} = C(t_j - t_k)$,
and obtain
\begin{align} \label{app_path9}
 \rho(A_2,t_2;A_1,t_1) &= 
 \int_{-\infty}^{\infty} \frac{{\rm d} q_1}{2\pi} \int_{-\infty}^{\infty} \frac{{\rm d} q_2}{2\pi} 
 \exp\left(  -  \frac{1}{2}  \sum_{j,k=1}^2 q_j q_k   I_{jk}
 + \imath \sum_{j=1}^2   q_j (A_j - \langle A_S \rangle)
    \right).
\end{align}
The Gaussian integrals over $q_1$ and $q_2$ can be done, after which we obtain the final result reported in Eq. \eqref{dist3},
\begin{align} \label{app_path10}
 \rho(A_2,t_2;A_1,t_1) = 
 \frac{  \exp \left(-  \sum_{j,k=1}^2  (A_j -  \langle  A_S \rangle) I_{jk}^{-1}  (A_k -  \langle  A_S \rangle)/2  \right)}
 { \sqrt{ \det 2 \pi I}},
\end{align}
which  is the two-point distribution of a general Gaussian process. Note that in Eq. \eqref{dist3} the Einstein summation convention  is used. 

As a  final step, we show how to marginalize the two-point distribution. 
This is most easily done on the level of Eq. \eqref{app_path9}
where the positions $A_1$ and $A_2$ appear linearly. By integration over one observable value we obtain
\begin{align} \label{app_path11}
 \rho(A_1,t_1) &=  \int_{-\infty}^{\infty} {\rm d} A_2
  \rho(A_2,t_2;A_1,t_1)=
   \int_{-\infty}^{\infty} \frac{{\rm d} q_1}{2\pi} 
  \exp\left(  -  \frac{q_1^2   C(0)  }{2}  
 +   q_1 (A_1 - \langle A_S \rangle)    \right)
 = \frac{    \exp\left(  -  \frac{(A_1 - \langle A_S \rangle)^2     }{2C(0)}      \right)}
 { \sqrt{2 \pi C(0)}},
\end{align}
which, expectedly,  is a Gaussian distribution with a variance corresponding to the 
equal-time correlation $C(0)$.

\end{widetext}

\bibliography{NonEq}

%apsrev4-2.bst 2019-01-14 (MD) hand-edited version of apsrev4-1.bst
%Control: key (0)
%Control: author (8) initials jnrlst
%Control: editor formatted (1) identically to author
%Control: production of article title (0) allowed
%Control: page (0) single
%Control: year (1) truncated
%Control: production of eprint (0) enabled
\begin{thebibliography}{78}%
\makeatletter
\providecommand \@ifxundefined [1]{%
 \@ifx{#1\undefined}
}%
\providecommand \@ifnum [1]{%
 \ifnum #1\expandafter \@firstoftwo
 \else \expandafter \@secondoftwo
 \fi
}%
\providecommand \@ifx [1]{%
 \ifx #1\expandafter \@firstoftwo
 \else \expandafter \@secondoftwo
 \fi
}%
\providecommand \natexlab [1]{#1}%
\providecommand \enquote  [1]{``#1''}%
\providecommand \bibnamefont  [1]{#1}%
\providecommand \bibfnamefont [1]{#1}%
\providecommand \citenamefont [1]{#1}%
\providecommand \href@noop [0]{\@secondoftwo}%
\providecommand \href [0]{\begingroup \@sanitize@url \@href}%
\providecommand \@href[1]{\@@startlink{#1}\@@href}%
\providecommand \@@href[1]{\endgroup#1\@@endlink}%
\providecommand \@sanitize@url [0]{\catcode `\\12\catcode `\$12\catcode
  `\&12\catcode `\#12\catcode `\^12\catcode `\_12\catcode `\%12\relax}%
\providecommand \@@startlink[1]{}%
\providecommand \@@endlink[0]{}%
\providecommand \url  [0]{\begingroup\@sanitize@url \@url }%
\providecommand \@url [1]{\endgroup\@href {#1}{\urlprefix }}%
\providecommand \urlprefix  [0]{URL }%
\providecommand \Eprint [0]{\href }%
\providecommand \doibase [0]{https://doi.org/}%
\providecommand \selectlanguage [0]{\@gobble}%
\providecommand \bibinfo  [0]{\@secondoftwo}%
\providecommand \bibfield  [0]{\@secondoftwo}%
\providecommand \translation [1]{[#1]}%
\providecommand \BibitemOpen [0]{}%
\providecommand \bibitemStop [0]{}%
\providecommand \bibitemNoStop [0]{.\EOS\space}%
\providecommand \EOS [0]{\spacefactor3000\relax}%
\providecommand \BibitemShut  [1]{\csname bibitem#1\endcsname}%
\let\auto@bib@innerbib\@empty
%</preamble>
\bibitem [{\citenamefont {Prigogine}(1947)}]{Prigogine1947}%
  \BibitemOpen
  \bibfield  {author} {\bibinfo {author} {\bibfnamefont {I.}~\bibnamefont
  {Prigogine}},\ }\href@noop {} {\emph {\bibinfo {title} {Etude thermodynamic
  des phenomene irreversibles}}}\ (\bibinfo  {publisher} {Desoer, Liege},\
  \bibinfo {year} {1947})\BibitemShut {NoStop}%
\bibitem [{\citenamefont {Prigogine}\ and\ \citenamefont
  {Mazur}(1953)}]{Mazur1953}%
  \BibitemOpen
  \bibfield  {author} {\bibinfo {author} {\bibfnamefont {I.}~\bibnamefont
  {Prigogine}}\ and\ \bibinfo {author} {\bibfnamefont {P.}~\bibnamefont
  {Mazur}},\ }\bibfield  {title} {\bibinfo {title} {Sur l'extension de la
  thermodynamique aux phenomenes irreversibles lies aux degres de liberte
  internes},\ }\href@noop {} {\bibfield  {journal} {\bibinfo  {journal}
  {Physica}\ }\textbf {\bibinfo {volume} {XIX}},\ \bibinfo {pages} {241}
  (\bibinfo {year} {1953})}\BibitemShut {NoStop}%
\bibitem [{\citenamefont {Lebowitz}(1959)}]{Lebowitz1959}%
  \BibitemOpen
  \bibfield  {author} {\bibinfo {author} {\bibfnamefont {J.~L.}\ \bibnamefont
  {Lebowitz}},\ }\bibfield  {title} {\bibinfo {title} {Stationary
  nonequilibrium {Gibbsian} ensembles},\ }\href@noop {} {\bibfield  {journal}
  {\bibinfo  {journal} {Phys. Rev.}\ }\textbf {\bibinfo {volume} {114}},\
  \bibinfo {pages} {1192} (\bibinfo {year} {1959})}\BibitemShut {NoStop}%
\bibitem [{\citenamefont {Zwanzig}(1960)}]{zwanzig_ensemble_1960}%
  \BibitemOpen
  \bibfield  {author} {\bibinfo {author} {\bibfnamefont {R.}~\bibnamefont
  {Zwanzig}},\ }\bibfield  {title} {\bibinfo {title} {Ensemble method in the
  theory of irreversibility},\ }\href {https://doi.org/10.1063/1.1731409}
  {\bibfield  {journal} {\bibinfo  {journal} {The Journal of Chemical Physics}\
  }\textbf {\bibinfo {volume} {33}},\ \bibinfo {pages} {1338} (\bibinfo {year}
  {1960})}\BibitemShut {NoStop}%
\bibitem [{\citenamefont {de~Groot}\ and\ \citenamefont
  {Mazur}(1962)}]{deGroot}%
  \BibitemOpen
  \bibfield  {author} {\bibinfo {author} {\bibfnamefont {S.~R.}\ \bibnamefont
  {de~Groot}}\ and\ \bibinfo {author} {\bibfnamefont {P.}~\bibnamefont
  {Mazur}},\ }\href@noop {} {\emph {\bibinfo {title} {Non-Equilibrium
  Thermodynamics}}}\ (\bibinfo  {publisher} {North-Holland Pub. Co.,
  Amsterdam},\ \bibinfo {year} {1962})\BibitemShut {NoStop}%
\bibitem [{\citenamefont {Grabert}\ \emph {et~al.}(1980)\citenamefont
  {Grabert}, \citenamefont {H{\"a}nggi},\ and\ \citenamefont
  {Talkner}}]{grabert_microdynamics_1980}%
  \BibitemOpen
  \bibfield  {author} {\bibinfo {author} {\bibfnamefont {H.}~\bibnamefont
  {Grabert}}, \bibinfo {author} {\bibfnamefont {P.}~\bibnamefont
  {H{\"a}nggi}},\ and\ \bibinfo {author} {\bibfnamefont {P.}~\bibnamefont
  {Talkner}},\ }\bibfield  {title} {\bibinfo {title} {Microdynamics and
  nonlinear stochastic processes of gross variables},\ }\href
  {https://doi.org/10.1007/BF01011337} {\bibfield  {journal} {\bibinfo
  {journal} {Journal of Statistical Physics}\ }\textbf {\bibinfo {volume}
  {22}},\ \bibinfo {pages} {537} (\bibinfo {year} {1980})}\BibitemShut
  {NoStop}%
\bibitem [{\citenamefont {Risken}(1984)}]{Risken}%
  \BibitemOpen
  \bibfield  {author} {\bibinfo {author} {\bibfnamefont {H.}~\bibnamefont
  {Risken}},\ }\href@noop {} {\emph {\bibinfo {title} {The Fokker-Planck
  Equation}}}\ (\bibinfo  {publisher} {Springer, Berlin},\ \bibinfo {year}
  {1984})\BibitemShut {NoStop}%
\bibitem [{\citenamefont {Zwanzig}(2001)}]{zwanzig_nonequilibrium_2001}%
  \BibitemOpen
  \bibfield  {author} {\bibinfo {author} {\bibfnamefont {R.}~\bibnamefont
  {Zwanzig}},\ }\href@noop {} {\emph {\bibinfo {title} {Nonequilibrium
  statistical mechanics}}}\ (\bibinfo  {publisher} {Oxford UnivPress},\
  \bibinfo {address} {Oxford [u.a.]},\ \bibinfo {year} {2001})\BibitemShut
  {NoStop}%
\bibitem [{\citenamefont {Mizuno}\ \emph {et~al.}(2007)\citenamefont {Mizuno},
  \citenamefont {Tardin}, \citenamefont {Schmidt},\ and\ \citenamefont
  {MacKintosh}}]{Schmidt07}%
  \BibitemOpen
  \bibfield  {author} {\bibinfo {author} {\bibfnamefont {D.}~\bibnamefont
  {Mizuno}}, \bibinfo {author} {\bibfnamefont {C.}~\bibnamefont {Tardin}},
  \bibinfo {author} {\bibfnamefont {C.~F.}\ \bibnamefont {Schmidt}},\ and\
  \bibinfo {author} {\bibfnamefont {F.~C.}\ \bibnamefont {MacKintosh}},\
  }\bibfield  {title} {\bibinfo {title} {Nonequilibrium mechanics of active
  cytoskeletal networks},\ }\href@noop {} {\bibfield  {journal} {\bibinfo
  {journal} {Science}\ }\textbf {\bibinfo {volume} {315}},\ \bibinfo {pages}
  {370} (\bibinfo {year} {2007})}\BibitemShut {NoStop}%
\bibitem [{\citenamefont {Gomez-Solano}\ \emph {et~al.}(2009)\citenamefont
  {Gomez-Solano}, \citenamefont {Petrosyan}, \citenamefont {Ciliberto},
  \citenamefont {Chetrite},\ and\ \citenamefont {Gawedzki}}]{Gawedzki2009}%
  \BibitemOpen
  \bibfield  {author} {\bibinfo {author} {\bibfnamefont {J.~R.}\ \bibnamefont
  {Gomez-Solano}}, \bibinfo {author} {\bibfnamefont {A.}~\bibnamefont
  {Petrosyan}}, \bibinfo {author} {\bibfnamefont {S.}~\bibnamefont
  {Ciliberto}}, \bibinfo {author} {\bibfnamefont {R.}~\bibnamefont
  {Chetrite}},\ and\ \bibinfo {author} {\bibfnamefont {K.}~\bibnamefont
  {Gawedzki}},\ }\bibfield  {title} {\bibinfo {title} {Experimental
  verification of a modified fluctuation-dissipation relation for a
  micron-sized particle in a nonequilibrium steady state},\ }\href@noop {}
  {\bibfield  {journal} {\bibinfo  {journal} {Phys. Rev. Lett.}\ }\textbf
  {\bibinfo {volume} {103}},\ \bibinfo {pages} {040601} (\bibinfo {year}
  {2009})}\BibitemShut {NoStop}%
\bibitem [{\citenamefont {Mehl}\ \emph {et~al.}(2010)\citenamefont {Mehl},
  \citenamefont {Blickle}, \citenamefont {Seifert},\ and\ \citenamefont
  {Bechinger}}]{Bechinger2010}%
  \BibitemOpen
  \bibfield  {author} {\bibinfo {author} {\bibfnamefont {J.}~\bibnamefont
  {Mehl}}, \bibinfo {author} {\bibfnamefont {V.}~\bibnamefont {Blickle}},
  \bibinfo {author} {\bibfnamefont {U.}~\bibnamefont {Seifert}},\ and\ \bibinfo
  {author} {\bibfnamefont {C.}~\bibnamefont {Bechinger}},\ }\bibfield  {title}
  {\bibinfo {title} {Experimental accessibility of generalized
  fluctuation-dissipation relations for nonequilibrium steady states},\
  }\href@noop {} {\bibfield  {journal} {\bibinfo  {journal} {Phys. Rev. E}\
  }\textbf {\bibinfo {volume} {82}},\ \bibinfo {pages} {032401} (\bibinfo
  {year} {2010})}\BibitemShut {NoStop}%
\bibitem [{\citenamefont {Theurkauff}\ \emph {et~al.}(2012)\citenamefont
  {Theurkauff}, \citenamefont {Cottin-Bizonne}, \citenamefont {Palacci},
  \citenamefont {Ybert},\ and\ \citenamefont {Bocquet}}]{Bocquet2012}%
  \BibitemOpen
  \bibfield  {author} {\bibinfo {author} {\bibfnamefont {I.}~\bibnamefont
  {Theurkauff}}, \bibinfo {author} {\bibfnamefont {C.}~\bibnamefont
  {Cottin-Bizonne}}, \bibinfo {author} {\bibfnamefont {J.}~\bibnamefont
  {Palacci}}, \bibinfo {author} {\bibfnamefont {C.}~\bibnamefont {Ybert}},\
  and\ \bibinfo {author} {\bibfnamefont {L.}~\bibnamefont {Bocquet}},\
  }\bibfield  {title} {\bibinfo {title} {Dynamic clustering in active colloidal
  suspensions with chemical signaling},\ }\href@noop {} {\bibfield  {journal}
  {\bibinfo  {journal} {Phys. Rev. Lett.}\ }\textbf {\bibinfo {volume} {108}},\
  \bibinfo {pages} {268303} (\bibinfo {year} {2012})}\BibitemShut {NoStop}%
\bibitem [{\citenamefont {Dinis}\ \emph {et~al.}(2012)\citenamefont {Dinis},
  \citenamefont {Martin}, \citenamefont {Barral}, \citenamefont {Prost},\ and\
  \citenamefont {Joanny}}]{Dinis2012}%
  \BibitemOpen
  \bibfield  {author} {\bibinfo {author} {\bibfnamefont {L.}~\bibnamefont
  {Dinis}}, \bibinfo {author} {\bibfnamefont {P.}~\bibnamefont {Martin}},
  \bibinfo {author} {\bibfnamefont {J.}~\bibnamefont {Barral}}, \bibinfo
  {author} {\bibfnamefont {J.}~\bibnamefont {Prost}},\ and\ \bibinfo {author}
  {\bibfnamefont {J.-F.}\ \bibnamefont {Joanny}},\ }\bibfield  {title}
  {\bibinfo {title} {Fluctuation-response theorem for the active noisy
  oscillator of the hair-cell bundle},\ }\href@noop {} {\bibfield  {journal}
  {\bibinfo  {journal} {Phys. Rev. Lett.}\ }\textbf {\bibinfo {volume} {109}},\
  \bibinfo {pages} {160602} (\bibinfo {year} {2012})}\BibitemShut {NoStop}%
\bibitem [{\citenamefont {Bohec}\ \emph {et~al.}(2013)\citenamefont {Bohec},
  \citenamefont {F.~Gallet}, \citenamefont {Maes}, \citenamefont {Safaverdi},
  \citenamefont {Visco},\ and\ \citenamefont {van Wijland}}]{Bohec2013}%
  \BibitemOpen
  \bibfield  {author} {\bibinfo {author} {\bibfnamefont {P.}~\bibnamefont
  {Bohec}}, \bibinfo {author} {\bibfnamefont {F.}~\bibnamefont {F.~Gallet}},
  \bibinfo {author} {\bibfnamefont {C.}~\bibnamefont {Maes}}, \bibinfo {author}
  {\bibfnamefont {S.}~\bibnamefont {Safaverdi}}, \bibinfo {author}
  {\bibfnamefont {P.}~\bibnamefont {Visco}},\ and\ \bibinfo {author}
  {\bibfnamefont {F.}~\bibnamefont {van Wijland}},\ }\bibfield  {title}
  {\bibinfo {title} {Probing active forces via a fluctuation-dissipation
  relation: Application to living cells},\ }\href@noop {} {\bibfield  {journal}
  {\bibinfo  {journal} {Europhys. Lett.}\ }\textbf {\bibinfo {volume} {102}},\
  \bibinfo {pages} {50005} (\bibinfo {year} {2013})}\BibitemShut {NoStop}%
\bibitem [{\citenamefont {Guo}\ \emph {et~al.}(2014)\citenamefont {Guo},
  \citenamefont {Ehrlicher}, \citenamefont {Jensen}, \citenamefont {Renz},
  \citenamefont {Moore}, \citenamefont {Goldman}, \citenamefont
  {Lippincott-Schwartz}, \citenamefont {MacKintosh},\ and\ \citenamefont
  {Weitz}}]{Weitz2014}%
  \BibitemOpen
  \bibfield  {author} {\bibinfo {author} {\bibfnamefont {M.}~\bibnamefont
  {Guo}}, \bibinfo {author} {\bibfnamefont {A.~J.}\ \bibnamefont {Ehrlicher}},
  \bibinfo {author} {\bibfnamefont {M.~H.}\ \bibnamefont {Jensen}}, \bibinfo
  {author} {\bibfnamefont {M.}~\bibnamefont {Renz}}, \bibinfo {author}
  {\bibfnamefont {J.~R.}\ \bibnamefont {Moore}}, \bibinfo {author}
  {\bibfnamefont {R.~D.}\ \bibnamefont {Goldman}}, \bibinfo {author}
  {\bibfnamefont {J.}~\bibnamefont {Lippincott-Schwartz}}, \bibinfo {author}
  {\bibfnamefont {F.~C.}\ \bibnamefont {MacKintosh}},\ and\ \bibinfo {author}
  {\bibfnamefont {D.~A.}\ \bibnamefont {Weitz}},\ }\href@noop {} {\bibfield
  {journal} {\bibinfo  {journal} {Cell}\ }\textbf {\bibinfo {volume} {158}},\
  \bibinfo {pages} {822} (\bibinfo {year} {2014})}\BibitemShut {NoStop}%
\bibitem [{\citenamefont {Krug}(1991)}]{Krug}%
  \BibitemOpen
  \bibfield  {author} {\bibinfo {author} {\bibfnamefont {J.}~\bibnamefont
  {Krug}},\ }\bibfield  {title} {\bibinfo {title} {Boundary-induced phase
  transitions in driven diffusive systems},\ }\href@noop {} {\bibfield
  {journal} {\bibinfo  {journal} {Phys. Rev. Lett.}\ }\textbf {\bibinfo
  {volume} {67}},\ \bibinfo {pages} {1882} (\bibinfo {year}
  {1991})}\BibitemShut {NoStop}%
\bibitem [{\citenamefont {Schmittmann}\ and\ \citenamefont {Zia}(1998)}]{Zia}%
  \BibitemOpen
  \bibfield  {author} {\bibinfo {author} {\bibfnamefont {B.}~\bibnamefont
  {Schmittmann}}\ and\ \bibinfo {author} {\bibfnamefont {R.~K.~P.}\
  \bibnamefont {Zia}},\ }\bibfield  {title} {\bibinfo {title} {Driven diffusive
  systems. an introduction and recent developments},\ }\href@noop {} {\bibfield
   {journal} {\bibinfo  {journal} {Phys. Rep.}\ }\textbf {\bibinfo {volume}
  {301}},\ \bibinfo {pages} {45} (\bibinfo {year} {1998})}\BibitemShut
  {NoStop}%
\bibitem [{\citenamefont {Derrida}\ \emph {et~al.}(2001)\citenamefont
  {Derrida}, \citenamefont {Lebowitz},\ and\ \citenamefont
  {Speer}}]{Derrida2001}%
  \BibitemOpen
  \bibfield  {author} {\bibinfo {author} {\bibfnamefont {B.}~\bibnamefont
  {Derrida}}, \bibinfo {author} {\bibfnamefont {J.~L.}\ \bibnamefont
  {Lebowitz}},\ and\ \bibinfo {author} {\bibfnamefont {E.~R.}\ \bibnamefont
  {Speer}},\ }\bibfield  {title} {\bibinfo {title} {Free energy functional for
  nonequilibrium systems: An exactly solvable case},\ }\href@noop {} {\bibfield
   {journal} {\bibinfo  {journal} {Phys. Rev. Lett.}\ }\textbf {\bibinfo
  {volume} {87}},\ \bibinfo {pages} {150601} (\bibinfo {year}
  {2001})}\BibitemShut {NoStop}%
\bibitem [{\citenamefont {Ilg}\ and\ \citenamefont
  {Barrat}(2007)}]{Barrat2007}%
  \BibitemOpen
  \bibfield  {author} {\bibinfo {author} {\bibfnamefont {P.}~\bibnamefont
  {Ilg}}\ and\ \bibinfo {author} {\bibfnamefont {J.~L.}\ \bibnamefont
  {Barrat}},\ }\bibfield  {title} {\bibinfo {title} {From single-particle to
  collective effective temperatures in an active fluid of self-propelled
  particles},\ }\href@noop {} {\bibfield  {journal} {\bibinfo  {journal}
  {Europhys. Lett.}\ }\textbf {\bibinfo {volume} {111}},\ \bibinfo {pages}
  {26001} (\bibinfo {year} {2007})}\BibitemShut {NoStop}%
\bibitem [{\citenamefont {Grosberg}\ and\ \citenamefont
  {Joanny}(2015)}]{Grosberg2015}%
  \BibitemOpen
  \bibfield  {author} {\bibinfo {author} {\bibfnamefont {A.~Y.}\ \bibnamefont
  {Grosberg}}\ and\ \bibinfo {author} {\bibfnamefont {J.~F.}\ \bibnamefont
  {Joanny}},\ }\bibfield  {title} {\bibinfo {title} {Nonequilibrium statistical
  mechanics of mixtures of particles in contact with different thermostats},\
  }\href@noop {} {\bibfield  {journal} {\bibinfo  {journal} {Phys. Rev. E}\
  }\textbf {\bibinfo {volume} {92}},\ \bibinfo {pages} {032118} (\bibinfo
  {year} {2015})}\BibitemShut {NoStop}%
\bibitem [{\citenamefont {Fodor}\ \emph {et~al.}(2016)\citenamefont {Fodor},
  \citenamefont {Nardini}, \citenamefont {Cates}, \citenamefont {Tailleur},
  \citenamefont {Visco},\ and\ \citenamefont {van Wijland}}]{Fodor2016}%
  \BibitemOpen
  \bibfield  {author} {\bibinfo {author} {\bibfnamefont {E.}~\bibnamefont
  {Fodor}}, \bibinfo {author} {\bibfnamefont {C.}~\bibnamefont {Nardini}},
  \bibinfo {author} {\bibfnamefont {M.~E.}\ \bibnamefont {Cates}}, \bibinfo
  {author} {\bibfnamefont {J.}~\bibnamefont {Tailleur}}, \bibinfo {author}
  {\bibfnamefont {P.}~\bibnamefont {Visco}},\ and\ \bibinfo {author}
  {\bibfnamefont {F.}~\bibnamefont {van Wijland}},\ }\bibfield  {title}
  {\bibinfo {title} {How far from equilibrium is active matter?},\ }\href@noop
  {} {\bibfield  {journal} {\bibinfo  {journal} {Phys. Rev. Lett.}\ }\textbf
  {\bibinfo {volume} {117}},\ \bibinfo {pages} {038103} (\bibinfo {year}
  {2016})}\BibitemShut {NoStop}%
\bibitem [{\citenamefont {Carlon}\ \emph {et~al.}(2018)\citenamefont {Carlon},
  \citenamefont {Orland}, \citenamefont {Sakaue},\ and\ \citenamefont
  {Vanderzande}}]{Carlon2018}%
  \BibitemOpen
  \bibfield  {author} {\bibinfo {author} {\bibfnamefont {E.}~\bibnamefont
  {Carlon}}, \bibinfo {author} {\bibfnamefont {H.}~\bibnamefont {Orland}},
  \bibinfo {author} {\bibfnamefont {T.}~\bibnamefont {Sakaue}},\ and\ \bibinfo
  {author} {\bibfnamefont {C.}~\bibnamefont {Vanderzande}},\ }\bibfield
  {title} {\bibinfo {title} {Effect of memory and active forces on transition
  path time distributions},\ }\href@noop {} {\bibfield  {journal} {\bibinfo
  {journal} {J. Phys. Chem. B}\ }\textbf {\bibinfo {volume} {122}},\ \bibinfo
  {pages} {11186} (\bibinfo {year} {2018})}\BibitemShut {NoStop}%
\bibitem [{\citenamefont {Lavacchi}\ \emph {et~al.}(2022)\citenamefont
  {Lavacchi}, \citenamefont {Daldrop},\ and\ \citenamefont
  {Netz}}]{Lavacchi2022}%
  \BibitemOpen
  \bibfield  {author} {\bibinfo {author} {\bibfnamefont {L.}~\bibnamefont
  {Lavacchi}}, \bibinfo {author} {\bibfnamefont {J.~O.}\ \bibnamefont
  {Daldrop}},\ and\ \bibinfo {author} {\bibfnamefont {R.~R.}\ \bibnamefont
  {Netz}},\ }\bibfield  {title} {\bibinfo {title} {Non-arrhenius barrier
  crossing dynamics of non-equilibrium non-markovian systems},\ }\href@noop {}
  {\bibfield  {journal} {\bibinfo  {journal} {Europhys. Lett.}\ }\textbf
  {\bibinfo {volume} {139}},\ \bibinfo {pages} {51001} (\bibinfo {year}
  {2022})}\BibitemShut {NoStop}%
\bibitem [{\citenamefont {Jarzynski}(2000)}]{Jarzynski2000}%
  \BibitemOpen
  \bibfield  {author} {\bibinfo {author} {\bibfnamefont {C.}~\bibnamefont
  {Jarzynski}},\ }\bibfield  {title} {\bibinfo {title} {Hamiltonian derivation
  of a detailed fluctuation theorem},\ }\href@noop {} {\bibfield  {journal}
  {\bibinfo  {journal} {Journal of Statistical Physics}\ }\textbf {\bibinfo
  {volume} {98}},\ \bibinfo {pages} {77} (\bibinfo {year} {2000})}\BibitemShut
  {NoStop}%
\bibitem [{\citenamefont {Hatano}\ and\ \citenamefont
  {Sasa}(2001)}]{Hatano2001}%
  \BibitemOpen
  \bibfield  {author} {\bibinfo {author} {\bibfnamefont {T.}~\bibnamefont
  {Hatano}}\ and\ \bibinfo {author} {\bibfnamefont {S.~I.}\ \bibnamefont
  {Sasa}},\ }\bibfield  {title} {\bibinfo {title} {Steady-state thermodynamics
  of langevin systems},\ }\href@noop {} {\bibfield  {journal} {\bibinfo
  {journal} {Phys. Rev. Lett.}\ }\textbf {\bibinfo {volume} {86}},\ \bibinfo
  {pages} {3463} (\bibinfo {year} {2001})}\BibitemShut {NoStop}%
\bibitem [{\citenamefont {Harada}\ and\ \citenamefont
  {Sasa}(2005)}]{Harada2005}%
  \BibitemOpen
  \bibfield  {author} {\bibinfo {author} {\bibfnamefont {T.}~\bibnamefont
  {Harada}}\ and\ \bibinfo {author} {\bibfnamefont {S.~I.}\ \bibnamefont
  {Sasa}},\ }\bibfield  {title} {\bibinfo {title} {Equality connecting energy
  dissipation with a violation of the fluctuation-response relation},\
  }\href@noop {} {\bibfield  {journal} {\bibinfo  {journal} {Phys. Rev. Lett.}\
  }\textbf {\bibinfo {volume} {95}},\ \bibinfo {pages} {130602} (\bibinfo
  {year} {2005})}\BibitemShut {NoStop}%
\bibitem [{\citenamefont {Seifert}(2005)}]{Seifert2005}%
  \BibitemOpen
  \bibfield  {author} {\bibinfo {author} {\bibfnamefont {U.}~\bibnamefont
  {Seifert}},\ }\bibfield  {title} {\bibinfo {title} {Entropy production along
  a stochastic trajectory and an integral fluctuation theorem},\ }\href@noop {}
  {\bibfield  {journal} {\bibinfo  {journal} {Phys. Rev. Lett.}\ }\textbf
  {\bibinfo {volume} {95}},\ \bibinfo {pages} {040602} (\bibinfo {year}
  {2005})}\BibitemShut {NoStop}%
\bibitem [{\citenamefont {Lapolla}\ and\ \citenamefont
  {Godec}(2020)}]{Godec2020}%
  \BibitemOpen
  \bibfield  {author} {\bibinfo {author} {\bibfnamefont {A.}~\bibnamefont
  {Lapolla}}\ and\ \bibinfo {author} {\bibfnamefont {A.}~\bibnamefont
  {Godec}},\ }\bibfield  {title} {\bibinfo {title} {Faster uphill relaxation in
  thermodynamically equidistant temperature quenches},\ }\href@noop {}
  {\bibfield  {journal} {\bibinfo  {journal} {Phys. Rev. Lett.}\ }\textbf
  {\bibinfo {volume} {125}},\ \bibinfo {pages} {110602} (\bibinfo {year}
  {2020})}\BibitemShut {NoStop}%
\bibitem [{\citenamefont {Prost}\ \emph {et~al.}(2009)\citenamefont {Prost},
  \citenamefont {Joanny},\ and\ \citenamefont {Parrondo}}]{Prost2009}%
  \BibitemOpen
  \bibfield  {author} {\bibinfo {author} {\bibfnamefont {J.}~\bibnamefont
  {Prost}}, \bibinfo {author} {\bibfnamefont {J.-F.}\ \bibnamefont {Joanny}},\
  and\ \bibinfo {author} {\bibfnamefont {J.~M.~R.}\ \bibnamefont {Parrondo}},\
  }\bibfield  {title} {\bibinfo {title} {Generalized fluctuation-dissipation
  theorem for steady-state systems},\ }\href@noop {} {\bibfield  {journal}
  {\bibinfo  {journal} {Phys. Rev. Lett.}\ }\textbf {\bibinfo {volume} {103}},\
  \bibinfo {pages} {090601} (\bibinfo {year} {2009})}\BibitemShut {NoStop}%
\bibitem [{\citenamefont {Baiesi}\ \emph {et~al.}(2009)\citenamefont {Baiesi},
  \citenamefont {Maes},\ and\ \citenamefont {Wynants}}]{Wynants2009}%
  \BibitemOpen
  \bibfield  {author} {\bibinfo {author} {\bibfnamefont {M.}~\bibnamefont
  {Baiesi}}, \bibinfo {author} {\bibfnamefont {C.}~\bibnamefont {Maes}},\ and\
  \bibinfo {author} {\bibfnamefont {B.}~\bibnamefont {Wynants}},\ }\bibfield
  {title} {\bibinfo {title} {Fluctuations and response of nonequilibrium
  states},\ }\href@noop {} {\bibfield  {journal} {\bibinfo  {journal} {Phys.
  Rev. Lett.}\ }\textbf {\bibinfo {volume} {103}},\ \bibinfo {pages} {010602}
  (\bibinfo {year} {2009})}\BibitemShut {NoStop}%
\bibitem [{\citenamefont {Seifert}\ and\ \citenamefont
  {Speck}(2010)}]{Seifert2010}%
  \BibitemOpen
  \bibfield  {author} {\bibinfo {author} {\bibfnamefont {U.}~\bibnamefont
  {Seifert}}\ and\ \bibinfo {author} {\bibfnamefont {T.}~\bibnamefont
  {Speck}},\ }\bibfield  {title} {\bibinfo {title} {Fluctuation-dissipation
  theorem in nonequilibrium steady states},\ }\href@noop {} {\bibfield
  {journal} {\bibinfo  {journal} {Europhys. Lett.}\ }\textbf {\bibinfo {volume}
  {89}},\ \bibinfo {pages} {10007} (\bibinfo {year} {2010})}\BibitemShut
  {NoStop}%
\bibitem [{\citenamefont {Willareth}\ \emph {et~al.}(2017)\citenamefont
  {Willareth}, \citenamefont {Sokolov}, \citenamefont {Roichman},\ and\
  \citenamefont {Lindner}}]{Lindner2017}%
  \BibitemOpen
  \bibfield  {author} {\bibinfo {author} {\bibfnamefont {L.}~\bibnamefont
  {Willareth}}, \bibinfo {author} {\bibfnamefont {I.~M.}\ \bibnamefont
  {Sokolov}}, \bibinfo {author} {\bibfnamefont {Y.}~\bibnamefont {Roichman}},\
  and\ \bibinfo {author} {\bibfnamefont {B.}~\bibnamefont {Lindner}},\
  }\bibfield  {title} {\bibinfo {title} {Generalized fluctuation-dissipation
  theorem as a test of the markovianity of a system},\ }\href@noop {}
  {\bibfield  {journal} {\bibinfo  {journal} {Europhys. Lett.}\ }\textbf
  {\bibinfo {volume} {118}},\ \bibinfo {pages} {20001} (\bibinfo {year}
  {2017})}\BibitemShut {NoStop}%
\bibitem [{\citenamefont {Netz}(2018)}]{Netz2018}%
  \BibitemOpen
  \bibfield  {author} {\bibinfo {author} {\bibfnamefont {R.~R.}\ \bibnamefont
  {Netz}},\ }\bibfield  {title} {\bibinfo {title} {Fluctuation-dissipation
  relation and stationary distribution of an exactly solvable many-particle
  model for active biomatter far from equilibrium},\ }\href@noop {} {\bibfield
  {journal} {\bibinfo  {journal} {J. Chem. Phys.}\ }\textbf {\bibinfo {volume}
  {148}},\ \bibinfo {pages} {185101} (\bibinfo {year} {2018})}\BibitemShut
  {NoStop}%
\bibitem [{\citenamefont {Netz}(2020)}]{Netz2020}%
  \BibitemOpen
  \bibfield  {author} {\bibinfo {author} {\bibfnamefont {R.~R.}\ \bibnamefont
  {Netz}},\ }\bibfield  {title} {\bibinfo {title} {Approach to equilibrium and
  nonequilibrium stationary distributions of interacting many-particle systems
  that are coupled to different heat baths},\ }\href@noop {} {\bibfield
  {journal} {\bibinfo  {journal} {Phys. Rev. E}\ }\textbf {\bibinfo {volume}
  {101}},\ \bibinfo {pages} {022120} (\bibinfo {year} {2020})}\BibitemShut
  {NoStop}%
\bibitem [{\citenamefont {Nakajima}(1958)}]{nakajima_quantum_1958}%
  \BibitemOpen
  \bibfield  {author} {\bibinfo {author} {\bibfnamefont {S.}~\bibnamefont
  {Nakajima}},\ }\bibfield  {title} {\bibinfo {title} {On {Quantum} {Theory} of
  {Transport} {PhenomenaSteady} {Diffusion}},\ }\href
  {https://doi.org/10.1143/PTP.20.948} {\bibfield  {journal} {\bibinfo
  {journal} {Progress of Theoretical Physics}\ }\textbf {\bibinfo {volume}
  {20}},\ \bibinfo {pages} {948} (\bibinfo {year} {1958})}\BibitemShut
  {NoStop}%
\bibitem [{\citenamefont {Zwanzig}(1961)}]{zwanzig_memory_1961}%
  \BibitemOpen
  \bibfield  {author} {\bibinfo {author} {\bibfnamefont {R.}~\bibnamefont
  {Zwanzig}},\ }\bibfield  {title} {\bibinfo {title} {Memory {Effects} in
  {Irreversible} {Thermodynamics}},\ }\href
  {https://doi.org/10.1103/PhysRev.124.983} {\bibfield  {journal} {\bibinfo
  {journal} {Physical Review}\ }\textbf {\bibinfo {volume} {124}},\ \bibinfo
  {pages} {983} (\bibinfo {year} {1961})}\BibitemShut {NoStop}%
\bibitem [{\citenamefont {Mori}(1965)}]{mori_transport_1965}%
  \BibitemOpen
  \bibfield  {author} {\bibinfo {author} {\bibfnamefont {H.}~\bibnamefont
  {Mori}},\ }\bibfield  {title} {\bibinfo {title} {Transport, {Collective}
  {Motion}, and {Brownian} {Motion}},\ }\href
  {https://doi.org/10.1143/PTP.33.423} {\bibfield  {journal} {\bibinfo
  {journal} {Progress of Theoretical Physics}\ }\textbf {\bibinfo {volume}
  {33}},\ \bibinfo {pages} {423} (\bibinfo {year} {1965})}\BibitemShut
  {NoStop}%
\bibitem [{\citenamefont {Ciccotti}\ and\ \citenamefont
  {Ryckaert}(1981)}]{Ciccotti1981}%
  \BibitemOpen
  \bibfield  {author} {\bibinfo {author} {\bibfnamefont {G.}~\bibnamefont
  {Ciccotti}}\ and\ \bibinfo {author} {\bibfnamefont {J.-P.}\ \bibnamefont
  {Ryckaert}},\ }\bibfield  {title} {\bibinfo {title} {{On the derivation of
  the gen- eralized Langevin equation for interacting Brownian particles}},\
  }\href@noop {} {\bibfield  {journal} {\bibinfo  {journal} {Journal of
  Statistical Physics}\ }\textbf {\bibinfo {volume} {26}},\ \bibinfo {pages}
  {73} (\bibinfo {year} {1981})}\BibitemShut {NoStop}%
\bibitem [{\citenamefont {Straub}\ \emph {et~al.}(1987)\citenamefont {Straub},
  \citenamefont {Borkovec},\ and\ \citenamefont {Berne}}]{Straub_1987}%
  \BibitemOpen
  \bibfield  {author} {\bibinfo {author} {\bibfnamefont {J.~E.}\ \bibnamefont
  {Straub}}, \bibinfo {author} {\bibfnamefont {M.}~\bibnamefont {Borkovec}},\
  and\ \bibinfo {author} {\bibfnamefont {B.~J.}\ \bibnamefont {Berne}},\
  }\bibfield  {title} {\bibinfo {title} {{Calculation of dynamic friction on
  intramolecular degrees of freedom}},\ }\href@noop {} {\bibfield  {journal}
  {\bibinfo  {journal} {The Journal of Physical Chemistry}\ }\textbf {\bibinfo
  {volume} {91}},\ \bibinfo {pages} {4995} (\bibinfo {year}
  {1987})}\BibitemShut {NoStop}%
\bibitem [{\citenamefont {Lange}\ and\ \citenamefont
  {Grubm{\"u}ller}(2006)}]{lange_collective_2006}%
  \BibitemOpen
  \bibfield  {author} {\bibinfo {author} {\bibfnamefont {O.~F.}\ \bibnamefont
  {Lange}}\ and\ \bibinfo {author} {\bibfnamefont {H.}~\bibnamefont
  {Grubm{\"u}ller}},\ }\bibfield  {title} {\bibinfo {title} {Collective
  {Langevin} dynamics of conformational motions in proteins},\ }\href
  {https://doi.org/10.1063/1.2199530} {\bibfield  {journal} {\bibinfo
  {journal} {The Journal of Chemical Physics}\ }\textbf {\bibinfo {volume}
  {124}},\ \bibinfo {pages} {214903} (\bibinfo {year} {2006})},\ \bibinfo
  {note} {publisher: American Institute of Physics}\BibitemShut {NoStop}%
\bibitem [{\citenamefont {Kinjo}\ and\ \citenamefont
  {Hyodo}(2007)}]{kinjo_equation_2007}%
  \BibitemOpen
  \bibfield  {author} {\bibinfo {author} {\bibfnamefont {T.}~\bibnamefont
  {Kinjo}}\ and\ \bibinfo {author} {\bibfnamefont {S.~A.}\ \bibnamefont
  {Hyodo}},\ }\bibfield  {title} {\bibinfo {title} {Equation of motion for
  coarse-grained simulation based on microscopic description},\ }\href
  {https://doi.org/10.1103/PhysRevE.75.051109} {\bibfield  {journal} {\bibinfo
  {journal} {Physical Review E}\ }\textbf {\bibinfo {volume} {75}},\ \bibinfo
  {pages} {051109} (\bibinfo {year} {2007})},\ \bibinfo {note} {publisher:
  American Physical Society}\BibitemShut {NoStop}%
\bibitem [{\citenamefont {Darve}\ \emph {et~al.}(2009)\citenamefont {Darve},
  \citenamefont {Solomon},\ and\ \citenamefont {Kia}}]{darve_computing_2009}%
  \BibitemOpen
  \bibfield  {author} {\bibinfo {author} {\bibfnamefont {E.}~\bibnamefont
  {Darve}}, \bibinfo {author} {\bibfnamefont {J.}~\bibnamefont {Solomon}},\
  and\ \bibinfo {author} {\bibfnamefont {A.}~\bibnamefont {Kia}},\ }\bibfield
  {title} {\bibinfo {title} {Computing generalized {Langevin} equations and
  generalized {Fokker}{\textendash}{Planck} equations},\ }\href
  {https://doi.org/10.1073/pnas.0902633106} {\bibfield  {journal} {\bibinfo
  {journal} {Proceedings of the National Academy of Sciences}\ }\textbf
  {\bibinfo {volume} {106}},\ \bibinfo {pages} {10884} (\bibinfo {year}
  {2009})}\BibitemShut {NoStop}%
\bibitem [{\citenamefont {Hij{\'o}n}\ \emph {et~al.}(2010)\citenamefont
  {Hij{\'o}n}, \citenamefont {Espa{\~n}ol}, \citenamefont {Vanden-Eijnden},\
  and\ \citenamefont {Delgado-Buscalioni}}]{hijon_morizwanzig_2010}%
  \BibitemOpen
  \bibfield  {author} {\bibinfo {author} {\bibfnamefont {C.}~\bibnamefont
  {Hij{\'o}n}}, \bibinfo {author} {\bibfnamefont {P.}~\bibnamefont
  {Espa{\~n}ol}}, \bibinfo {author} {\bibfnamefont {E.}~\bibnamefont
  {Vanden-Eijnden}},\ and\ \bibinfo {author} {\bibfnamefont {R.}~\bibnamefont
  {Delgado-Buscalioni}},\ }\bibfield  {title} {\bibinfo {title}
  {Mori{\textendash}{Zwanzig} formalism as a practical computational tool},\
  }\href {https://doi.org/10.1039/B902479B} {\bibfield  {journal} {\bibinfo
  {journal} {Faraday Discussions}\ }\textbf {\bibinfo {volume} {144}},\
  \bibinfo {pages} {301} (\bibinfo {year} {2010})},\ \bibinfo {note}
  {publisher: Royal Society of Chemistry}\BibitemShut {NoStop}%
\bibitem [{\citenamefont {Izvekov}(2013)}]{izvekov_microscopic_2013}%
  \BibitemOpen
  \bibfield  {author} {\bibinfo {author} {\bibfnamefont {S.}~\bibnamefont
  {Izvekov}},\ }\bibfield  {title} {\bibinfo {title} {Microscopic derivation of
  particle-based coarse-grained dynamics},\ }\href
  {https://doi.org/10.1063/1.4795091} {\bibfield  {journal} {\bibinfo
  {journal} {The Journal of Chemical Physics}\ }\textbf {\bibinfo {volume}
  {138}},\ \bibinfo {pages} {134106} (\bibinfo {year} {2013})},\ \bibinfo
  {note} {publisher: American Institute of Physics}\BibitemShut {NoStop}%
\bibitem [{\citenamefont {Lee}\ \emph {et~al.}(2019)\citenamefont {Lee},
  \citenamefont {Ahn},\ and\ \citenamefont {Darve}}]{lee_multi_2019}%
  \BibitemOpen
  \bibfield  {author} {\bibinfo {author} {\bibfnamefont {H.~S.}\ \bibnamefont
  {Lee}}, \bibinfo {author} {\bibfnamefont {S.-H.}\ \bibnamefont {Ahn}},\ and\
  \bibinfo {author} {\bibfnamefont {E.~F.}\ \bibnamefont {Darve}},\ }\bibfield
  {title} {\bibinfo {title} {The multi-dimensional generalized langevin
  equation for conformational motion of proteins},\ }\href@noop {} {\bibfield
  {journal} {\bibinfo  {journal} {The Journal of chemical physics}\ }\textbf
  {\bibinfo {volume} {150}},\ \bibinfo {pages} {174113} (\bibinfo {year}
  {2019})}\BibitemShut {NoStop}%
\bibitem [{\citenamefont {Ayaz}\ \emph {et~al.}(2022)\citenamefont {Ayaz},
  \citenamefont {Scalfi}, \citenamefont {Dalton},\ and\ \citenamefont
  {Netz}}]{Ayaz2022}%
  \BibitemOpen
  \bibfield  {author} {\bibinfo {author} {\bibfnamefont {C.}~\bibnamefont
  {Ayaz}}, \bibinfo {author} {\bibfnamefont {L.}~\bibnamefont {Scalfi}},
  \bibinfo {author} {\bibfnamefont {B.~A.}\ \bibnamefont {Dalton}},\ and\
  \bibinfo {author} {\bibfnamefont {R.~R.}\ \bibnamefont {Netz}},\ }\bibfield
  {title} {\bibinfo {title} {Generalized langevin equation with a nonlinear
  potential of mean force and nonlinear memory friction from a hybrid
  projection scheme},\ }\href@noop {} {\bibfield  {journal} {\bibinfo
  {journal} {Physical Review E}\ }\textbf {\bibinfo {volume} {105}},\ \bibinfo
  {pages} {054138} (\bibinfo {year} {2022})}\BibitemShut {NoStop}%
\bibitem [{\citenamefont {Vroylandt}\ \emph {et~al.}(2022)\citenamefont
  {Vroylandt}, \citenamefont {Gouden{\`e}ge}, \citenamefont {Monmarch{\'e}},
  \citenamefont {Pietrucci},\ and\ \citenamefont
  {Rotenberg}}]{vroylandt_likelihood_2022}%
  \BibitemOpen
  \bibfield  {author} {\bibinfo {author} {\bibfnamefont {H.}~\bibnamefont
  {Vroylandt}}, \bibinfo {author} {\bibfnamefont {L.}~\bibnamefont
  {Gouden{\`e}ge}}, \bibinfo {author} {\bibfnamefont {P.}~\bibnamefont
  {Monmarch{\'e}}}, \bibinfo {author} {\bibfnamefont {F.}~\bibnamefont
  {Pietrucci}},\ and\ \bibinfo {author} {\bibfnamefont {B.}~\bibnamefont
  {Rotenberg}},\ }\bibfield  {title} {\bibinfo {title} {Likelihood-based
  non-markovian models from molecular dynamics},\ }\href@noop {} {\bibfield
  {journal} {\bibinfo  {journal} {Proceedings of the National Academy of
  Sciences}\ }\textbf {\bibinfo {volume} {119}},\ \bibinfo {pages}
  {e2117586119} (\bibinfo {year} {2022})}\BibitemShut {NoStop}%
\bibitem [{\citenamefont {Plotkin}\ and\ \citenamefont
  {Wolynes}(1998)}]{plotkin_non-markovian_1998}%
  \BibitemOpen
  \bibfield  {author} {\bibinfo {author} {\bibfnamefont {S.~S.}\ \bibnamefont
  {Plotkin}}\ and\ \bibinfo {author} {\bibfnamefont {P.~G.}\ \bibnamefont
  {Wolynes}},\ }\bibfield  {title} {\bibinfo {title} {Non-{Markovian}
  {Configurational} {Diffusion} and {Reaction} {Coordinates} for {Protein}
  {Folding}},\ }\href {https://doi.org/10.1103/PhysRevLett.80.5015} {\bibfield
  {journal} {\bibinfo  {journal} {Physical Review Letters}\ }\textbf {\bibinfo
  {volume} {80}},\ \bibinfo {pages} {5015} (\bibinfo {year} {1998})},\ \bibinfo
  {note} {publisher: American Physical Society}\BibitemShut {NoStop}%
\bibitem [{\citenamefont {Satija}\ and\ \citenamefont
  {Makarov}(2019)}]{satija_generalized_2019}%
  \BibitemOpen
  \bibfield  {author} {\bibinfo {author} {\bibfnamefont {R.}~\bibnamefont
  {Satija}}\ and\ \bibinfo {author} {\bibfnamefont {D.~E.}\ \bibnamefont
  {Makarov}},\ }\bibfield  {title} {\bibinfo {title} {Generalized {Langevin}
  {Equation} as a {Model} for {Barrier} {Crossing} {Dynamics} in {Biomolecular}
  {Folding}},\ }\href {https://doi.org/10.1021/acs.jpcb.8b11137} {\bibfield
  {journal} {\bibinfo  {journal} {The Journal of Physical Chemistry B}\
  }\textbf {\bibinfo {volume} {123}},\ \bibinfo {pages} {802} (\bibinfo {year}
  {2019})},\ \bibinfo {note} {publisher: American Chemical Society}\BibitemShut
  {NoStop}%
\bibitem [{\citenamefont {Ayaz}\ \emph {et~al.}(2021)\citenamefont {Ayaz},
  \citenamefont {Tepper}, \citenamefont {Br{\"u}nig}, \citenamefont {Kappler},
  \citenamefont {Daldrop},\ and\ \citenamefont
  {Netz}}]{ayaz_non-markovian_2021}%
  \BibitemOpen
  \bibfield  {author} {\bibinfo {author} {\bibfnamefont {C.}~\bibnamefont
  {Ayaz}}, \bibinfo {author} {\bibfnamefont {L.}~\bibnamefont {Tepper}},
  \bibinfo {author} {\bibfnamefont {F.~N.}\ \bibnamefont {Br{\"u}nig}},
  \bibinfo {author} {\bibfnamefont {J.}~\bibnamefont {Kappler}}, \bibinfo
  {author} {\bibfnamefont {J.~O.}\ \bibnamefont {Daldrop}},\ and\ \bibinfo
  {author} {\bibfnamefont {R.~R.}\ \bibnamefont {Netz}},\ }\bibfield  {title}
  {\bibinfo {title} {Non-{Markovian} modeling of protein folding},\ }\bibfield
  {journal} {\bibinfo  {journal} {Proceedings of the National Academy of
  Sciences}\ }\textbf {\bibinfo {volume} {118}},\ \href
  {https://doi.org/10.1073/pnas.2023856118} {10.1073/pnas.2023856118} (\bibinfo
  {year} {2021}),\ \bibinfo {note} {publisher: National Academy of Sciences
  Section: Physical Sciences}\BibitemShut {NoStop}%
\bibitem [{\citenamefont {Dalton}\ \emph {et~al.}(2023)\citenamefont {Dalton},
  \citenamefont {Ayaz}, \citenamefont {Kiefer}, \citenamefont {Klimek},
  \citenamefont {Tepper},\ and\ \citenamefont {Netz}}]{Dalton_2023}%
  \BibitemOpen
  \bibfield  {author} {\bibinfo {author} {\bibfnamefont {B.~A.}\ \bibnamefont
  {Dalton}}, \bibinfo {author} {\bibfnamefont {C.}~\bibnamefont {Ayaz}},
  \bibinfo {author} {\bibfnamefont {H.}~\bibnamefont {Kiefer}}, \bibinfo
  {author} {\bibfnamefont {A.}~\bibnamefont {Klimek}}, \bibinfo {author}
  {\bibfnamefont {L.}~\bibnamefont {Tepper}},\ and\ \bibinfo {author}
  {\bibfnamefont {R.~R.}\ \bibnamefont {Netz}},\ }\bibfield  {title} {\bibinfo
  {title} {Fast protein folding is governed by memory-dependent friction},\
  }\href@noop {} {\bibfield  {journal} {\bibinfo  {journal} {Proceedings of the
  National Academy of Sciences}\ }\textbf {\bibinfo {volume} {120}},\ \bibinfo
  {pages} {e2220068120} (\bibinfo {year} {2023})}\BibitemShut {NoStop}%
\bibitem [{\citenamefont {Bagchi}\ and\ \citenamefont
  {Oxtoby}(1983)}]{Bagchi_1983}%
  \BibitemOpen
  \bibfield  {author} {\bibinfo {author} {\bibfnamefont {B.}~\bibnamefont
  {Bagchi}}\ and\ \bibinfo {author} {\bibfnamefont {D.~W.}\ \bibnamefont
  {Oxtoby}},\ }\bibfield  {title} {\bibinfo {title} {{The effect of frequency
  dependent friction on isomerization dynamics in solution}},\ }\href@noop {}
  {\bibfield  {journal} {\bibinfo  {journal} {The Journal of Chemical Physics}\
  }\textbf {\bibinfo {volume} {78}},\ \bibinfo {pages} {2735} (\bibinfo {year}
  {1983})}\BibitemShut {NoStop}%
\bibitem [{\citenamefont {Straub}\ \emph {et~al.}(1986)\citenamefont {Straub},
  \citenamefont {Borkovec},\ and\ \citenamefont {Berne}}]{Straub_1986}%
  \BibitemOpen
  \bibfield  {author} {\bibinfo {author} {\bibfnamefont {J.~E.}\ \bibnamefont
  {Straub}}, \bibinfo {author} {\bibfnamefont {M.}~\bibnamefont {Borkovec}},\
  and\ \bibinfo {author} {\bibfnamefont {B.~J.}\ \bibnamefont {Berne}},\
  }\bibfield  {title} {\bibinfo {title} {{Non‐Markovian activated rate
  processes: Comparison of current theories with numerical simulation data}},\
  }\href@noop {} {\bibfield  {journal} {\bibinfo  {journal} {The Journal of
  Chemical Physics}\ }\textbf {\bibinfo {volume} {84}},\ \bibinfo {pages}
  {1788} (\bibinfo {year} {1986})}\BibitemShut {NoStop}%
\bibitem [{\citenamefont {Pollak}\ \emph {et~al.}(1989)\citenamefont {Pollak},
  \citenamefont {Grabert},\ and\ \citenamefont {H{\"{a}}nggi}}]{Pollak_1989}%
  \BibitemOpen
  \bibfield  {author} {\bibinfo {author} {\bibfnamefont {E.}~\bibnamefont
  {Pollak}}, \bibinfo {author} {\bibfnamefont {H.}~\bibnamefont {Grabert}},\
  and\ \bibinfo {author} {\bibfnamefont {P.}~\bibnamefont {H{\"{a}}nggi}},\
  }\bibfield  {title} {\bibinfo {title} {{Theory of activated rate processes
  for arbitrary frequency dependent friction: Solution of the turnover
  problem}},\ }\href@noop {} {\bibfield  {journal} {\bibinfo  {journal} {The
  Journal of Chemical Physics}\ }\textbf {\bibinfo {volume} {91}},\ \bibinfo
  {pages} {4073} (\bibinfo {year} {1989})}\BibitemShut {NoStop}%
\bibitem [{\citenamefont {Br{\"{u}}nig}\ \emph
  {et~al.}(2022{\natexlab{a}})\citenamefont {Br{\"{u}}nig}, \citenamefont
  {Daldrop},\ and\ \citenamefont {Netz}}]{Brunig_2022d}%
  \BibitemOpen
  \bibfield  {author} {\bibinfo {author} {\bibfnamefont {F.~N.}\ \bibnamefont
  {Br{\"{u}}nig}}, \bibinfo {author} {\bibfnamefont {J.~O.}\ \bibnamefont
  {Daldrop}},\ and\ \bibinfo {author} {\bibfnamefont {R.~R.}\ \bibnamefont
  {Netz}},\ }\bibfield  {title} {\bibinfo {title} {{Pair-Reaction Dynamics in
  Water: Competition of Memory, Potential Shape, and Inertial Effects}},\
  }\href@noop {} {\bibfield  {journal} {\bibinfo  {journal} {The Journal of
  Physical Chemistry B}\ }\textbf {\bibinfo {volume} {126}},\ \bibinfo {pages}
  {10295} (\bibinfo {year} {2022}{\natexlab{a}})}\BibitemShut {NoStop}%
\bibitem [{\citenamefont {Mitterwallner}\ \emph {et~al.}(2020)\citenamefont
  {Mitterwallner}, \citenamefont {Schreiber}, \citenamefont {Daldrop},
  \citenamefont {Rädler},\ and\ \citenamefont {Netz}}]{Mitterwallner_2020}%
  \BibitemOpen
  \bibfield  {author} {\bibinfo {author} {\bibfnamefont {B.~G.}\ \bibnamefont
  {Mitterwallner}}, \bibinfo {author} {\bibfnamefont {C.}~\bibnamefont
  {Schreiber}}, \bibinfo {author} {\bibfnamefont {J.~O.}\ \bibnamefont
  {Daldrop}}, \bibinfo {author} {\bibfnamefont {J.~O.}\ \bibnamefont
  {Rädler}},\ and\ \bibinfo {author} {\bibfnamefont {R.~R.}\ \bibnamefont
  {Netz}},\ }\bibfield  {title} {\bibinfo {title} {{Non-Markovian data-driven
  modeling of single-cell motility}},\ }\href@noop {} {\bibfield  {journal}
  {\bibinfo  {journal} {Physical Review E}\ }\textbf {\bibinfo {volume}
  {101}},\ \bibinfo {pages} {032408} (\bibinfo {year} {2020})}\BibitemShut
  {NoStop}%
\bibitem [{\citenamefont {Tuckerman}\ and\ \citenamefont
  {Berne}(1993)}]{Tuckerman1993}%
  \BibitemOpen
  \bibfield  {author} {\bibinfo {author} {\bibfnamefont {M.}~\bibnamefont
  {Tuckerman}}\ and\ \bibinfo {author} {\bibfnamefont {B.}~\bibnamefont
  {Berne}},\ }\bibfield  {title} {\bibinfo {title} {{Vibrational relaxation in
  simple fluids: Comparison of theory and simulation}},\ }\href@noop {}
  {\bibfield  {journal} {\bibinfo  {journal} {The Journal of Chemical Physics}\
  }\textbf {\bibinfo {volume} {98}},\ \bibinfo {pages} {7301} (\bibinfo {year}
  {1993})}\BibitemShut {NoStop}%
\bibitem [{\citenamefont {Gottwald}\ \emph {et~al.}(2015)\citenamefont
  {Gottwald}, \citenamefont {Ivanov},\ and\ \citenamefont
  {Kühn}}]{Gottwald2015}%
  \BibitemOpen
  \bibfield  {author} {\bibinfo {author} {\bibfnamefont {F.}~\bibnamefont
  {Gottwald}}, \bibinfo {author} {\bibfnamefont {S.~D.}\ \bibnamefont
  {Ivanov}},\ and\ \bibinfo {author} {\bibfnamefont {O.}~\bibnamefont
  {Kühn}},\ }\bibfield  {title} {\bibinfo {title} {{Applicability of the
  Caldeira–Leggett model to vibrational spectroscopy in solution}},\
  }\href@noop {} {\bibfield  {journal} {\bibinfo  {journal} {The Journal of
  Physical Chemistry Letters}\ }\textbf {\bibinfo {volume} {6}},\ \bibinfo
  {pages} {2722} (\bibinfo {year} {2015})}\BibitemShut {NoStop}%
\bibitem [{\citenamefont {Br{\"{u}}nig}\ \emph
  {et~al.}(2022{\natexlab{b}})\citenamefont {Br{\"{u}}nig}, \citenamefont
  {Geburtig}, \citenamefont {von Canal}, \citenamefont {Kappler},\ and\
  \citenamefont {Netz}}]{Brunig_2022a}%
  \BibitemOpen
  \bibfield  {author} {\bibinfo {author} {\bibfnamefont {F.~N.}\ \bibnamefont
  {Br{\"{u}}nig}}, \bibinfo {author} {\bibfnamefont {O.}~\bibnamefont
  {Geburtig}}, \bibinfo {author} {\bibfnamefont {A.}~\bibnamefont {von Canal}},
  \bibinfo {author} {\bibfnamefont {J.}~\bibnamefont {Kappler}},\ and\ \bibinfo
  {author} {\bibfnamefont {R.~R.}\ \bibnamefont {Netz}},\ }\bibfield  {title}
  {\bibinfo {title} {{Time-Dependent Friction Effects on Vibrational Infrared
  Frequencies and Line Shapes of Liquid Water}},\ }\href@noop {} {\bibfield
  {journal} {\bibinfo  {journal} {The Journal of Physical Chemistry B}\
  }\textbf {\bibinfo {volume} {126}},\ \bibinfo {pages} {1579} (\bibinfo {year}
  {2022}{\natexlab{b}})}\BibitemShut {NoStop}%
\bibitem [{\citenamefont {Herrera-Delgado}\ \emph {et~al.}(2020)\citenamefont
  {Herrera-Delgado}, \citenamefont {Briscoe},\ and\ \citenamefont
  {Sollich}}]{Sollich2020}%
  \BibitemOpen
  \bibfield  {author} {\bibinfo {author} {\bibfnamefont {E.}~\bibnamefont
  {Herrera-Delgado}}, \bibinfo {author} {\bibfnamefont {J.}~\bibnamefont
  {Briscoe}},\ and\ \bibinfo {author} {\bibfnamefont {P.}~\bibnamefont
  {Sollich}},\ }\bibfield  {title} {\bibinfo {title} {Tractable nonlinear
  memory functions as a tool to capture and explain dynamical behaviors},\
  }\href@noop {} {\bibfield  {journal} {\bibinfo  {journal} {Physical Review
  Research}\ }\textbf {\bibinfo {volume} {2}},\ \bibinfo {pages} {043069}
  (\bibinfo {year} {2020})}\BibitemShut {NoStop}%
\bibitem [{\citenamefont {Chorin}\ \emph {et~al.}(2000)\citenamefont {Chorin},
  \citenamefont {Hald},\ and\ \citenamefont {Kupferman}}]{chorin_optimal_2000}%
  \BibitemOpen
  \bibfield  {author} {\bibinfo {author} {\bibfnamefont {A.~J.}\ \bibnamefont
  {Chorin}}, \bibinfo {author} {\bibfnamefont {O.~H.}\ \bibnamefont {Hald}},\
  and\ \bibinfo {author} {\bibfnamefont {R.}~\bibnamefont {Kupferman}},\
  }\bibfield  {title} {\bibinfo {title} {Optimal prediction and the
  {Mori}{\textendash}{Zwanzig} representation of irreversible processes},\
  }\href {https://doi.org/10.1073/pnas.97.7.2968} {\bibfield  {journal}
  {\bibinfo  {journal} {Proceedings of the National Academy of Sciences}\
  }\textbf {\bibinfo {volume} {97}},\ \bibinfo {pages} {2968} (\bibinfo {year}
  {2000})},\ \bibinfo {note} {publisher: National Academy of Sciences Section:
  Physical Sciences}\BibitemShut {NoStop}%
\bibitem [{\citenamefont {Robertson}(1966)}]{Robertson1966}%
  \BibitemOpen
  \bibfield  {author} {\bibinfo {author} {\bibfnamefont {B.}~\bibnamefont
  {Robertson}},\ }\bibfield  {title} {\bibinfo {title} {Equations of motion in
  nonequilibriurn statistical mechanics},\ }\href@noop {} {\bibfield  {journal}
  {\bibinfo  {journal} {Physical Review}\ }\textbf {\bibinfo {volume} {144}},\
  \bibinfo {pages} {151} (\bibinfo {year} {1966})}\BibitemShut {NoStop}%
\bibitem [{\citenamefont {Nordholm}\ and\ \citenamefont
  {Zwanzig}(1975)}]{Zwanzig1975}%
  \BibitemOpen
  \bibfield  {author} {\bibinfo {author} {\bibfnamefont {S.}~\bibnamefont
  {Nordholm}}\ and\ \bibinfo {author} {\bibfnamefont {R.}~\bibnamefont
  {Zwanzig}},\ }\bibfield  {title} {\bibinfo {title} {A systematic derivation
  of exact generalized brownian motion theory},\ }\href@noop {} {\bibfield
  {journal} {\bibinfo  {journal} {Journal of Statistical Physics}\ }\textbf
  {\bibinfo {volume} {13}},\ \bibinfo {pages} {347} (\bibinfo {year}
  {1975})}\BibitemShut {NoStop}%
\bibitem [{\citenamefont {Picard}\ and\ \citenamefont
  {Willis}(1977)}]{Picard1977}%
  \BibitemOpen
  \bibfield  {author} {\bibinfo {author} {\bibfnamefont {R.~H.}\ \bibnamefont
  {Picard}}\ and\ \bibinfo {author} {\bibfnamefont {C.~R.}\ \bibnamefont
  {Willis}},\ }\bibfield  {title} {\bibinfo {title} {Time-dependent
  projection-operator approach to master equations for coupled systems. ii.
  systems with correlations},\ }\href@noop {} {\bibfield  {journal} {\bibinfo
  {journal} {Physical Review A}\ }\textbf {\bibinfo {volume} {16}},\ \bibinfo
  {pages} {1625} (\bibinfo {year} {1977})}\BibitemShut {NoStop}%
\bibitem [{\citenamefont {Uchiyama}\ and\ \citenamefont
  {Shibata}(1999)}]{Uchiyama1999}%
  \BibitemOpen
  \bibfield  {author} {\bibinfo {author} {\bibfnamefont {C.}~\bibnamefont
  {Uchiyama}}\ and\ \bibinfo {author} {\bibfnamefont {F.}~\bibnamefont
  {Shibata}},\ }\bibfield  {title} {\bibinfo {title} {Unified projection
  operator formalism in nonequilibrium statistical mechanics},\ }\href@noop {}
  {\bibfield  {journal} {\bibinfo  {journal} {Physical Review E}\ }\textbf
  {\bibinfo {volume} {60}},\ \bibinfo {pages} {2636} (\bibinfo {year}
  {1999})}\BibitemShut {NoStop}%
\bibitem [{\citenamefont {Koide}(2002)}]{Koide2002}%
  \BibitemOpen
  \bibfield  {author} {\bibinfo {author} {\bibfnamefont {T.}~\bibnamefont
  {Koide}},\ }\bibfield  {title} {\bibinfo {title} {Derivation of transport
  equations using the time-dependent projection operator method},\ }\href@noop
  {} {\bibfield  {journal} {\bibinfo  {journal} {Progress of Theoretical
  Physics}\ }\textbf {\bibinfo {volume} {60}},\ \bibinfo {pages} {525}
  (\bibinfo {year} {2002})}\BibitemShut {NoStop}%
\bibitem [{\citenamefont {Latz}(2002)}]{Latz2002}%
  \BibitemOpen
  \bibfield  {author} {\bibinfo {author} {\bibfnamefont {A.}~\bibnamefont
  {Latz}},\ }\bibfield  {title} {\bibinfo {title} {Non-equilibrium
  projection-operator for a quenched thermostatted system},\ }\href@noop {}
  {\bibfield  {journal} {\bibinfo  {journal} {Journal of Statistical Physics}\
  }\textbf {\bibinfo {volume} {109}},\ \bibinfo {pages} {607} (\bibinfo {year}
  {2002})}\BibitemShut {NoStop}%
\bibitem [{\citenamefont {Meyer}\ \emph {et~al.}(2017)\citenamefont {Meyer},
  \citenamefont {Voigtmann},\ and\ \citenamefont
  {Schilling}}]{meyer_non-stationary_2017}%
  \BibitemOpen
  \bibfield  {author} {\bibinfo {author} {\bibfnamefont {H.}~\bibnamefont
  {Meyer}}, \bibinfo {author} {\bibfnamefont {T.}~\bibnamefont {Voigtmann}},\
  and\ \bibinfo {author} {\bibfnamefont {T.}~\bibnamefont {Schilling}},\
  }\bibfield  {title} {\bibinfo {title} {On the non-stationary generalized
  {Langevin} equation},\ }\href {https://doi.org/10.1063/1.5006980} {\bibfield
  {journal} {\bibinfo  {journal} {The Journal of Chemical Physics}\ }\textbf
  {\bibinfo {volume} {147}},\ \bibinfo {pages} {214110} (\bibinfo {year}
  {2017})},\ \bibinfo {note} {publisher: American Institute of
  Physics}\BibitemShut {NoStop}%
\bibitem [{\citenamefont {Cui}\ and\ \citenamefont {Zaccone}(2018)}]{Cui2018}%
  \BibitemOpen
  \bibfield  {author} {\bibinfo {author} {\bibfnamefont {B.}~\bibnamefont
  {Cui}}\ and\ \bibinfo {author} {\bibfnamefont {A.}~\bibnamefont {Zaccone}},\
  }\bibfield  {title} {\bibinfo {title} {Generalized langevin equation and
  fluctuation-dissipation theorem for particle-bath systems in external
  oscillating fields},\ }\href@noop {} {\bibfield  {journal} {\bibinfo
  {journal} {Physical Review E}\ }\textbf {\bibinfo {volume} {97}},\ \bibinfo
  {pages} {060102(R)} (\bibinfo {year} {2018})}\BibitemShut {NoStop}%
\bibitem [{\citenamefont {te~Vrugt}\ and\ \citenamefont
  {Wittkowski}(2019)}]{Vrugt2019}%
  \BibitemOpen
  \bibfield  {author} {\bibinfo {author} {\bibfnamefont {M.}~\bibnamefont
  {te~Vrugt}}\ and\ \bibinfo {author} {\bibfnamefont {R.}~\bibnamefont
  {Wittkowski}},\ }\bibfield  {title} {\bibinfo {title} {Mori-zwanzig
  projection operator formalism for far-from-equilibrium systems with
  time-dependent hamiltonians},\ }\href@noop {} {\bibfield  {journal} {\bibinfo
   {journal} {Physical Review E}\ }\textbf {\bibinfo {volume} {99}},\ \bibinfo
  {pages} {062118} (\bibinfo {year} {2019})}\BibitemShut {NoStop}%
\bibitem [{\citenamefont {Meyer}\ \emph {et~al.}(2020)\citenamefont {Meyer},
  \citenamefont {Pelagejcev},\ and\ \citenamefont
  {Schilling}}]{meyer_non-markovian_2020}%
  \BibitemOpen
  \bibfield  {author} {\bibinfo {author} {\bibfnamefont {H.}~\bibnamefont
  {Meyer}}, \bibinfo {author} {\bibfnamefont {P.}~\bibnamefont {Pelagejcev}},\
  and\ \bibinfo {author} {\bibfnamefont {T.}~\bibnamefont {Schilling}},\
  }\bibfield  {title} {\bibinfo {title} {Non-{Markovian} out-of-equilibrium
  dynamics: {A} general numerical procedure to construct time-dependent memory
  kernels for coarse-grained observables},\ }\href
  {https://doi.org/10.1209/0295-5075/128/40001} {\bibfield  {journal} {\bibinfo
   {journal} {EPL (Europhysics Letters)}\ }\textbf {\bibinfo {volume} {128}},\
  \bibinfo {pages} {40001} (\bibinfo {year} {2020})},\ \bibinfo {note}
  {publisher: IOP Publishing}\BibitemShut {NoStop}%
\bibitem [{\citenamefont {Dyson}(1949)}]{dyson_radiation_1949}%
  \BibitemOpen
  \bibfield  {author} {\bibinfo {author} {\bibfnamefont {F.~J.}\ \bibnamefont
  {Dyson}},\ }\bibfield  {title} {\bibinfo {title} {The {Radiation} {Theories}
  of {Tomonaga}, {Schwinger}, and {Feynman}},\ }\href
  {https://doi.org/10.1103/PhysRev.75.486} {\bibfield  {journal} {\bibinfo
  {journal} {Physical Review}\ }\textbf {\bibinfo {volume} {75}},\ \bibinfo
  {pages} {486} (\bibinfo {year} {1949})},\ \bibinfo {note} {publisher:
  American Physical Society}\BibitemShut {NoStop}%
\bibitem [{\citenamefont {Feynman}(1951)}]{feynman_operator_1951}%
  \BibitemOpen
  \bibfield  {author} {\bibinfo {author} {\bibfnamefont {R.~P.}\ \bibnamefont
  {Feynman}},\ }\bibfield  {title} {\bibinfo {title} {An {Operator} {Calculus}
  {Having} {Applications} in {Quantum} {Electrodynamics}},\ }\href
  {https://doi.org/10.1103/PhysRev.84.108} {\bibfield  {journal} {\bibinfo
  {journal} {Physical Review}\ }\textbf {\bibinfo {volume} {84}},\ \bibinfo
  {pages} {108} (\bibinfo {year} {1951})},\ \bibinfo {note} {publisher:
  American Physical Society}\BibitemShut {NoStop}%
\bibitem [{\citenamefont {Carof}\ \emph {et~al.}(2014)\citenamefont {Carof},
  \citenamefont {Vuilleumier},\ and\ \citenamefont
  {Rotenberg}}]{carof_two_2014}%
  \BibitemOpen
  \bibfield  {author} {\bibinfo {author} {\bibfnamefont {A.}~\bibnamefont
  {Carof}}, \bibinfo {author} {\bibfnamefont {R.}~\bibnamefont {Vuilleumier}},\
  and\ \bibinfo {author} {\bibfnamefont {B.}~\bibnamefont {Rotenberg}},\
  }\bibfield  {title} {\bibinfo {title} {Two algorithms to compute projected
  correlation functions in molecular dynamics simulations},\ }\href
  {https://doi.org/10.1063/1.4868653} {\bibfield  {journal} {\bibinfo
  {journal} {The Journal of Chemical Physics}\ }\textbf {\bibinfo {volume}
  {140}},\ \bibinfo {pages} {124103} (\bibinfo {year} {2014})},\ \bibinfo
  {note} {publisher: American Institute of Physics}\BibitemShut {NoStop}%
\bibitem [{\citenamefont {Jung}\ \emph {et~al.}(2017)\citenamefont {Jung},
  \citenamefont {Hanke},\ and\ \citenamefont {Schmid}}]{jung_iterative_2017}%
  \BibitemOpen
  \bibfield  {author} {\bibinfo {author} {\bibfnamefont {G.}~\bibnamefont
  {Jung}}, \bibinfo {author} {\bibfnamefont {M.}~\bibnamefont {Hanke}},\ and\
  \bibinfo {author} {\bibfnamefont {F.}~\bibnamefont {Schmid}},\ }\bibfield
  {title} {\bibinfo {title} {Iterative {Reconstruction} of {Memory}
  {Kernels}},\ }\href {https://doi.org/10.1021/acs.jctc.7b00274} {\bibfield
  {journal} {\bibinfo  {journal} {Journal of Chemical Theory and Computation}\
  }\textbf {\bibinfo {volume} {13}},\ \bibinfo {pages} {2481} (\bibinfo {year}
  {2017})},\ \bibinfo {note} {publisher: American Chemical Society}\BibitemShut
  {NoStop}%
\bibitem [{\citenamefont {Daldrop}\ \emph {et~al.}(2017)\citenamefont
  {Daldrop}, \citenamefont {Kowalik},\ and\ \citenamefont
  {Netz}}]{daldrop_external_2017}%
  \BibitemOpen
  \bibfield  {author} {\bibinfo {author} {\bibfnamefont {J.~O.}\ \bibnamefont
  {Daldrop}}, \bibinfo {author} {\bibfnamefont {B.~G.}\ \bibnamefont
  {Kowalik}},\ and\ \bibinfo {author} {\bibfnamefont {R.~R.}\ \bibnamefont
  {Netz}},\ }\bibfield  {title} {\bibinfo {title} {External {Potential}
  {Modifies} {Friction} of {Molecular} {Solutes} in {Water}},\ }\href
  {https://doi.org/10.1103/PhysRevX.7.041065} {\bibfield  {journal} {\bibinfo
  {journal} {Physical Review X}\ }\textbf {\bibinfo {volume} {7}},\ \bibinfo
  {pages} {041065} (\bibinfo {year} {2017})},\ \bibinfo {note} {publisher:
  American Physical Society}\BibitemShut {NoStop}%
\bibitem [{\citenamefont {Daldrop}\ \emph {et~al.}(2018)\citenamefont
  {Daldrop}, \citenamefont {Kappler}, \citenamefont {Br{\"u}nig},\ and\
  \citenamefont {Netz}}]{daldrop_butane_2018}%
  \BibitemOpen
  \bibfield  {author} {\bibinfo {author} {\bibfnamefont {J.~O.}\ \bibnamefont
  {Daldrop}}, \bibinfo {author} {\bibfnamefont {J.}~\bibnamefont {Kappler}},
  \bibinfo {author} {\bibfnamefont {F.~N.}\ \bibnamefont {Br{\"u}nig}},\ and\
  \bibinfo {author} {\bibfnamefont {R.~R.}\ \bibnamefont {Netz}},\ }\bibfield
  {title} {\bibinfo {title} {Butane dihedral angle dynamics in water is
  dominated by internal friction},\ }\href
  {https://doi.org/10.1073/pnas.1722327115} {\bibfield  {journal} {\bibinfo
  {journal} {Proceedings of the National Academy of Sciences}\ }\textbf
  {\bibinfo {volume} {115}},\ \bibinfo {pages} {5169} (\bibinfo {year}
  {2018})},\ \bibinfo {note} {publisher: National Academy of Sciences Section:
  Biological Sciences}\BibitemShut {NoStop}%
\bibitem [{\citenamefont {Klippenstein}\ and\ \citenamefont {van~der
  Vegt}(2021)}]{klippenstein_cross-correlation_2021}%
  \BibitemOpen
  \bibfield  {author} {\bibinfo {author} {\bibfnamefont {V.}~\bibnamefont
  {Klippenstein}}\ and\ \bibinfo {author} {\bibfnamefont {N.~F.~A.}\
  \bibnamefont {van~der Vegt}},\ }\bibfield  {title} {\bibinfo {title}
  {Cross-correlation corrected friction in (generalized) {Langevin} models},\
  }\href {https://doi.org/10.1063/5.0049324} {\bibfield  {journal} {\bibinfo
  {journal} {The Journal of Chemical Physics}\ }\textbf {\bibinfo {volume}
  {154}},\ \bibinfo {pages} {191102} (\bibinfo {year} {2021})},\ \bibinfo
  {note} {publisher: American Institute of Physics}\BibitemShut {NoStop}%
\end{thebibliography}%

\end{document}